\documentclass[10pt,journal,compsoc,nofonttune]{IEEEtran}

\usepackage{amsmath,amsfonts}
\usepackage{array}
\usepackage[font=footnotesize,labelfont=sf,textfont=sf]{subfig}
\usepackage[font=small]{caption}
\usepackage{textcomp}
\usepackage{stfloats}
\usepackage{url}
\usepackage{verbatim}
\usepackage{graphicx}

\usepackage{url}
\usepackage[utf8]{inputenc}
\usepackage{xcolor}
\usepackage{xspace}
\usepackage{epsfig}
\usepackage{balance}
\usepackage{booktabs}
\usepackage{tabularx}
\usepackage{cite}
\usepackage{mathtools}
\usepackage{soul}
\usepackage{lipsum}
\usepackage{bm}
\usepackage{enumerate}
\usepackage{enumitem}
\usepackage[sort&compress,numbers]{natbib}

\usepackage[]{algorithmic}
\usepackage{algorithm}

\usepackage[acronyms,nonumberlist,nopostdot,nomain,nogroupskip,acronymlists={hidden}]{glossaries}
\newglossary[algh]{hidden}{acrh}{acnh}{Hidden Acronyms}
\glsdisablehyper

\usepackage{tikz}
\usepackage{pgfplots}
\usepackage{glossaries}

\pgfplotsset{compat=newest}
\pgfplotsset{plot coordinates/math parser=false}
\newlength\fheight
\newlength\fwidth
\usetikzlibrary{plotmarks,patterns,decorations.pathreplacing,backgrounds,calc,arrows,arrows.meta,spy,matrix,scopes}
\usepgfplotslibrary{patchplots,groupplots}
\usepackage{tikzscale}
\usepackage[draft]{hyperref}
\newacronym{rmr}{RMR}{RIC Message Router}
\newacronym{3gpp}{3GPP}{3rd Generation Partnership Project}
\newacronym{4g}{4G}{4th generation}
\newacronym{5g}{5G}{Fifth generation}
\newacronym{6g}{6G}{Sixth generation}
\newacronym{5gc}{5GC}{5G Core}
\newacronym{abr}{ABR}{Adaptive Bitrate Streaming}
\newacronym{adc}{ADC}{Analog to Digital Converter}
\newacronym{aerpaw}{AERPAW}{Aerial Experimentation and Research Platform for Advanced Wireless}
\newacronym{ai}{AI}{Artificial Intelligence}
\newacronym{aimd}{AIMD}{Additive Increase Multiplicative Decrease}
\newacronym{am}{AM}{Acknowledged Mode}
\newacronym{amc}{AMC}{Adaptive Modulation and Coding}
\newacronym{amf}{AMF}{Access and Mobility Management Function}
\newacronym{aops}{AOPS}{Adaptive Order Prediction Scheduling}
\newacronym{api}{API}{Application Programming Interface}
\newacronym{apn}{APN}{Access Point Name}
\newacronym{aqm}{AQM}{Active Queue Management}
\newacronym{arc}{ARC}{Aerial RAN CoLab}
\newacronym{arc-ota}{ARC-OTA}{Aerial RAN CoLab Over-the-Air}
\newacronym{asic}{ASIC}{Application-Specific Integrated Circuit}
\newacronym{ausf}{AUSF}{Authentication Server Function}
\newacronym{avc}{AVC}{Advanced Video Coding}
\newacronym{awgn}{AGWN}{Additive White Gaussian Noise}
\newacronym{balia}{BALIA}{Balanced Link Adaptation Algorithm}
\newacronym{bbu}{BBU}{Base Band Unit}
\newacronym{bdp}{BDP}{Bandwidth-Delay Product}
\newacronym{ber}{BER}{Bit Error Rate}
\newacronym{bf}{BF}{Beamforming}
\newacronym{bler}{BLER}{Block Error Rate}
\newacronym{bom}{BoM}{Bill of Materials}
\newacronym{brr}{BRR}{Bayesian Ridge Regressor}
\newacronym{bsr}{BSR}{Buffer Status Report}
\newacronym{bs}{BS}{Base Station}
\newacronym{bpsk}{BPSK}{Binary Phase-shift keying}
\newacronym{bss}{BSS}{Business Support System}
\newacronym{ca}{CA}{Carrier Aggregation}
\newacronym{caas}{CaaS}{Connectivity-as-a-Service}
\newacronym{cast}{\textit{CaST}}{Channel emulation generator and Sounder Toolchain}
\newacronym{cb}{CB}{Code Block}
\newacronym{cbrs}{CBRS}{Citizen Broadband Radio Service}
\newacronym{cc}{CC}{Congestion Control}
\newacronym{ccid}{CCID}{Congestion Control ID}
\newacronym{cco}{CC}{Carrier Component}
\newacronym{cdd}{CDD}{Cyclic Delay Diversity}
\newacronym{cdf}{CDF}{Cumulative Distribution Function}
\newacronym{cdn}{CDN}{Content Distribution Network}
\newacronym{cir}{CIR}{Channel Impulse Response}
\newacronym{cn}{CN}{Core Network}
\newacronym{codel}{CoDel}{Controlled Delay Management}
\newacronym{comac}{COMAC}{Converged Multi-Access and Core}
\newacronym{cord}{CORD}{Central Office Re-architected as a Datacenter}
\newacronym{cornet}{CORNET}{COgnitive Radio NETwork}
\newacronym{cosmos}{COSMOS}{Cloud Enhanced Open Software Defined Mobile Wireless Testbed for City-Scale Deployment}
\newacronym{cots}{COTS}{Commercial Off-the-Shelf}
\newacronym{cp}{CP}{Control Plane}
\newacronym{cpu}{CPU}{Central Processing Unit}
\newacronym{cqi}{CQI}{Channel Quality Information}
\newacronym{cr}{CR}{Cognitive Radio}
\newacronym{cran}{CRAN}{Cloud \gls{ran}}
\newacronym{crs}{CRS}{Cell Reference Signal}
\newacronym{csi}{CSI}{Channel State Information}
\newacronym{csirs}{CSI-RS}{Channel State Information - Reference Signal}
\newacronym{cu}{CU}{Central Unit}
\newacronym{d2tcp}{D$^2$TCP}{Deadline-aware Data center TCP}
\newacronym{d3}{D$^3$}{Deadline-Driven Delivery}
\newacronym{dac}{DAC}{Digital to Analog Converter}
\newacronym{dag}{DAG}{Directed Acyclic Graph}
\newacronym{darpa}{DARPA}{Defense Advanced Research Projects Agency}
\newacronym{dapp}{dApp}{Distributed RAN Application}
\newacronym{das}{DAS}{Distributed Antenna System}
\newacronym{dash}{DASH}{Dynamic Adaptive Streaming over HTTP}
\newacronym{dc}{DC}{Dual Connectivity}
\newacronym{dccp}{DCCP}{Datagram Congestion Control Protocol}
\newacronym{dce}{DCE}{Direct Code Execution}
\newacronym{dci}{DCI}{Downlink Control Information}
\newacronym{dcl}{DCL}{Dear Colleague Letter}
\newacronym{dctcp}{DCTCP}{Data Center TCP}
\newacronym{dl}{DL}{Downlink}
\newacronym{dmr}{DMR}{Deadline Miss Ratio}
\newacronym{dmrs}{DMRS}{DeModulation Reference Signal}
\newacronym{dpu}{DPU}{Data Processing Unit}
\newacronym{drlcc}{DRL-CC}{Deep Reinforcement Learning Congestion Control}
\newacronym{drs}{DRS}{Discovery Reference Signal}
\newacronym{dsp}{DSP}{Digital Signal Processing}
\newacronym{du}{DU}{Distributed Unit}
\newacronym{e2e}{E2E}{end-to-end}
\newacronym{e2sm}{E2SM}{E2 Service Model}
\newacronym{ecaas}{ECaaS}{Edge-Cloud-as-a-Service}
\newacronym{ecn}{ECN}{Explicit Congestion Notification}
\newacronym{edf}{EDF}{Earliest Deadline First}
\newacronym{eirp}{EIRP}{Effective Isotropic Radiated Power}
\newacronym{em}{EM}{Electro-Magnetic}
\newacronym{embb}{eMBB}{Enhanced Mobile Broadband}
\newacronym{empower}{EMPOWER}{EMpowering transatlantic PlatfOrms for advanced WirEless Research}
\newacronym{enb}{eNB}{evolved Node Base}
\newacronym{endc}{EN-DC}{E-UTRAN-\gls{nr} \gls{dc}}
\newacronym{epc}{EPC}{Evolved Packet Core}
\newacronym{eps}{EPS}{Evolved Packet System}
\newacronym{es}{ES}{Edge Server}
\newacronym{etsi}{ETSI}{European Telecommunications Standards Institute}
\newacronym[firstplural=Estimated Times of Arrival (ETAs)]{eta}{ETA}{Estimated Time of Arrival}
\newacronym{eutran}{E-UTRAN}{Evolved Universal Terrestrial Access Network}
\newacronym{faas}{FaaS}{Function-as-a-Service}
\newacronym{fapi}{FAPI}{Functional Application Platform Interface}
\newacronym{fcc}{FCC}{Federal Communications Commission}
\newacronym{fdd}{FDD}{Frequency Division Duplexing}
\newacronym{fdm}{FDM}{Frequency Division Multiplexing}
\newacronym{fdma}{FDMA}{Frequency Division Multiple Access}
\newacronym{fed4fire}{FED4FIRE+}{Federation 4 Future Internet Research and Experimentation Plus}
\newacronym{fir}{FIR}{Finite Impulse Response}
\newacronym{fit}{FIT}{Future \acrlong{iot}}
\newacronym{fpga}{FPGA}{Field Programmable Gate Array}
\newacronym{fr2}{FR2}{Frequency Range 2}
\newacronym{frand}{FRAND}{Fair, Reasonable, And Non-Discriminatory}
\newacronym{fs}{FS}{Fast Switching}
\newacronym{fscc}{FSCC}{Flow Sharing Congestion Control}
\newacronym{ftp}{FTP}{File Transfer Protocol}
\newacronym{fw}{FW}{Flow Window}
\newacronym{ga128}{Ga}{Golay Sequence type A}
\newacronym{ge}{GE}{Gaussian Elimination}
\newacronym{gh}{GH}{Grace Hopper}
\newacronym{glfsr}{GLFSR}{Galois Linear Feedback Shift Register}
\newacronym{gnb}{gNB}{Next Generation Node Base}
\newacronym{gold}{Gold}{Gold}
\newacronym{gop}{GOP}{Group of Pictures}
\newacronym{gpr}{GPR}{Gaussian Process Regressor}
\newacronym{gpu}{GPU}{Graphics Processing Unit}
\newacronym{gtp}{GTP}{GPRS Tunneling Protocol}
\newacronym{gtpc}{GTP-C}{GPRS Tunnelling Protocol Control Plane}
\newacronym{gtpu}{GTP-U}{GPRS Tunnelling Protocol User Plane}
\newacronym{gtpv2c}{GTPv2-C}{\gls{gtp} v2 - Control}
\newacronym{gw}{GW}{Gateway}
\newacronym{harq}{HARQ}{Hybrid Automatic Repeat reQuest}
\newacronym{hdr}{HDR}{High Dynamic Range}
\newacronym{hetnet}{HetNet}{Heterogeneous Network}
\newacronym{hh}{HH}{Hard Handover}
\newacronym{hol}{HOL}{Head-of-Line}
\newacronym{hqf}{HQF}{Highest-quality-first}
\newacronym{hss}{HSS}{Home Subscription Server}
\newacronym{http}{HTTP}{HyperText Transfer Protocol}
\newacronym{ia}{IA}{Initial Access}
\newacronym{iab}{IAB}{Integrated Access and Backhaul}
\newacronym{ic}{IC}{Incident Command}
\newacronym{ietf}{IETF}{Internet Engineering Task Force}
\newacronym{ifw}{IFW}{Interference Free Window}
\newacronym{imsi}{IMSI}{International Mobile Subscriber Identity}
\newacronym{imt}{IMT}{International Mobile Telecommunication}
\newacronym{io}{I/O}{Input/Output}
\newacronym{iot}{IoT}{Internet of Things}
\newacronym{ip}{IP}{Internet Protocol}
\newacronym{ipc}{IPC}{Inter-Process Communication}
\newacronym{iq}{IQ}{In-phase and Quadrature}
\newacronym{itu}{ITU}{International Telecommunication Union}
\newacronym{kpi}{KPI}{Key Performance Indicator}
\newacronym{kpm}{KPM}{Key Performance Measurement}
\newacronym{kvm}{KVM}{Kernel-based Virtual Machine}
\newacronym{ldpc}{LDPC}{Low-Density Parity-Check}
\newacronym{leo}{LEO}{Low Earth Orbit}
\newacronym{los}{LOS}{Line-of-Sight}
\newacronym{ls}{LS}{Loosely Synchronised}
\newacronym{lsm}{LSM}{Link-to-System Mapping}
\newacronym{lstm}{LSTM}{Long Short Term Memory}
\newacronym{lte}{LTE}{Long Term Evolution}
\newacronym{lxc}{LXC}{Linux Container}
\newacronym{m2m}{M2M}{Machine to Machine}
\newacronym{mac}{MAC}{Medium Access Control}
\newacronym{manet}{MANET}{Mobile Ad Hoc Network}
\newacronym{mano}{MANO}{Management and Orchestration}
\newacronym{mc}{MC}{Multi-Connectivity}
\newacronym{mcc}{MCC}{Mobile Cloud Computing}
\newacronym{mchem}{MCHEM}{Massive Channel Emulator}
\newacronym{mcs}{MCS}{Modulation and Coding Scheme}
\newacronym{mec}{MEC}{Multi-access Edge Computing}
\newacronym{mec2}{MEC}{Mobile Edge Cloud}
\newacronym{mfc}{MFC}{Mobile Fog Computing}
\newacronym{mi}{MI}{Mutual Information}
\newacronym{mib}{MIB}{Master Information Block}
\newacronym{miesm}{MIESM}{Mutual Information Based Effective SINR}
\newacronym{mimo}{MIMO}{Multiple Input, Multiple Output}
\newacronym{mgen}{MGEN}{Multi-Generator}
\newacronym{ml}{ML}{Machine Learning}
\newacronym{mlr}{MLR}{Maximum-local-rate}
\newacronym[plural=\gls{mme}s,firstplural=Mobility Management Entities (MMEs)]{mme}{MME}{Mobility Management Entity}
\newacronym{mmtc}{mMTC}{Massive Machine-Type Communications}
\newacronym{mmwave}{mmWave}{millimeter wave}
\newacronym{mpdccp}{MP-DCCP}{Multipath Datagram Congestion Control Protocol}
\newacronym{mptcp}{MPTCP}{Multipath TCP}
\newacronym{mr}{MR}{Maximum Rate}
\newacronym{mrdc}{MR-DC}{Multi \gls{rat} \gls{dc}}
\newacronym{mse}{MSE}{Mean Square Error}
\newacronym{mss}{MSS}{Maximum Segment Size}
\newacronym{mt}{MT}{Mobile Termination}
\newacronym{mtd}{MTD}{Machine-Type Device}
\newacronym{mtu}{MTU}{Maximum Transmission Unit}
\newacronym{mumimo}{MU-MIMO}{Multi-user \gls{mimo}}
\newacronym{mvno}{MVNO}{Mobile Virtual Network Operator}
\newacronym{nalu}{NALU}{Network Abstraction Layer Unit}
\newacronym{nas}{NAS}{Network Attached Storage}
\newacronym{nbiot}{NB-IoT}{Narrow Band IoT}
\newacronym{nfv}{NFV}{Network Function Virtualization}
\newacronym{nfvi}{NFVI}{Network Function Virtualization Infrastructure}
\newacronym{nic}{NIC}{Network Interface Card}
\newacronym{nlos}{NLOS}{Non-Line-of-Sight}
\newacronym{now}{NOW}{Non Overlapping Window}
\newacronym{nrdz}{NRDZ}{National Radio Dynamic Zone}
\newacronym{nsf}{NSF}{National Science Foundation}
\newacronym{nsm}{NSM}{Network Service Mesh}
\newacronym[type=hidden]{nr}{NR}{New Radio}
\newacronym{nrf}{NRF}{Network Repository Function}
\newacronym{nsa}{NSA}{Non Stand Alone}
\newacronym{nse}{NSE}{Network Slicing Engine}
\newacronym{nssf}{NSSF}{Network Slice Selection Function}
\newacronym{ntp}{NTP}{Network Time Protocol}
\newacronym{nvipc}{NVIPC}{NVIDIA Inter-Process Communication}
\newacronym{o2i}{O2I}{Outdoor to Indoor}
\newacronym{oai}{OAI}{OpenAirInterface}
\newacronym{oaic}{OAIC}{Open AI Cellular}
\newacronym{oaicn}{OAI-CN}{\gls{oai} \acrlong{cn}}
\newacronym{oairan}{OAI-RAN}{\acrlong{oai} \acrlong{ran}}
\newacronym{oam}{OAM}{Operations, Administration and Maintenance}
\newacronym[plural=\gls{obu}s,firstplural=Onboard Units (OBUs)]{obu}{OBU}{Onboard Unit}
\newacronym{ofdm}{OFDM}{Orthogonal Frequency Division Multiplexing}
\newacronym{olia}{OLIA}{Opportunistic Linked Increase Algorithm}
\newacronym{omec}{OMEC}{Open Mobile Evolved Core}
\newacronym{onap}{ONAP}{Open Network Automation Platform}
\newacronym{onf}{ONF}{Open Networking Foundation}
\newacronym{onos}{ONOS}{Open Networking Operating System}
\newacronym{oom}{OOM}{\gls{onap} Operations Manager}
\newacronym{opnfv}{OPNFV}{Open Platform for \gls{nfv}}
\newacronym{orbit}{ORBIT}{Open-Access Research Testbed for Next-Generation Wireless Networks}
\newacronym{os}{OS}{Operating System}
\newacronym{osc}{OSC}{O-RAN Software Community}
\newacronym{osm}{OSM}{Open Street Map}
\newacronym{oss}{OSS}{Operations Support System}
\newacronym{ota}{OTA}{Over-the-Air}
\newacronym{p5g}{P5G}{Private 5G}
\newacronym{pa}{PA}{Position-aware}
\newacronym{pase}{PASE}{Prioritization, Arbitration, and Self-adjusting Endpoints}
\newacronym{pawr}{PAWR}{Platforms for Advanced Wireless Research}
\newacronym{pbch}{PBCH}{Physical Broadcast Channel}
\newacronym{pci}{PCI}{Peripheral Component Interconnect}
\newacronym{pcef}{PCEF}{Policy and Charging Enforcement Function}
\newacronym{pcfich}{PCFICH}{Physical Control Format Indicator Channel}
\newacronym{pcrf}{PCRF}{Policy and Charging Rules Function}
\newacronym{pdcch}{PDCCH}{Physical Downlink Control Channel}
\newacronym{pdcp}{PDCP}{Packet Data Convergence Protocol}
\newacronym{pdsch}{PDSCH}{Physical Downlink Shared Channel}
\newacronym{pdu}{PDU}{Packet Data Unit}
\newacronym{pdp}{PDP}{Power Delay Profile}
\newacronym{pf}{PF}{Proportional Fair}
\newacronym{pgw}{PGW}{Packet Gateway}
\newacronym{ph}{PH}{Power Headroom}
\newacronym{phich}{PHICH}{Physical Hybrid ARQ Indicator Channel}
\newacronym{phy}{PHY}{Physical}
\newacronym{pl}{PL}{Path Loss}
\newacronym{pmch}{PMCH}{Physical Multicast Channel}
\newacronym{pmi}{PMI}{Precoding Matrix Indicators}
\newacronym{powder}{POWDER}{Platform for Open Wireless Data-driven Experimental Research}
\newacronym{ppo}{PPO}{Proximal Policy Optimization}
\newacronym{ppp}{PPP}{Poisson Point Process}
\newacronym{prach}{PRACH}{Physical Random Access Channel}
\newacronym{prb}{PRB}{Physical Resource Block}
\newacronym{psnr}{PSNR}{Peak Signal to Noise Ratio}
\newacronym{pss}{PSS}{Primary Synchronization Signal}
\newacronym{ptp}{PTP}{Precision Timing Protocol}
\newacronym{pucch}{PUCCH}{Physical Uplink Control Channel}
\newacronym{pusch}{PUSCH}{Physical Uplink Shared Channel}
\newacronym{qam}{QAM}{Quadrature Amplitude Modulation}
\newacronym{qci}{QCI}{\gls{qos} Class Identifier}
\newacronym{qoe}{QoE}{Quality of Experience}
\newacronym{qos}{QoS}{Quality of Service}
\newacronym{qtgui}{QT-GUI}{QT Graphical User Interface}
\newacronym{qsfp28}{QSFP28}{Quad Small Form-factor Pluggable 28}
\newacronym{quic}{QUIC}{Quick UDP Internet Connections}
\newacronym{rach}{RACH}{Random Access Channel}
\newacronym{ran}{RAN}{Radio Access Network}
\newacronym[firstplural=Radio Access Technologies (RATs)]{rat}{RAT}{Radio Access Technology}
\newacronym{rcn}{RCN}{Research Coordination Network}
\newacronym{rec}{REC}{Radio Edge Cloud}
\newacronym{red}{RED}{Random Early Detection}
\newacronym{renew}{RENEW}{Reconfigurable Eco-system for Next-generation End-to-end Wireless}
\newacronym{rf}{RF}{Radio Frequency}
\newacronym{rfc}{RFC}{Request for Comments}
\newacronym{rfr}{RFR}{Random Forest Regressor}
\newacronym{ric}{RIC}{RAN Intelligent Controller}
\newacronym{near-rt-ric}{near-RT-RIC}{near-RT-\gls{ric}}
\newacronym{rlc}{RLC}{Radio Link Control}
\newacronym{rlf}{RLF}{Radio Link Failure}
\newacronym{rlnc}{RLNC}{Random Linear Network Coding}
\newacronym{rmse}{RMSE}{Root Mean Squared Error}
\newacronym{rnis}{RNIS}{Radio Network Information Service}
\newacronym{rr}{RR}{Round Robin}
\newacronym{rrc}{RRC}{Radio Resource Control}
\newacronym{rrm}{RRM}{Radio Resource Management}
\newacronym{rru}{RRU}{Remote Radio Unit}
\newacronym{rs}{RS}{Remote Server}
\newacronym{rsrp}{RSRP}{Reference Signal Received Power}
\newacronym{rsrq}{RSRQ}{Reference Signal Received Quality}
\newacronym{rss}{RSS}{Received Signal Strength}
\newacronym{rssi}{RSSI}{Received Signal Strength Indicator}
\newacronym{rsu}{RSU}{Road-Side Unit}
\newacronym{rtt}{RTT}{Round Trip Time}
\newacronym{ru}{RU}{Radio Unit}
\newacronym{rw}{RW}{Receive Window}
\newacronym{rx}{RX}{Receiver}
\newacronym{s1ap}{S1AP}{S1 Application Protocol}
\newacronym{sa}{SA}{standalone}
\newacronym{sack}{SACK}{Selective Acknowledgment}
\newacronym{sap}{SAP}{Service Access Point}
\newacronym{sas}{SAS}{Spectrum Access System}
\newacronym{sc2}{SC2}{Spectrum Collaboration Challenge}
\newacronym{scef}{SCEF}{Service Capability Exposure Function}
\newacronym{sch}{SCH}{Secondary Cell Handover}
\newacronym{scoot}{SCOOT}{Split Cycle Offset Optimization Technique}
\newacronym{sfp+}{SFP+}{Small Form-factor Pluggable Plus}
\newacronym{scf}{SCF}{Small Cell Forum}\newacronym{sctp}{SCTP}{Stream Control Transmission Protocol}
\newacronym{sdap}{SDAP}{Service Data Adaptation Protocol}
\newacronym{sd}{SD}{Standard Deviation}
\newacronym{sdk}{SDK}{Software Development Kit}
\newacronym{sdm}{SDM}{Space Division Multiplexing}
\newacronym{sdma}{SDMA}{Spatial Division Multiple Access}
\newacronym{sdn}{SDN}{Software-defined Networking}
\newacronym{sdr}{SDR}{Software-defined Radio}
\newacronym{seba}{SEBA}{SDN-Enabled Broadband Access}
\newacronym{sgsn}{SGSN}{Serving GPRS Support Node}
\newacronym{sgw}{SGW}{Service Gateway}
\newacronym{si}{SI}{Study Item}
\newacronym{sib}{SIB}{Secondary Information Block}
\newacronym{sinr}{SINR}{Signal to Interference plus Noise Ratio}
\newacronym{sip}{SIP}{Session Initiation Protocol}
\newacronym{siso}{SISO}{Single Input, Single Output}
\newacronym{sla}{SLA}{Service Level Agreement}
\newacronym{sm}{SM}{Saturation Mode}
\newacronym{smf}{SMF}{Session Management Function}
\newacronym{smo}{SMO}{Service Management and Orchestration}
\newacronym{sms}{SMS}{Short Message Service}
\newacronym{smsgmsc}{SMS-GMSC}{\gls{sms}-Gateway}
\newacronym{snr}{SNR}{Signal-to-Noise-Ratio}
\newacronym{son}{SON}{Self-Organizing Network}
\newacronym{sptcp}{SPTCP}{Single Path TCP}
\newacronym{srb}{SRB}{Service Radio Bearer}
\newacronym{srn}{SRN}{Standard Radio Node}
\newacronym{srs}{SRS}{Sounding Reference Signal}
\newacronym{ss}{SS}{Synchronization Signal}
\newacronym{sss}{SSS}{Secondary Synchronization Signal}
\newacronym{st}{ST}{Spanning Tree}
\newacronym{svc}{SVC}{Scalable Video Coding}
\newacronym{synce}{SyncE}{Synchronous Ethernet}
\newacronym{tb}{TB}{Transport Block}
\newacronym{tcp}{TCP}{Transmission Control Protocol}
\newacronym{tdd}{TDD}{Time Division Duplexing}
\newacronym{tdm}{TDM}{Time Division Multiplexing}
\newacronym{tdma}{TDMA}{Time Division Multiple Access}
\newacronym{tfl}{TfL}{Transport for London}
\newacronym{tfrc}{TFRC}{TCP-Friendly Rate Control}
\newacronym{tft}{TFT}{Traffic Flow Template}
\newacronym{tgen}{TGEN}{Traffic Generator}
\newacronym{tip}{TIP}{Telecom Infra Project}
\newacronym{tm}{TM}{Transparent Mode}
\newacronym{to}{TO}{Telco Operator}
\newacronym{toa}{ToA}{Time of Arrival}
\newacronym{tr}{TR}{Technical Report}
\newacronym{trp}{TRP}{Transmitter Receiver Pair}
\newacronym{ts}{TS}{Technical Specification}
\newacronym{tti}{TTI}{Transmission Time Interval}
\newacronym{ttt}{TTT}{Time-to-Trigger}
\newacronym{tx}{TX}{Transmitter}
\newacronym{uas}{UAS}{Unmanned Aerial System}
\newacronym{uav}{UAV}{Unmanned Aerial Vehicle}
\newacronym{uci}{UCI}{Uplink Control Indication}
\newacronym{udm}{UDM}{Unified Data Management}
\newacronym{udp}{UDP}{User Datagram Protocol}
\newacronym{udr}{UDR}{Unified Data Repository}
\newacronym{ue}{UE}{User Equipment}
\newacronym{uhd}{UHD}{\gls{usrp} Hardware Driver}
\newacronym{ul}{UL}{Uplink}
\newacronym{um}{UM}{Unacknowledged Mode}
\newacronym{uml}{UML}{Unified Modeling Language}
\newacronym{upa}{UPA}{Uniform Planar Array}
\newacronym{upf}{UPF}{User Plane Function}
\newacronym{urllc}{URLLC}{Ultra Reliable and Low Latency Communication}
\newacronym{usa}{U.S.}{United States}
\newacronym{usim}{USIM}{Universal Subscriber Identity Module}
\newacronym{usrp}{USRP}{Universal Software Radio Peripheral}
\newacronym{utc}{UTC}{Urban Traffic Control}
\newacronym{vim}{VIM}{Virtualization Infrastructure Manager}
\newacronym{vlan}{VLAN}{Virtual Local Area Network}
\newacronym{vm}{VM}{Virtual Machine}
\newacronym{vnf}{VNF}{Virtual Network Function}
\newacronym{volte}{VoLTE}{Voice over \gls{lte}}
\newacronym{voltha}{VOLTHA}{Virtual OLT HArdware Abstraction}
\newacronym{vr}{VR}{Virtual Reality}
\newacronym{vran}{vRAN}{Virtualized \gls{ran}}
\newacronym{vss}{VSS}{Video Streaming Server}
\newacronym{wbf}{WBF}{Wired Bias Function}
\newacronym{wf}{WF}{Wired-first}
\newacronym{wi}{WI}{Wireless InSite}
\newacronym{wlan}{WLAN}{Wireless Local Area Network}
\newacronym{xapp}{xApp}{RAN Application}
\newacronym{pnf}{PNF}{Physical Network Function}
\newacronym{drl}{DRL}{Deep Reinforcement Learning}
\newacronym{mtc}{MTC}{Machine-type Communications}
\newacronym{v2x}{V2X}{Vehicle-to-everything}

\newif\ifexttikz
\exttikzfalse

\ifexttikz
  \usetikzlibrary{external}
\fi

\newif\ifoverleaf
\ifthenelse{\equal{\jobname}{\detokenize{output}}}
    {\overleaftrue}
    {\overleaffalse}%

\newcommand{\testbed}{X5G\xspace}

\newcommand{\blue}[1]{{#1}}

\ifexttikz
\else
\usepackage{tikzpagenodes,etoolbox}
\usetikzlibrary{calc}
\usepackage[contents={}]{background}
\AddEverypageHook{%
\ifnumequal{\thepage}{1}{%
    \tikz[remember picture,overlay]{%
        \node[draw,
        minimum width=1.03\textwidth,
        text width=0.95\textwidth,
        font=\scriptsize
        ]
        at ($(current page header area) - (0,5pt)$)
        {%
        This work has been accepted for publication on IEEE Transactions on Mobile Computing.\\
        ©2025 IEEE. Personal use of this material is permitted. Permission from IEEE must be obtained for all other uses, in any current or future media, including reprinting/republishing this material for advertising or promotional purposes, creating new collective works, for resale or redistribution to servers or lists, or reuse of any copyrighted component of this work in other works.
        };
    }%
}{}
}
\fi

\begin{document}
\bstctlcite{BSTcontrol}





\title{X5G: An Open, Programmable, Multi-vendor, End-to-end, Private 5G O-RAN Testbed with NVIDIA ARC and OpenAirInterface}

\author{\IEEEauthorblockN{
Davide Villa,
Imran Khan, 
Florian Kaltenberger, 
Nicholas Hedberg, 
R\'{u}ben Soares da Silva,\\
Stefano Maxenti,
Leonardo Bonati, 
Anupa Kelkar,
Chris Dick,
Eduardo Baena,\\
\hspace{1.8em}Josep M. Jornet,
Tommaso Melodia,
Michele Polese, 
and Dimitrios Koutsonikolas
}
\thanks{This is a revised and substantially extended version of~\cite{villa2024x5g}, which appeared in the Proceedings of the 2nd Workshop on Next-generation Open and Programmable Radio Access Networks (NG-OPERA) 2024.}
\thanks{This work was partially supported by the U.S.\ National Science Foundation under grant CNS-2117814, and by the U.S. National Telecommunications and Information Administration (NTIA)'s Public Wireless Supply Chain Innovation Fund (PWSCIF) under Award No. 25-60-IF054.}
\thanks{D. Villa, I. Khan, F. Kaltenberger, S. Maxenti, L. Bonati, E. Baena, J. M. Jornet, T. Melodia, M. Polese, and D. Koutsonikolas are with the Institute for the Wireless Internet of Things, Northeastern University, Boston, MA. Email: \{villa.d, khan.i, f.kaltenberger, maxenti.s, l.bonati, e.baena, j.jornet, melodia, m.polese, d.koutsonikolas\}@northeastern.edu}
\thanks{F. Kaltenberger is also with Eurecom, Sophia Antipolis, France. Email: florian.kaltenberger@eurecom.fr}
\thanks{N. Hedberg, A. Kelkar, and C. Dick are with NVIDIA, Inc., Santa Clara, CA. Email: \{nhedberg, anupak, cdick\}@nvidia.com}
\thanks{R. Soares da Silva is with Allbesmart, Castelo Branco, Portugal. Email: rsilva@allbesmart.pt}
\thanks{D. Villa and I. Khan are co-primary authors.}
}

\makeatletter
\patchcmd{\@maketitle}
  {\addvspace{0.5\baselineskip}\egroup}
  {\addvspace{-1.5\baselineskip}\egroup}
  {}
  {}
\makeatother

\IEEEoverridecommandlockouts

\maketitle

\glsunset{nr}

\begin{abstract}
As \gls{5g} cellular systems transition to softwarized, programmable, and intelligent networks, it becomes fundamental to enable public and private \gls{5g} deployments that are (i) primarily based on software components while (ii) maintaining or exceeding the performance of traditional monolithic systems and (iii) enabling programmability through bespoke configurations and optimized deployments. 
This requires hardware acceleration to scale the \gls{phy} layer performance, programmable elements in the \gls{ran} and intelligent controllers at the edge, careful planning of the \gls{rf} environment, as well as end-to-end integration and testing. 
In this paper, we describe how we developed the programmable \testbed testbed, addressing these challenges through the deployment of the first 8-node network based on the integration of NVIDIA \blue{\gls{arc-ota}}, \gls{oai}, and a near-real-time \gls{ric}. 
The Aerial \gls{sdk} provides the \gls{phy} layer, accelerated on \gls{gpu}, with the higher layers from the \gls{oai} open-source project interfaced with the \gls{phy} through the \gls{scf} \gls{fapi}. An E2 agent provides connectivity to the \gls{osc} near-real-time \gls{ric}.
%
We discuss software integration, network infrastructure, and a digital twin framework for \gls{rf} planning. We then profile the performance with up to 4~\gls{cots} smartphones for each base station with iPerf and video streaming applications, \blue{as well as up to 25 emulated \glspl{ue}}, measuring a cell rate higher than \blue{$1.65$\:Gbps} in downlink and \blue{$143$\:Mbps} in uplink.
\end{abstract}

\begin{IEEEkeywords} 
Private 5G; Multi-vendor; GPU acceleration; O-RAN.
\end{IEEEkeywords}

\maketitle


\glsresetall
\glsunset{usrp}
\glsunset{uhd}
\glsunset{mimo}
\glsunset{arc}


\section{Introduction}
\label{sec:intro}


The evolution of the \gls{ran} in \gls{5g} networks has led to key performance improvements in cell and user data rates, now at hundreds of Mbps on average, and in air interface latency~\cite{aarayanan2022comparative}, thanks to specifications developed within the \gls{3gpp}. 
From an architectural point of view, 
\gls{5g} deployments are also becoming more open, intelligent, programmable, and based on software~\cite{polese2023understanding}, through activities led by the O-RAN ALLIANCE, which is developing the network architecture for Open \gls{ran}.
These elements have the potential to transform how we deploy and manage wireless mobile networks~\cite{bonati2023neutran}, leveraging intelligent control, with \gls{ran} optimization and automation exercised via closed-loop data-driven control; softwarization, with the components of the end-to-end protocol stack defined through software rather than with dedicated hardware; and disaggregation, with the \gls{5g} \gls{ran} layers distributed across different network functions, i.e., the \gls{cu}, the \gls{du}, and the \gls{ru}. 

Open and programmable networks are often associated with lower capital and operational expenditures, facilitated by the increasing robustness and diversity of the telecom supply chain~\cite{dellOroRAN}, now also including open-source projects~\cite{kaltenberger2020openairinterface,gomez2016srs} and vendors focused on specific components of the disaggregated \gls{ran}. This, and increased spectrum availability in dedicated or shared bands, has opened opportunities to deploy private \gls{5g} systems, complementing public \gls{5g} networks with more agile and dynamic deployments for site-specific use cases (e.g., events, warehouse automation, industrial control, etc).

While the transition to disaggregated, software-based, and programmable networks comes with significant benefits, there are also several challenges that need to be addressed before Open \gls{ran} systems can align their performance or improve over traditional cellular systems.
First of all, the radio domain still exhibits a low degree of automation and zero-touch provisioning for the \gls{ran} configuration, complicating the successful deployment of end-to-end cellular systems. Second, the diverse vendor ecosystem comes with challenges related to interoperability and end-to-end integration across several products, potentially from different vendors~\cite{5gtesting,bahl2023accelerating,tang2023ai}. Third, the \gls{dsp} at the \gls{phy} layer of the stack is a computationally complex element, using about $90$\% of the available compute when run on general-purpose CPUs, and thus introducing a burden on the software-based and virtualized \gls{5g} stack components. Finally, there are still open questions in terms of how the intelligent and data-driven control loops can be implemented with \gls{ai} and \gls{ml} solutions that generalize well across a multitude of cellular network scenarios~\cite{fiandrino2023explora}.
These challenges call for a concerted effort across different communities (including hardware, \gls{dsp}, software, DevOps, \gls{ai}/\gls{ml}) that aims to design and deploy open, programmable, multi-vendor cellular networks and testbeds that can support private \gls{5g} requirements and use cases with the stability and performance of production-level systems.
%
\begin{figure*}[t]
    \centering
    \includegraphics[width=\linewidth]{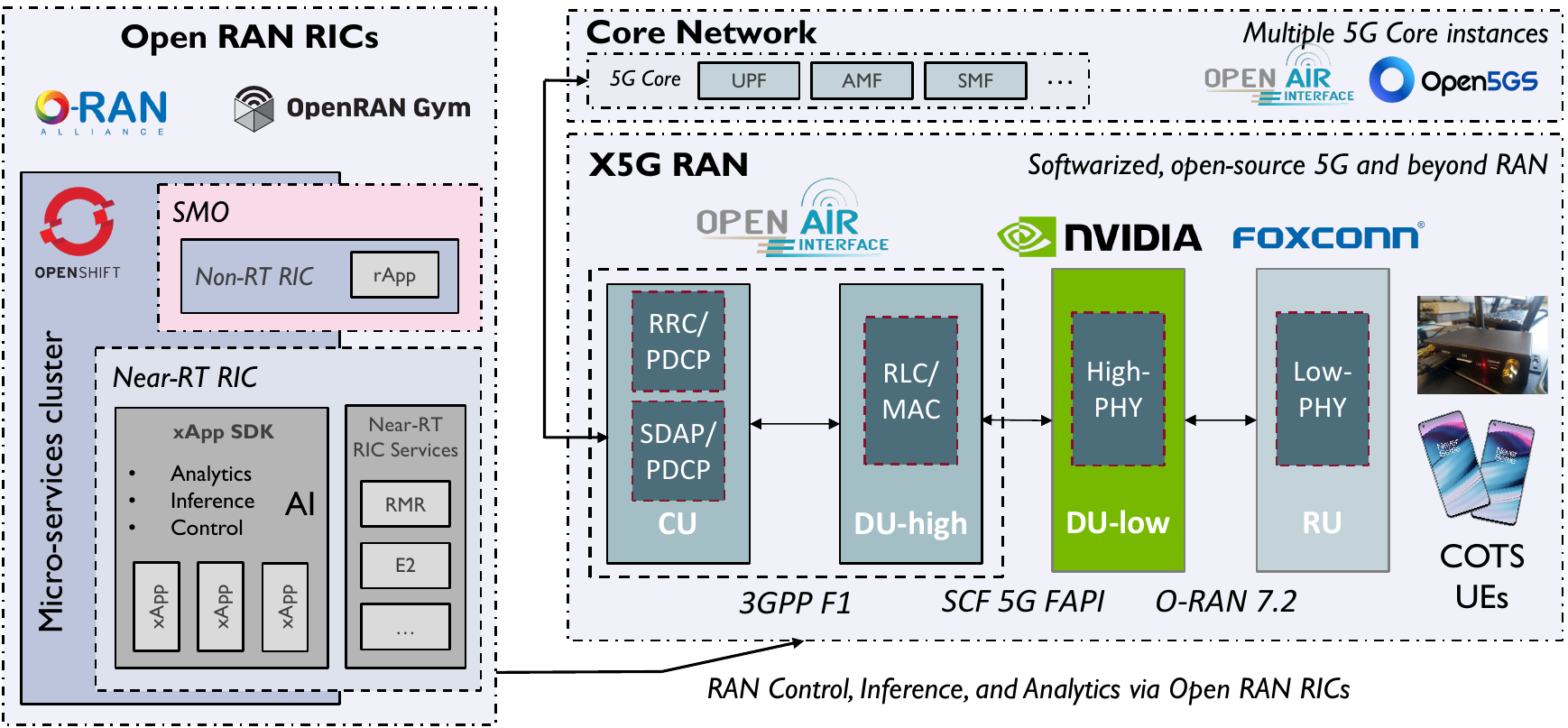}
    \caption{\testbed end-to-end programmable testbed overview.}
    \label{fig:x5g-e2e-overview}
    \vspace{-14pt}
\end{figure*}

In this paper, we introduce \testbed, a private \gls{5g} network testbed deployed at Northeastern University in Boston, MA, and based on multiple programmable and open-source components from the physical layer all the way up to the \gls{cn}, as shown in Figure~\ref{fig:x5g-e2e-overview}.
%
We discuss in detail the integration of a \gls{phy} layer implemented on \gls{gpu} (i.e., NVIDIA Aerial) with \gls{oai} for the higher layers of the 5G stack~\cite{kaltenberger2024driving}. This integration is based on the \gls{scf} \gls{fapi}, which regulates the interaction between the \gls{phy} and \gls{mac} layers.
\blue{This paper extends our recent work~\cite{villa2024x5g} by introducing new Open \gls{ran} elements and experimental results, including: (i) the integration of a near-real-time \gls{ric} from the \gls{osc} on an OpenShift cluster; (ii) the validation of additional \glspl{cn}, such as Open5GS and a commercial core from A5G; (iii) enhancements to the hardware architecture, including a more robust networking infrastructure and additional \gls{ran} servers; (iv) the evaluation of \testbed under diverse operational conditions, such as stress testing its performance ensuring reliability for a \gls{p5g} network; and (v) a more comprehensive related work section.}

\testbed leverages the inline acceleration of demanding \gls{phy} tasks on \gls{gpu}, hardware that is well equipped with massive parallelization of \gls{dsp} operations, enabling scalability and the embedding of \gls{ai}/\gls{ml} in the \gls{ran}.
The \testbed infrastructure \blue{is continuously expanding through the integration of an increasing number of components from various vendors, manufacturers, and open-source projects---such as NVIDIA, \gls{oai}, OpenShift, Keysight, \gls{osc}, Open5GS, and Foxconn---thereby creating a truly multi-vendor network architecture. It currently comprises more than 8~\gls{ran}} servers for the NVIDIA/\gls{oai} \gls{cu} and \gls{du} (known as NVIDIA \gls{arc-ota} and referred to as \gls{arc} in this paper), \blue{several \glspl{ru} from different vendors} that can be installed in a lab space, as well as a Keysight \gls{ru} emulator for further testing and profiling, O-RAN 7.2 fronthaul and timing hardware, \blue{along with} multiple \gls{5g} \glspl{cn}. The system delivers \glspl{kpi} representative of \gls{5g} sub-6 GHz systems, with cell throughput north of \blue{$1.65$\:Gbps with up to 25} connected \glspl{ue} and a $100$\:MHz carrier \blue{bandwidth}. 

The tools we developed, integrated, and deployed on \testbed can be readily used for the development of intelligent use cases for \gls{5g} and beyond, thanks to the combination of 
NVIDIA \gls{arc}, \gls{oai}, and the \gls{osc} projects. \blue{As a result, this combination offers performance improvements} over most open-source, non-accelerated solutions while maintaining the openness and code accessibility typical of Open \gls{ran} systems\blue{, further enhanced by the seamless integration of \glspl{gpu}.
\testbed provides researchers with the necessary capabilities to develop, test, and evaluate a wide range of \gls{ai}/\gls{ml} and \gls{ran} solutions on a production-ready platform, including spectrum sharing techniques~\cite{lacava2025dApps}, secure cellular networks~\cite{groen2024securing, groen2024timesafe}, resource optimization~\cite{Cheng2024oranslice}, interference detection and mitigation, handover strategies, and the development of intelligent and autonomous networks.}
In addition, documentation and tutorials allow for the replication and bootstrapping of the testbed and its functionalities across research institutions \blue{and beyond. In fact, the value propositions of a platform similar to \testbed, with its openness, multi-vendor support, and GPU-accelerated capabilities, have been demonstrated for industrial stakeholders, as shown by SoftBank and Fujitsu in~\cite{fujitsu1}.}

The rest of the paper is organized as follows. Section~\ref{sec:software} introduces the software frameworks we developed and integrated to enable \testbed. 
Section~\ref{sec:arc-hardware} concerns the deployment and configuration of the \testbed network infrastructure. 
Section~\ref{sec:ray-tracing} describes an \acrshort{rf} planning study to determine an optimal location for deploying the \glspl{ru}.
System performance is evaluated in Section~\ref{sec:exp-results} through various use case scenarios with multiple \gls{cots} \glspl{ue} and applications.
Section~\ref{sec:related-work} compares \testbed with the state of the art.
Section~\ref{sec:conclusions} draws conclusions and outlines our future work.


\glsreset{cn}
\section{\testbed Software}
\label{sec:software}

This section describes the software components of \testbed, also shown in Figure~\ref{fig:x5g-e2e-overview}.
These components can be divided into three main groups: (i) a full-stack programmable \gls{gnb} (\testbed \gls{ran})\blue{;} (ii) the Open \gls{ran} \glspl{ric} deployed on a micro-services cluster based on OpenShift\blue{;} and (iii) various \glspl{cn} deployed in a micro-services-based architecture essential for the effective functioning of the \gls{5g} network.

\glsunset{gnb}
\subsection{Full-stack Programmable \gls{ran} with NVIDIA Aerial and OpenAirInterface}
\label{sec:arc}


\glsreset{arc}





\begin{figure}[t]
    \centering
    \includegraphics[width=\linewidth]{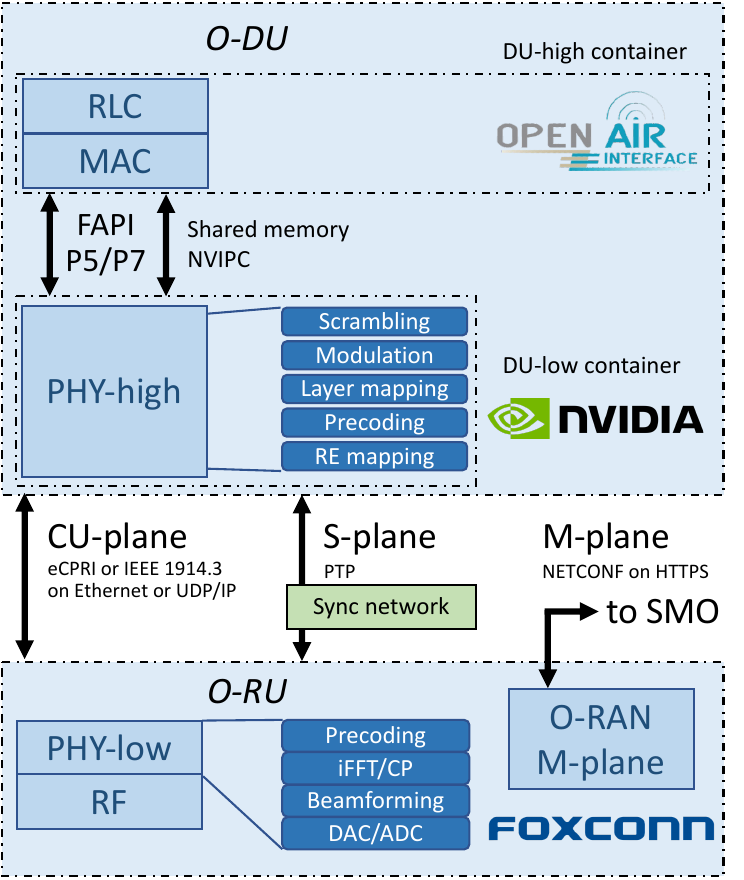}
    \caption{Architecture of the lower layers of the X5G \gls{ran} following O-RAN specifications and consisting of: (i) a Foxconn O-RU; (ii) an O-DU-low based on NVIDIA Aerial SDK; (iii) an O-DU-high based on OpenAirInterface with their corresponding interfaces.}
    \label{fig:ARC_archi}
\end{figure}

The right part of Figure~\ref{fig:x5g-e2e-overview} shows a detailed breakdown of the architecture of the \testbed \gls{ran}, which follows the basic O-RAN architecture split into \gls{cu}, \gls{du}, and \gls{ru}. The \gls{du} is further split into a DU-low, implementing Layer 1 (\gls{phy}, or L1) functionalities, and into a DU-high, implementing Layer 2 (\gls{mac} and \gls{rlc}, or L2) ones. As shown in Figure~\ref{fig:ARC_archi}, DU-low and DU-high communicate over the 5G \gls{fapi} interface specified by the \gls{scf}~\cite{SCF2020FAPI}. The DU-low is implemented using the NVIDIA Aerial \gls{sdk}~\cite{aerialsdk} on in-line \gls{gpu} accelerator cards, whereas DU-high and CU are implemented by \gls{oai} on general-purpose \glspl{cpu}. We deploy each function in separate Docker containers, sharing a dedicated memory space for the inter-process communication library that enables the \gls{fapi} interface. In our setup, we also combine the \gls{cu} and the \gls{du}-high into a combined L2/L3 \gls{gnb} Docker container, but the F1 split has also been deployed and tested.

The \gls{fapi} interface between the DU-high and DU-low defines two sets of procedures: configuration and slot procedures. Configuration procedures handle the management of the \gls{phy} layer and happen infrequently, e.g., when the \gls{gnb} stack is bootstrapped or reconfigured. On the contrary, slot procedures happen in every slot (i.e., every $500$\:$\mu$s for a $30$\:kHz subcarrier spacing) and determine the structure of each \gls{dl} and \gls{ul} slot. In our case, L1 serves as the primary and L2 as the subordinate. Upon the reception of a slot indication message from L1, L2 sends either an \gls{ul} or \gls{dl} request to dictate the required actions for the \gls{phy} layer in each slot. Additionally, L1 might transmit other indicators to L2, signaling the receipt of data related to \gls{rach}, \gls{uci}, \gls{srs}, checksums, or user plane activities.

In our implementation, we use \gls{fapi} version 222.10.02 with a few exceptions as outlined in the NVIDIA Aerial release notes \cite{aerialsdk-website}. 
The transport mechanism for \gls{fapi} messages is specified in the networked \gls{fapi} (nFAPI) specification~\cite{SCF2021nFAPI}, which assumes that messages are transported over a network. However, in our implementation, the L1 and L2 Docker containers communicate through the \gls{nvipc} library. This tool provides a robust shared memory communication framework specifically designed to meet the real-time performance demand of the data exchanges between \gls{mac} and \gls{phy} layers. In our implementation, we choose to transport the messages using little-endian with zero padding to $32$\:bits. The \gls{nvipc} library is also capable of tracing the \gls{fapi} messages and exporting them to a \texttt{pcap} file that can be analyzed offline with tools such as Wireshark.

The NVIDIA physical layer in the DU-low implements the O-RAN Open Fronthaul interface, also known as the O-RAN 7.2 interface~\cite{ORA2023}, to communicate directly with the O-RU, in our case manufactured by Foxconn. This interface transports frequency domain \gls{iq} samples (with optional block floating point compression) over a switched network, allowing for flexible deployments. The interface includes synchronization, control, and user planes. The synchronization plane, or S-plane, is based on PTPv2. We use synchronization architecture option 3~\cite{oran-wg4-fronthaul-cus}, where the fronthaul switch provides timing to both DU and RU. The interface also includes a management plane, although \blue{our system currently does not support it}.

\begin{table}[b]
    \begin{center}
    \footnotesize
    \caption{\testbed \acrshort{arc} deployment main features.}
    \label{table:testbeds-features}
    \begin{tabularx}{\columnwidth}{
        >{\raggedright\arraybackslash\hsize=\hsize}X
        >{\raggedright\arraybackslash\hsize=\hsize}X }
        \toprule
        Feature & Description \\
        \midrule
        3GPP Release & 15\\
        Frequency Band & n78 (FR1, TDD)\\
        Carrier Frequency & $3.75$\:GHz\\
        Bandwidth \blue{($\beta$)} & $100$\:MHz\\
        Subcarrier spacing \blue{($\Delta f$)} & $30$\:kHz\\
        \blue{Resource Block size ($\chi$)} & \blue{12 subcarriers} \\
        \blue{Modulation order ($Q_{m}$)} & \blue{8 (256-\gls{qam})} \\
        TDD config & DDDSU, DDDDDDSUUU$^*$ \\
        Number of antennas used & \blue{4 TX, 4 RX}\\
        MIMO config \blue{($L_{DL}, L_{UL}$)} & \blue{4 layers DL, 1 layer UL}\\
        Max theoretical cell throughput$^{**}$ \blue{($T_{DL}, T_{UL}$)} & \blue{$1.64$\:Gbps DL, $204$\:Mbps UL}\\
        \bottomrule
    \end{tabularx}
    \end{center}
    $^*$Currently the special slot is unused due to limitations in Foxconn radios.
    
    $^{**}$The single-user maximum theoretical DL throughput can currently only be reached in the DDDDDDSUUU TDD configuration. In the DDDSU TDD configuration, it is limited to $350$\:Mbps since we can schedule a maximum of 2~DL slots per user in one TDD period, as only 2~ACK/NACK feedback bits are available per user.
\end{table}
\glsunset{arc}

Table~\ref{table:testbeds-features} summarizes the main features and operational parameters of the \gls{arc} deployment in the \testbed testbed. The protocol stack is aligned with 3GPP Release 15 and uses the 5G n78 \gls{tdd} band and numerology~1. The DDDSU \gls{tdd} pattern, which repeats every $2.5$\:ms, includes three downlink slots, one special slot (which is not used due to limitations in the Foxconn \glspl{ru}), and an uplink slot. The uplink slot format implemented in \gls{oai} carries only two feedback bits for ACK/NACK per \gls{ue}, thus allowing only the scheduling of two downlink slots per \gls{ue}, eventually limiting the single \gls{ue} throughput. \blue{Alternative \gls{tdd} patterns, including DDDDDDSUUU and DDDDDDDSUU, repeating every $5$\:ms, are also already in use} to provide additional ACK/NACK bits for reporting from the \glspl{ue} \blue{and mitigate this limitation}.

\blue{To compute the maximum theoretical cell throughput in downlink ($T_{DL}$) and uplink ($T_{UL}$), we first derive a few additional parameters from Table~\ref{table:testbeds-features}. The number of resource blocks ($N_{RB}$) is computed using
}
\begin{align}\label{eq:nrb}
    \blue{N_{RB} = \frac{\beta}{\chi \cdot \Delta f} = 273.}
\end{align}
\blue{By default, the number of \gls{ofdm} symbols per slot ($N_{sym}$) is 14. The number of slots per second ($N_{slot}$) is inversely proportional to the slot duration, which for numerology $\mu = 1$ is  $0.5$~ms. Hence, $N_{slot} = 1s/0.5$ms$ = 2000$ slots/second.
The maximum theoretical cell throughput for downlink and uplink is given by
}
\begin{align}\label{eq:maxth}
    \blue{T_{DL,UL} = N_{RB} \cdot \chi \cdot N_{sym} \cdot N_{slot} \cdot Q_m \cdot R \cdot L_{DL,UL} \cdot \eta,}
\end{align}
\blue{where $R$ is the effective code rate, which can approach 0.93 (as specified in the 3GPP standard~\cite{3gpp5gnr}), and $\eta$ is the fraction of time allocated for downlink or uplink operations based on the chosen TDD pattern.
Considering the DDDDDDSUUU pattern to circumvent current \gls{oai} limitations on the ACK/NACK feedback bits, 60\% of time is allocated for downlink and 30\% for uplink since the special slot is unused due to Foxconn \gls{ru} constraints.
Consequently, the resulting theoretical peak cell throughput is $1.64$~Gbps for downlink ($T_{DL}$) and $204$~Mbps for uplink ($T_{UL}$).
These values do not account for overheads typical of real networks---such as \gls{dmrs}, \gls{pucch}, and \gls{pdcch}---which may further reduce net throughput.
As shown by the experimental results in Section~\ref{sec:exp-peak}, \testbed peak performance nearly reaches the theoretical downlink throughput, while the uplink is still under improvement.
}


\subsection{Integration with the \gls{osc} Near-RT RIC}
\label{sec:e2}

%
\begin{figure}[t]
    \centering
    \includegraphics[width=\linewidth]{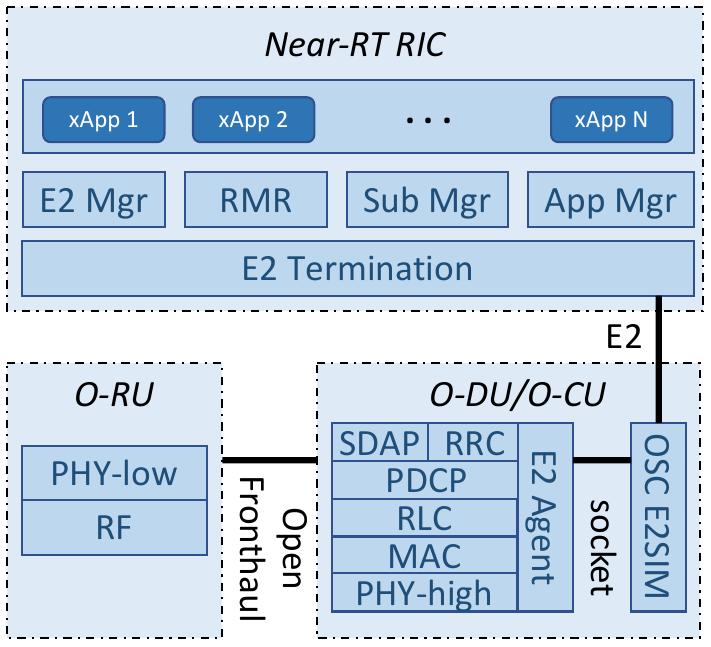}
    \caption{Integration of the \gls{osc} Near-RT \gls{ric} in the OpenShift cluster with the \testbed \gls{ran}.}
    \label{fig:ARC_e2if}
\end{figure}
One of the key components of an O-RAN deployment is the Near-Real-Time (or Near-RT) \gls{ric}, and the intelligent applications hosted therein, namely xApps. These can implement closed-control loops at timescales between $10$\:ms and $1$\:s to provide optimization and monitoring of the \gls{ran}~\cite{abdalla2022toward,mungari2021rl}. In the \blue{current} \testbed setup, we deploy the ``E'' release of the \gls{osc} Near-RT \gls{ric} on a RedHat OpenShift cluster~\cite{bonati2023neutran}, which manages the lifecycle of edge-computing workloads instantiated as containerized applications.
%
The Near-RT \gls{ric} and the \gls{arc} \gls{ran} are connected through the O-RAN E2 interface (see Figure~\ref{fig:ARC_e2if}), based on \gls{sctp} and an O-RAN-defined application protocol (E2AP) with multiple service models implementing the semantic of the interface (e.g., control, reporting, etc)~\cite{polese2023understanding}.
%
%
On the \gls{gnb} side, we integrate an E2 agent based on the \textit{e2sim} software library~\cite{moro2023nfv,e2sim}, which is used to transmit the metrics collected by the \gls{oai}~\gls{gnb} to the \gls{ric} via the \gls{kpm} E2 service model.
These metrics are then processed by xApps deployed on the \gls{ric}, and used to compute some control action (e.g., through \gls{ai}/\gls{ml} agents) that is sent to the \gls{ran} through the E2 interface and processed by the \textit{e2sim} agent.

\begin{figure}[t]
    \centering
    \includegraphics[width=\linewidth]{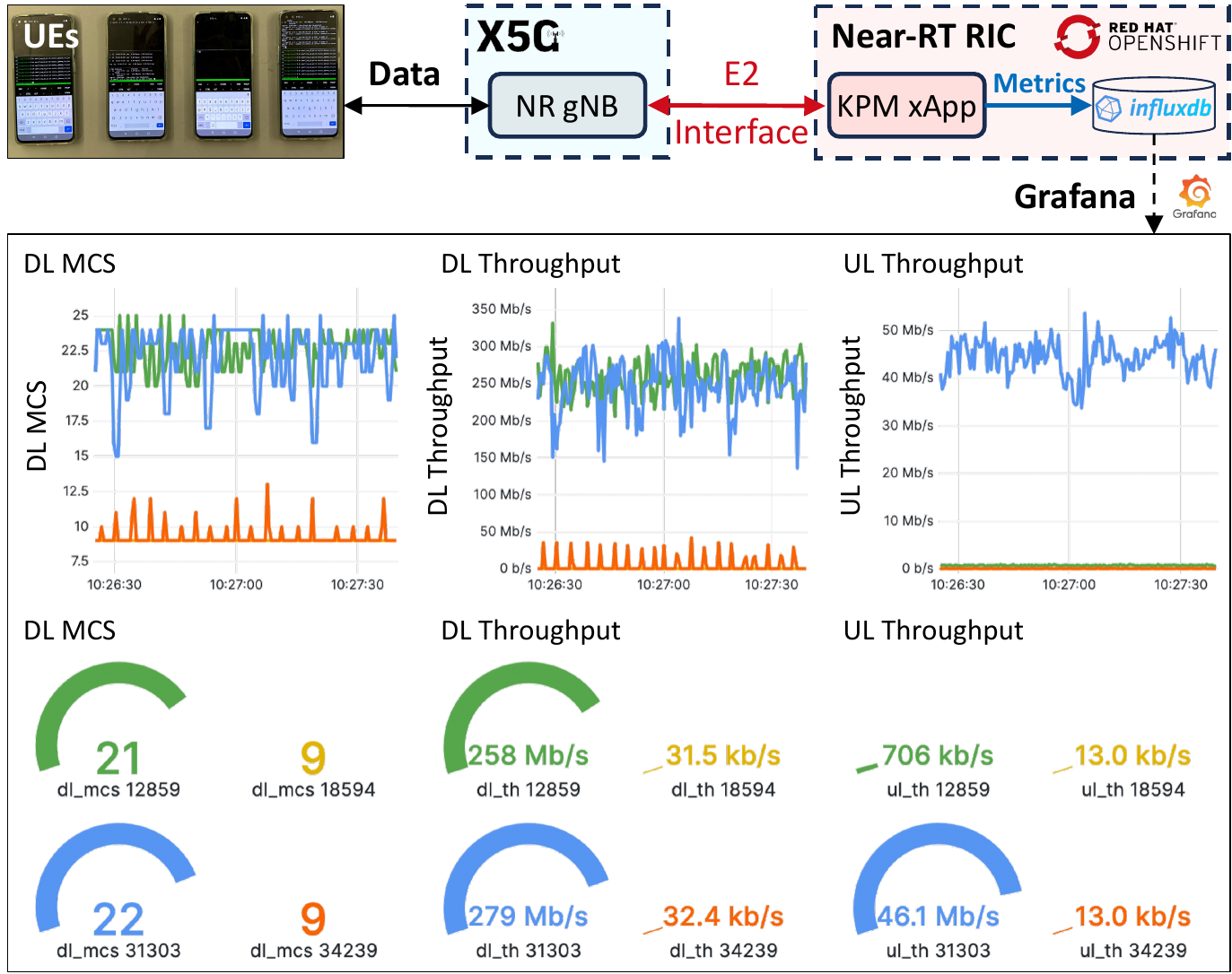}
    \caption{\gls{kpm} xApp example architecture \blue{including an \testbed \gls{gnb} with four connected \glspl{ue}, each performing a different operation (ping, video streaming, \gls{dl} test, and \gls{dl}/\gls{ul} tests), and a \gls{kpm} xApp that pushes \gls{ue} metrics into an Influx database, which are then visualized in a Grafana dashboard.}}
    \label{fig:kpm_xapp}
\end{figure}

As an example, Figure~\ref{fig:kpm_xapp} shows the architecture of a \gls{kpm} xApp integrated with the \testbed testbed.
%
%
This xApp receives metrics from the E2 agent in the \gls{gnb}, including throughput, number of \glspl{ue}, and \gls{rsrp}, and stores them in an InfluxDB database~\cite{influxdb}. The database is then queried to display the \gls{ran} performance on a Grafana dashboard~\cite{grafana} (see Figure~\ref{fig:kpm_xapp}).
This setup creates a user-friendly observation point for monitoring network performance and demonstrates the effective integration of the near-RT \gls{ric} in our configuration.
A tutorial on how to deploy and run this xApp in \testbed or on a similar testbed can be found on the OpenRAN Gym website~\cite{openrangymwebsite}, which hosts an open-source project and framework for collaborative research in the O-RAN ecosystem~\cite{bonati2022openrangym-pawr}.

\blue{The metrics collected by a \gls{kpm} xApp can then be leveraged by a second xApp or an rApp to perform smart closed-loop \gls{ran} controls at runtime, based on an arbitrary optimization strategy or specific requirements.
ORANSlice---an open-source, network-slicing-enabled Open \gls{ran} system that leverages open-source \gls{ran} frameworks such as \gls{oai}~\cite{Cheng2024oranslice}---was successfully integrated and tested in \testbed, enabling near-real-time slicing control of the resources allocated by a \gls{gnb} to multiple slices of the network, according to different policies set by the network manager.
Figure~\ref{fig:slicing_xapp_demo} presents the effects of various network policies applied by an ORANSlice slicing xApp in a \testbed\ \gls{gnb}. Figure~\ref{fig:slicing_xapp_th} shows the \gls{dl} throughput results for the two slices (slice 1 in blue and slice 2 in orange), with a single \gls{ue} per slice connected and transmitting $50$\:Mbps of \gls{dl} \gls{udp} data, according to the policy shown in Figure~\ref{fig:slicing_xapp_policy}. The slicing xApp switches between three policies: (0) no-priority, where all slices share all resources, so both \glspl{ue} achieve the target throughput of $50$\:Mbps; (1) prioritize slice 1: where 98\% of resources are reserved for the first slice and 2\% for the second one, causing the latter performance to drop to only $6$\:Mbps; (2) prioritize slice 2, where the opposite behavior of policy 1 is observed, with slice 1 now unable to achieve the target throughput.
In this example, the policy is applied arbitrarily as a proof-of-concept for the network slicing control capabilities of \testbed, while more intelligent strategies employing \gls{ai}/\gls{ml} components can be easily integrated into the decision process.
Additional applications, including the emerging dApps~\cite{lacava2025dApps}, are currently being integrated into \testbed to fully leverage its openness and programmability, further demonstrating the benefits of smart closed-loop control within the O-RAN ecosystem.}


\begin{figure}[t]
\centering
    \subfloat[DL Throughput]{
    \label{fig:slicing_xapp_th}
    \centering
    \setlength\fwidth{\linewidth}
    \setlength\fheight{.35\linewidth}
    \input{figures/fig_tex/slicing_xapp_th}
    \setlength\abovecaptionskip{.05cm}}
    \hfill    
    \subfloat[Policy]{
    \label{fig:slicing_xapp_policy}
    \centering
    \setlength\fwidth{.995\linewidth}
    \setlength\fheight{.28\linewidth}
\begin{tikzpicture}
\pgfplotsset{every tick label/.append style={font=\scriptsize}}

\definecolor{darkslategray38}{RGB}{38,38,38}
\definecolor{lightgray204}{RGB}{204,204,204}
\definecolor{steelblue31119180}{RGB}{31,119,180}
\definecolor{darkorange25512714}{RGB}{255,127,14}

\begin{axis}[
width=1\fwidth, 
height=1.05\fheight, 
axis line style={color=black},
legend cell align={center},
legend style={
  fill opacity=0.8,
  draw opacity=1,
  text opacity=1,
  at={(0.5, 1.08)},
  anchor=south,
  draw=lightgray204,
  font=\scriptsize,
},
tick align=inside,
tick pos=left,
x grid style={lightgray204},
xlabel=\textcolor{darkslategray38}{Time [s]},
label style={font=\scriptsize},
xmajorgrids,
xmin=0, xmax=420,
xtick style={color=darkslategray38},
y grid style={lightgray204},
ymajorgrids,
ymin=-1, ymax=3,
ytick={0,1,2},
ytick style={color=darkslategray38},
ylabel={Policy},
legend columns=2,
]

\addplot [thick, black]
table {%
0         0
1.28      0
2.56      0
3.84      0
5.12      0
7.68      0
8.96      0
10.24     0
11.52     0
12.80     0
14.08     0
15.36     0
16.64     0
17.92     0
19.20     0
20.48     0
21.76     0
23.04     0
24.32     0
25.60     0
26.88     0
28.16     0
29.44     0
30.72     0
32.00     0
33.28     0
34.56     0
35.84     0
37.12     0
38.40     0
39.68     0
40.96     0
42.24     0
43.52     0
44.80     0
46.08     0
47.36     0
48.64     0
49.92     0
51.20     0
52.48     0
53.76     0
55.04     0
56.32     0
57.60     0
58.88     0
60.16     1
61.44     1
62.72     1
64.00     1
65.28     1
66.56     1
67.84     1
69.12     1
70.40     1
71.68     1
72.96     1
74.24     1
75.52     1
76.80     1
78.08     1
79.36     1
80.64     1
81.92     1
83.20     1
84.48     1
85.76     1
87.04     1
88.32     1
89.60     1
90.88     1
92.16     1
93.44     1
94.72     1
96.00     1
97.28     1
98.56     1
99.84     1
101.12    1
102.40    1
103.68    1
104.96    1
106.24    1
107.52    1
108.80    1
110.08    1
111.36    1
112.64    1
113.92    1
115.20    1
116.48    1
117.76    1
119.04    1
120.32    1
121.60    2
122.88    2
124.16    2
125.44    2
126.72    2
128.00    2
129.28    2
130.56    2
131.84    2
133.12    2
134.40    2
135.68    2
136.96    2
138.24    2
139.52    2
140.80    2
142.08    2
143.36    2
144.64    2
145.92    2
147.20    2
148.48    2
149.76    2
151.04    2
152.32    2
153.60    2
154.88    2
156.16    2
157.44    2
158.72    2
160.00    2
161.28    2
162.56    2
163.84    2
165.12    2
166.40    2
167.68    2
168.96    2
170.24    2
171.52    2
172.80    2
174.08    2
175.36    2
176.64    2
177.92    2
179.20    2
180.48    0
181.76    0
183.04    0
184.32    0
185.60    0
186.88    0
188.16    0
189.44    0
190.72    0
192.00    0
193.28    0
194.56    0
195.84    0
197.12    0
198.40    0
199.68    0
200.96    0
202.24    0
203.52    0
204.80    0
206.08    0
207.36    0
208.64    0
209.92    0
211.20    0
212.48    0
213.76    0
215.04    0
216.32    0
217.60    0
218.88    0
220.16    0
221.44    0
222.72    0
224.00    0
225.28    0
226.56    0
227.84    0
229.12    0
230.40    0
231.68    0
232.96    0
234.24    0
235.52    0
236.80    0
238.08    0
239.36    0
240.64    0
241.92    0
243.20    0
244.48    0
245.76    0
247.04    1
248.32    1
249.60    1
250.88    1
252.16    1
253.44    1
254.72    1
256.00    1
257.28    1
258.56    1
259.84    1
261.12    1
262.40    1
263.68    1
264.96    1
266.24    1
267.52    1
268.80    1
270.08    1
271.36    1
272.64    1
273.92    1
275.20    1
276.48    1
277.76    1
279.04    1
280.32    1
281.60    1
282.88    1
284.16    1
285.44    1
286.72    1
288.00    1
289.28    1
290.56    1
291.84    1
293.12    1
294.40    1
295.68    1
296.96    1
298.24    1
299.52    1
300.80    1
302.08    1
303.36    1
304.64    2
305.92    2
307.20    2
308.48    2
309.76    2
311.04    2
312.32    2
313.60    2
314.88    2
316.16    2
317.44    2
318.72    2
320.00    2
321.28    2
322.56    2
323.84    2
325.12    2
326.40    2
327.68    2
328.96    2
330.24    2
331.52    2
332.80    2
334.08    2
335.36    2
336.64    2
337.92    2
339.20    2
340.48    2
341.76    2
343.04    2
344.32    2
345.60    2
346.88    2
348.16    2
349.44    2
350.72    2
352.00    2
353.28    2
354.56    2
355.84    2
357.12    2
358.40    2
359.68    2
360.96    0
362.24    0
363.52    0
364.80    0
366.08    0
367.36    0
368.64    0
369.92    0
371.20    0
372.48    0
373.76    0
375.04    0
376.32    0
377.60    0
378.88    0
380.16    0
381.44    0
382.72    0
384.00    0
385.28    0
386.56    0
387.84    0
389.12    0
390.40    0
391.68    0
392.96    0
394.24    0
395.52    0
396.80    0
398.08    0
399.36    0
400.64    0
401.92    0
403.20    0
404.48    0
405.76    0
407.04    0
408.32    0
409.60    0
410.88    0
412.16    0
413.44    0
414.72    0
416.00    0
417.28    0
418.56    0
419.84    0
};

\end{axis}

\end{tikzpicture}
    \setlength\abovecaptionskip{.05cm}}
\caption{\blue{Slicing xApp example showing: (a) \gls{dl} throughput for two different slices, each with a single \gls{ue} connected and pushing $50$\:Mbps of \gls{udp} traffic; (b) the network policy applied by the slicing xApp, switching between no-priority (0), prioritize slice 1 (1), and prioritize slice 2 (2).}}
\label{fig:slicing_xapp_demo}
\end{figure}

\subsection{Core Network}

The \testbed testbed facilitates the integration and testing of different \glspl{cn} from various vendors and projects. We leverage virtualization to deploy all the necessary micro-services, e.g., \gls{amf}, \gls{smf}, \gls{upf}, in the OpenShift cluster that also supports the Near-RT \gls{ric}.
%
We have successfully tested and integrated the \testbed \gls{ran} with two open-source core network implementations, i.e., the \gls{5g} \glspl{cn} from \gls{oai}~\cite{kaltenberger2024driving}, as also discussed in~\cite{villa2024x5g}, and, in this paper, also with Open5GS~\cite{open5gs_website} \blue{and the CoreSIM software from Keysight~\cite{keysight_coresim}}. As part of our ongoing efforts, we plan to incorporate additional cores, including the commercial core from A5G~\cite{a5gnetworks}.
%
%

\subsection{X5G Software Licensing and Tutorials}

\testbed, including the Aerial \gls{phy}, the \gls{oai} higher layers, as well as the \gls{osc} \gls{ric}, is open and can be extended with custom features and functionalities. The NVIDIA \gls{arc} framework is documented on the NVIDIA portal~\cite{aerialsdk-website}, which is accessible through NVIDIA's 6G developer program. As mentioned in Section~\ref{sec:e2}, the step-by-step integration between the \gls{osc} \gls{ric} and the \gls{arc} stack through the \testbed E2 agent is discussed in a tutorial on the OpenRAN Gym website~\cite{openrangymwebsite,bonati2022openrangym-pawr}.

The components implemented by \gls{oai} are published under the \gls{oai} public license v1.1  created by the \gls{oai} Software Alliance (OSA) in 2017~\cite{oai}.
This license is a modified Apache v2.0 License, with an additional clause that allows contributors to make patent licenses available to third parties under \gls{frand} terms, similar to \gls{3gpp} for commercial exploitation, to allow contributions from companies holding intellectual property in related areas. The usage of \gls{oai} code is free for non-commercial/academic research purposes. The Aerial \gls{sdk} is available through an early adopter program~\cite{aerialsdk-website}. The \gls{osc} software is published under the Apache v2.0 License.

\section{\testbed Infrastructure}
\label{sec:arc-hardware}

\begin{figure*}[t]
    \centering
    \includegraphics[width=1\linewidth]{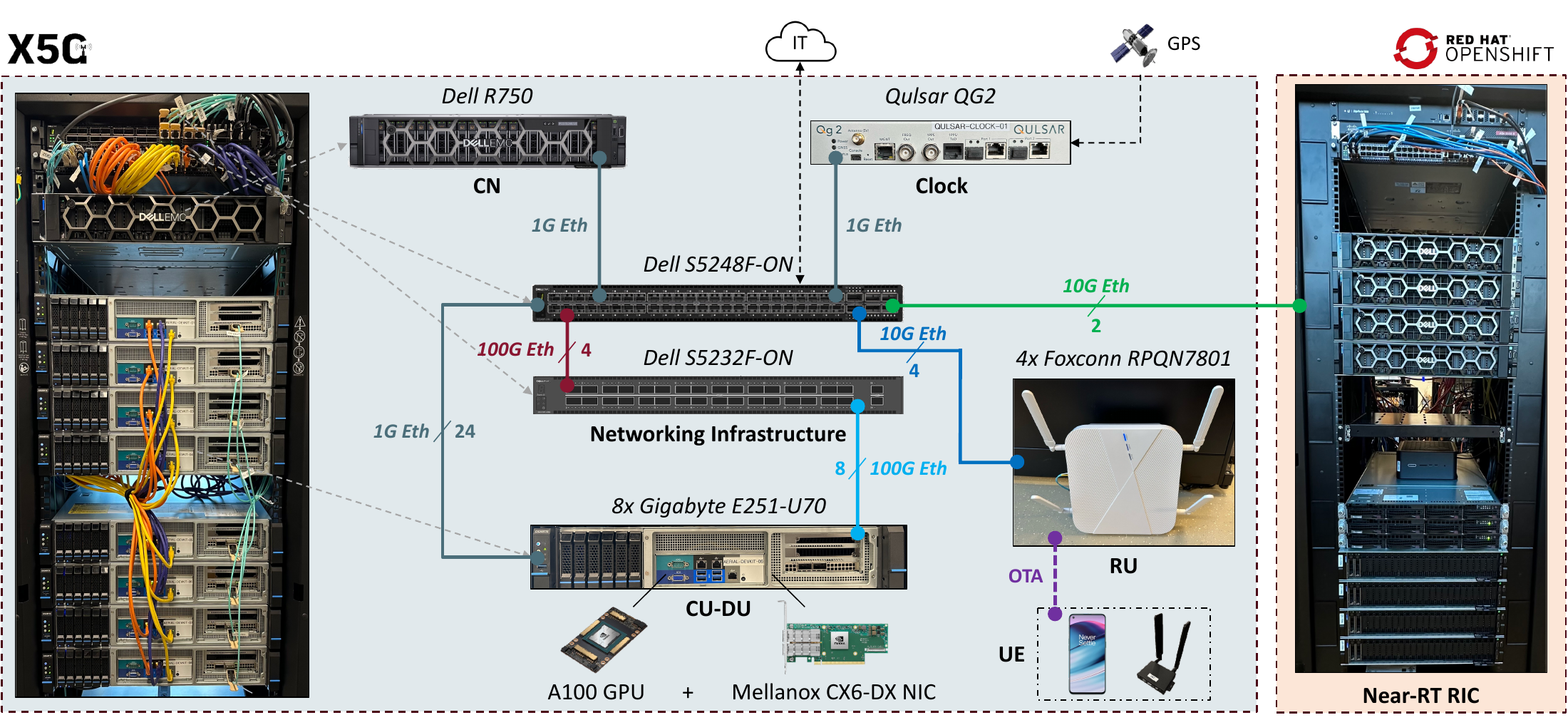}
    \caption{Hardware and architecture infrastructure of the \testbed deployment at Northeastern University.}
    \label{fig:arc-hardware}
\end{figure*}

This section describes the \testbed physical deployment that is currently located on the Northeastern University campus in Boston, MA.\footnote{\testbed website: \url{https://x5g.org}.} The deployment includes a server room with a dedicated rack for the private 5G system and an indoor laboratory open space area with benches and experimental equipment that provide a realistic \gls{rf} environment with rich scattering and obstacles.
Figure~\ref{fig:arc-hardware} illustrates the hardware infrastructure that we deployed to support the \testbed operations. This includes synchronization and networking infrastructures, radio nodes, eight \gls{arc} servers with integrated \gls{du} and \gls{cu}, and additional compute infrastructure for the \gls{ric} and \gls{cn} deployments. This infrastructure, which will be described next, has been leveraged to provide connectivity for up to eight concurrent \gls{cots} \glspl{ue}, such as OnePlus smartphones (AC Nord 2003) and Sierra Wireless boards (EM9191)~\cite{sierrawireless}.

\textbf{Synchronization Infrastructure.} The synchronization infrastructure consists of a Qulsar (now VIAVI) QG-2 device acting as grandmaster clock. The QG-2 unit is connected to a GPS antenna for precise class-6 timing and generates both \gls{ptp} and \gls{synce} signals to provide frequency, phase, and time synchronization compliant with the ITU-T G.8265.1, G.8275.1, and G.8275.2 profiles. It sends the synchronization packets to the networking infrastructure through a $1$\:Gbps Ethernet connection. The networking infrastructure then offers full on-path support, which is necessary to distribute phase synchronization throughout the \testbed platform.

\textbf{Networking Infrastructure.} The networking infrastructure provides connectivity between all the components of the \testbed platform. It features fronthaul and backhaul capabilities through the use of two Dell switches (S5248F-ON and S5232F-ON) interconnected via four $100$\:Gbps cables in a port channel configuration. This configuration allows for the aggregation of multiple physical links into a single logical one to increase bandwidth and provide redundancy in case some of them fail. All switch ports are sliced into different \glspl{vlan} to allow the proper coexistence of the various types of traffic (i.e., fronthaul, backhaul, management).
The Dell S5248F-ON switch primarily provides backhaul capabilities to the network and acts as a boundary clock in the synchronization plane, receiving \gls{ptp} signals from the synchronization infrastructure. This switch includes 48~\glsname{sfp+} ports: 12 ports are dedicated to the \glspl{ru} and receive \gls{ptp} synchronization packets, 10 are used to connect to the OpenShift cluster and service network, 10 are used for the out-of-band management network, and 16 connect to the \gls{cn} and the Internet. Additionally, the switch includes 6~\glsname{qsfp28} ports, 4 of which interconnect with the second switch.
The Dell S5232F-ON switch mainly provides fronthaul connectivity to the \glspl{gnb}. It includes 32~\glsname{qsfp28} ports: 8 ports connect to the Mellanox cards of the \gls{arc} nodes via $100$\:Gbps fiber links, and 4 connect to the Dell S5248F-ON switch. The latter also acts as a boundary clock, receiving the synchronization messages from the S5232F-ON and delivering them to the \glspl{gnb}.


\textbf{\gls{ru}.} We deployed \blue{eight} Foxconn RPQN 4T4R \glspl{ru}, operating in the \blue{n78} band, with additional units being tested in the lab, and the Keysight RuSIM emulator. The Foxconn units have 4 externally mounted antennas, each antenna with a $5$\:dBi gain, and $24$\:dBm of transmit power. The \gls{ota} transmissions are regulated as part of the Northeastern University \gls{fcc} Innovation Zone~\cite{FCC-IZ-Boston}, with an additional transmit attenuation of $20$\:dB per port to comply with transmit power limits and guarantee the coexistence of multiple in-band \glspl{ru} in the same environment. As we will discuss in Section~\ref{sec:exp-results}, we leverage two of these \glspl{ru} for the experimental analysis presented in this work. These \glspl{ru} are deployed following \gls{rf} planning procedures discussed in Section~\ref{sec:ray-tracing}. 
Plans are in place to procure \gls{cbrs} \glspl{ru} and deploy them in outdoor locations. 
\glspl{ru} from additional vendors are being tested and integrated as part of our future works. Finally, we also tested and integrated the \gls{arc} stack with the Keysight RuSIM emulator, which supports the termination of the fronthaul interface on the \gls{ru} side and exposes multiple \glspl{ru} to the \gls{ran} stack, for troubleshooting, conformance testing, and performance testing~\cite{keysight_rusim}. 


\textbf{\gls{cu} and \gls{du}.} The 8~\gls{arc} nodes that execute the containerized \gls{cu}/\gls{du} workloads are deployed on Gigabyte E251-U70 servers with 24-core Intel Xeon Gold 6240R CPU and $96$\:GB of RAM.
The servers---which come in a half rack chassis for deployment in \gls{ran} and edge scenarios---are equipped with a Broadcom PEX 8747 \gls{pci} switch that 
enables direct connectivity between cards installed in two dedicated \gls{pci} slots without the need for interactions with the CPU.
Specifically, the two \gls{pci} slots host an NVIDIA A100 \gls{gpu}, which supports the computational operations of the NVIDIA Aerial \gls{phy} layer, as well as a Mellanox ConnectX-6 Dx \gls{nic}. The latter, which is used for the fronthaul interface, connects to the fronthaul part of the networking infrastructure via a \acrshort{qsfp28} port and $100$\:Gbps fiber-optic cable.
In this way, the \gls{nic} can offload or receive packets directly from the GPU, thus enabling low-latency packet processing.
Finally, each server is connected to the backhaul part of the networking infrastructure through three $1$\:Gbps Ethernet links, which provide connectivity with the OpenShift cluster (and thus the Near-RT \gls{ric} and the core networks), the management infrastructure, and the Internet.
\blue{In addition to the Gigabyte servers, seven \gls{gh} machines---one of the latest NVIDIA high-computing ARM-based devices---are currently being integrated into \testbed to run the \gls{arc} \gls{cu}/\gls{du} worloads. Each \gls{gh} combines a 72-core NVIDIA Grace CPU Superchip and an NVIDIA H100 Tensor Core GPU, linked through NVIDIA NVLink-C2C technology, which ensures seamless data sharing with up to 900 GB/s of bandwidth. It also features $480$\:GB of RAM, two BlueField-3 \glspl{dpu}, and ConnectX-7 \glspl{nic}. This configuration provides significantly higher computational capabilities compared to the Gigabyte servers, enabling the efficient support of concurrent \gls{ran} and \gls{ai}/\gls{ml} workloads.}

\textbf{Additional Compute.} We leverage additional servers that are part of the OpenShift cluster and are used to instantiate the various \glspl{cn} and the Near-RT \gls{ric}. 
The OpenShift cluster includes three Dell R740 servers acting as control-plane nodes and two Microway Navion Dual servers as worker nodes. The OpenShift rack is linked to the \testbed rack through two $10$ Gbps connections, one dedicated to OpenShift operations, and the other for the out-of-band management.
Additionally, a Dell R750 server with $56$\:cores and $256$\:GB RAM is available for the deployment and testing of additional core network elements. This server connects to the networking infrastructure via a $1$\:Gbps Ethernet link and has access to the Internet through the Northeastern University network.




\section{RF Planning with Ray-tracing}
\label{sec:ray-tracing}


In this section, we present \gls{rf} planning procedures to identify suitable locations for the \gls{ru} deployment. 
\blue{This approach leverages an exhaustive search within a ray-tracing-based digital twin framework, with the objective of maximizing the \glspl{ru} coverage while minimizing the overall interference.
The study is conducted only once during the system deployment phase and remains valid as long as no significant changes occur in the environment.}
%
%
We perform ray-tracing in a detailed \blue{digitized} representation of our indoor laboratory space in the Northeastern University ISEC building in Boston, MA, to achieve high fidelity between the real-world environment and the \blue{digital} one. We carefully study how to deploy 2 \glspl{ru} by considering the \gls{sinr} between the \glspl{ru} and the \glspl{ue} as the objective function in the optimization problem. We \blue{limit} the optimization space by using a grid of 24 possible \gls{ru} locations and 52 \gls{ue} test points\blue{, enabling an exhaustive search approach instead of a formal integer optimization problem, since these constraints keep the computation manageable.} 

\begin{figure}[b]
    \centering
    \subfloat[Site viewer top view.]
    {\label{fig:siteviewertop}
    \includegraphics[width=0.99\linewidth]{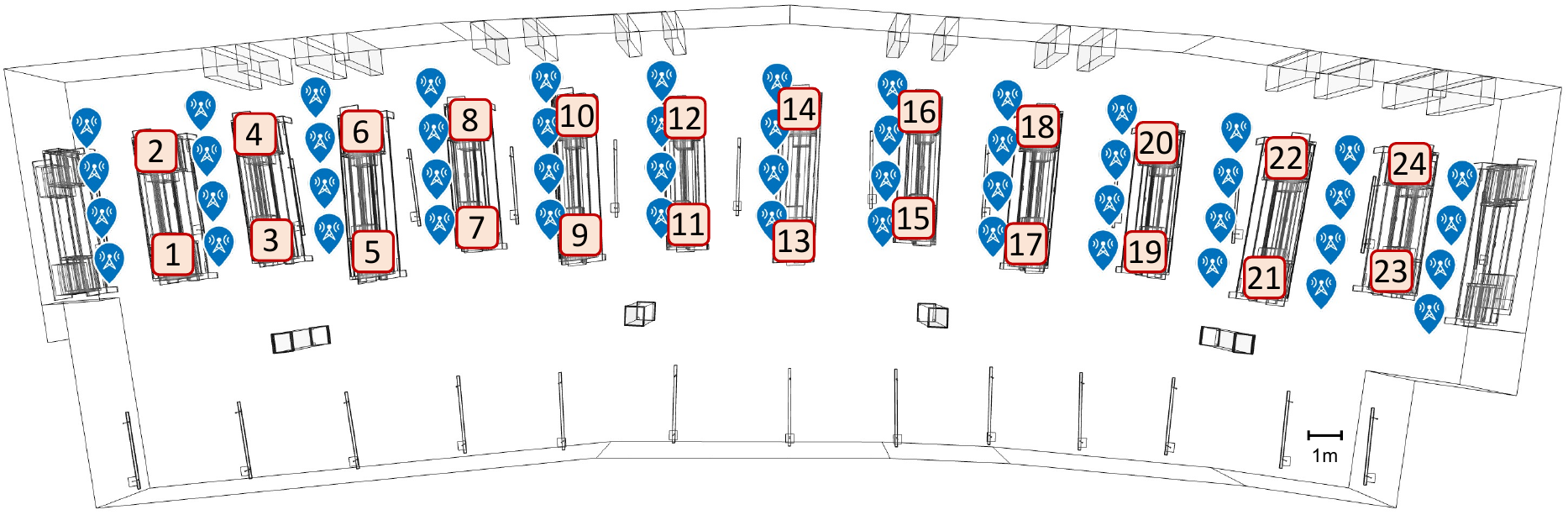}}
    \hfill
    \subfloat[Site viewer side view.]
    {
    \label{fig:siteviewerside}
    \includegraphics[width=0.99\linewidth]{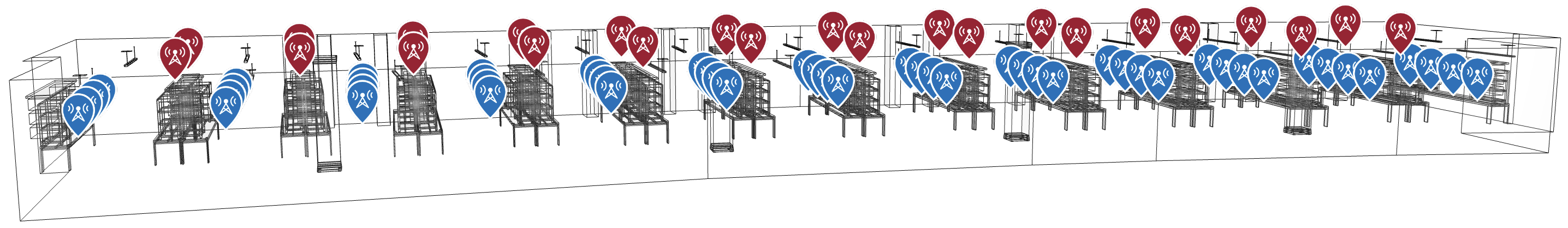}}        
    \caption{Site viewer with \gls{ru} (red squares) and \gls{ue} (blue icons) locations.}
    \label{fig:siteviewer}
\end{figure}

First, we leverage the 3D representation of our laboratory space, \blue{created as part of the digital twin framework developed in~\cite{villa2024dt} using the SketchUp modeling software.}
We then import the model in the MATLAB ray-tracing software and define the locations of \glspl{ru} and \glspl{ue} as shown in Figure~\ref{fig:siteviewertop} (from a top perspective) and in Figure~\ref{fig:siteviewerside} (from a side view). The 24 possible \glspl{ru} locations (2 for each bench) are shown in red, while the 52 test points for the \glspl{ue} (arranged in a $4 \times 13$ grid) are in blue.
%
Tables~\ref{table:testbeds-features} and~\ref{table:raytracing-parameters} summarize the parameters used in our ray-tracing model. For the deployment planning purpose, we consider the \glspl{ru} as transmitter nodes (TX) and the \glspl{ue} as receiver ones (RX), i.e., we tailor our deployment to downlink transmissions.
%
\begin{table}[h]
    \centering
    \footnotesize
    \caption{Parameters of the MATLAB ray-tracing study to determine \gls{ru} locations.}
    \label{table:raytracing-parameters}
    \begin{tabularx}{0.95\columnwidth}{
        >{\raggedright\arraybackslash\hsize=1.0\hsize}X
        >{\raggedright\arraybackslash\hsize=1.0\hsize}X }
        \toprule
        Parameter & Value \\
        \midrule

        \gls{ru} antenna spacing & $0.25$\:m \\
        \gls{ru} antenna TX power ($P_{RU}$) & $24$\:dBm \\
        \gls{ru} antenna gain ($G_{RU}$) & $5$\:dBi \\
        \gls{ru} antenna pattern & Isotropic \\
        \gls{ru} TX attenuation ($A_{RU}$) & $[0-50]$\:dB \\
        \blue{Set} of \gls{ru} locations (\blue{$\mathcal{R}$}) & $24$ in a $2 \times 12$ grid \\
        \gls{ru} height & $2.2$\:m \\

        \gls{ue} number of antennas & 2 \\
        \gls{ue} antenna spacing & $0.07$\:m \\
        \gls{ue} antenna gain ($G_{UE}$) & $1.1$\:dBi \\
        \gls{ue} noise figure ($F_{UE}$) & $5$\:dB \\
        \blue{Set} of \glspl{ue} locations (\blue{$\mathcal{U}$}) & $52$ in a $4 \times 13$ grid \\
        \gls{ue} height & $0.8$\:m \\

        Environment material & Wood \\
        Max number of reflections & $3$ \\
        Max diffraction order & $1$ \\
        Ray-tracing method & Shooting and bouncing rays \\
        \bottomrule
    \end{tabularx}
\end{table}

The ray-tracer generates a $24 \times 52$ matrix $\mathbf{C}$ where each entry $c_{i,j}$ corresponds to the channel information between $RU_i$ with \blue{$i \in \mathcal{R}$, $ \mathcal{R} = {1,...,24}$, and $UE_j$ with $j \in \mathcal{U}$, $ \mathcal{U} = {1,...,52}$}. We use this to derive relevant parameters such as the thermal noise (\blue{$\mathcal{N}$}) and the path loss (\blue{$\mathcal{L}$}) to compute 
%
the \gls{rssi} $\blue{\mathcal{S}}_{i,j}$ for $UE_j$ connected to $RU_i$, as follows:
%
%
\begin{align}\label{eq:rssi}\small
    \blue{\mathcal{S}}_{i,j} = P_{RU,i} + G_{RU,i} - A_{RU,i} - \blue{\mathcal{L}}_{i,j} + G_{UE,j},
\end{align}
where $P_{RU,i}$, $G_{RU,i}$, and $A_{RU,i}$ are the antenna TX power, gain, and attenuation of $RU_i$, respectively.
%
Then, considering the linear representation of $\hat{\blue{\mathcal{S}}}_{i,j}$, the \gls{sinr} $\Gamma_{i,j}$ is
\begin{align}\label{eq:sinr}
    \Gamma_{i,j} = \frac{\hat{\blue{\mathcal{S}}}_{i,j}}{\blue{\mathcal{N}} F_{UE,i} + \sum\limits_{u=1, u \neq i}^M \hat{\blue{\mathcal{S}}}_{u,j}},
\end{align}
%
%
%
where 
$M$ is the number of \glspl{ru} being deployed, $\blue{\mathcal{N}}$ is the thermal noise\blue{, and $F_{UE,i}$ is the noise figure of $UE_i$}.
%
The \gls{sinr} $\Gamma_{i,j}$ considers the interference to the signal from $RU_i$ to $UE_j$ due to downlink transmissions of all other $M - 1$ \glspl{ru} being deployed.

In our \gls{rf} planning, we deploy two \glspl{ru} (i.e., $M=2$). In the following study, we consider scenarios where the first \gls{ru} serves one \gls{ue} from the test locations, while we assume that the second \gls{ru} creates interference with the first, even without being assigned any \gls{ue} from the list.
%
%
We test all possible combinations of the 24~\gls{ru} test locations, which, following the combinatorial equation of choosing 24 elements ($n$) in groups of~2 ($r$) as $C(n, r) = \frac{n!}{r!(n-r)!}$, results in a total of 276 pairs.
\blue{The proposed approach for determining the optimal \gls{ru} locations and the maximum average \gls{sinr} ($\Phi_{\max}(\Gamma)$), called score, is presented in Algorithm~\ref{algo:rfplanning}.
It takes as input the set of \gls{ru} locations ($\mathcal{R}$), the set of \gls{ue} test points ($\mathcal{U}$), and the \gls{sinr} matrix $\mathbf{\Gamma}$. Then, it performs an exhaustive search, testing all pairs of \gls{ru} against all \glspl{ue} to determine the optimal \gls{ru} pair $(p^*, q^*)$ with the best maximum average \gls{sinr} $\Phi_{\max}(\Gamma)$.}
%
%
%
%
%
\begin{algorithm}
\caption{\blue{Exhaustive Search Algorithm for RF Planning}}
\begin{algorithmic}[1]\label{algo:rfplanning}
\REQUIRE \blue{Set of RU locations ($\mathcal{R}$), set of UE test points ($\mathcal{U}$), precomputed SINR matrix $\mathbf{\Gamma}$}
\ENSURE \blue{Optimal RU pair $(p^*, q^*)$ and maximum average SINR $\Phi_{\max}(\Gamma)$}
\blue{
\STATE Initialize $bestScore \gets -\infty$ and $bestPair \gets \text{None}$
\FORALL{RU pairs $(p, q) \in \binom{\mathcal{R}}{2}$}
    \STATE $sumSINR \gets 0$
    \FORALL{UE $j \in \mathcal{U}$}
        \STATE Compute $\Gamma_{p,j}$ (RU $p$ serving, RU $q$ interfering)
        \STATE Compute $\Gamma_{q,j}$ (RU $q$ serving, RU $p$ interfering)
        \STATE $sinrMax \gets \max(\Gamma_{p,j}, \Gamma_{q,j})$
        \STATE $sumSINR \gets sumSINR + sinrMax$
    \ENDFOR
    \STATE $avgSINR \gets sumSINR / |\mathcal{U}|$
    \IF{$avgSINR > bestScore$}
        \STATE $bestScore \gets avgSINR$
        \STATE $bestPair \gets (p, q)$
    \ENDIF
\ENDFOR
\RETURN $bestPair, bestScore$
}
\end{algorithmic}
\end{algorithm}

We test this algorithm with different values of the attenuation $A_{RU}$, from $0$ to $50$\:dB in $10$\:dB increments.
\begin{figure}[hbt]
  \centering
  \subfloat[$0$\:dB attenuation]{\label{fig:heatmap-scores-0db}\includegraphics[width=0.49\linewidth]{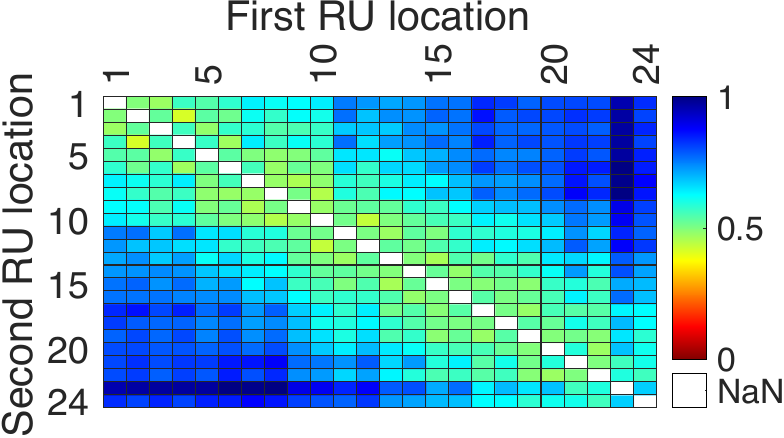}}
  \hfill
  \subfloat[$10$\:dB attenuation]{\label{fig:heatmap-scores-10db}\includegraphics[width=0.49\linewidth]{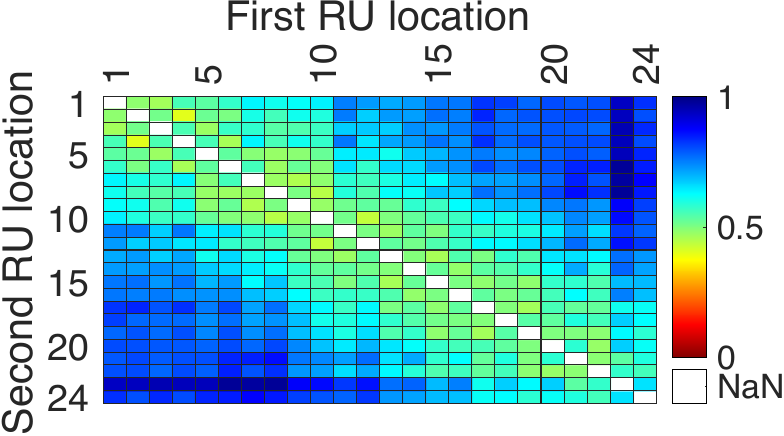}}
  \hfill
  \subfloat[$20$\:dB attenuation]
  {\label{fig:heatmap-scores-20db}\includegraphics[width=0.49\linewidth]{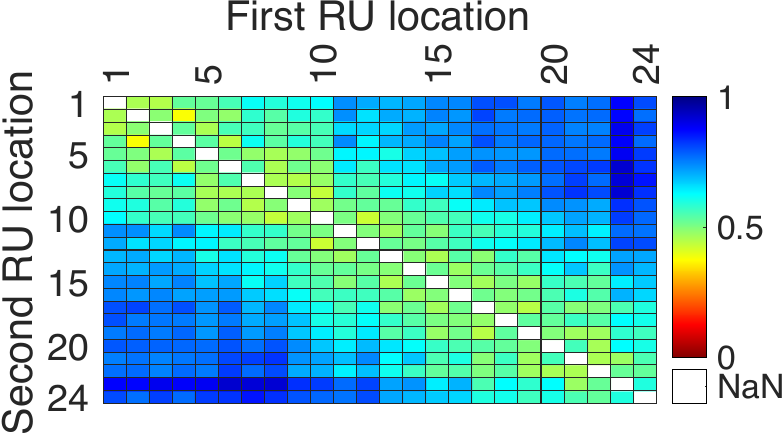}}
  \hfill
  \subfloat[$30$\:dB attenuation]
  {\label{fig:heatmap-scores-30db}\includegraphics[width=0.49\linewidth]{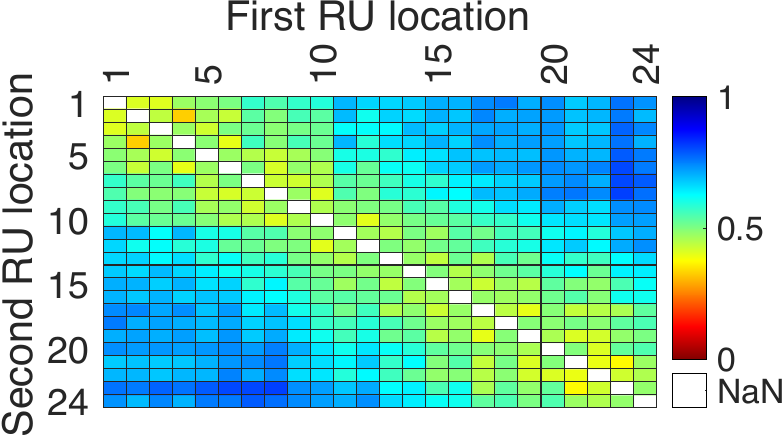}}
  \hfill
  \subfloat[$40$\:dB attenuation]{\label{fig:heatmap-scores-40db}\includegraphics[width=0.49\linewidth]{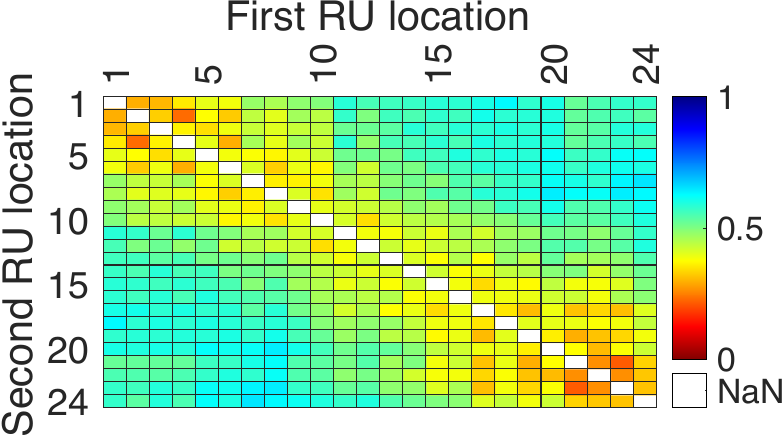}}
  \hfill
  \subfloat[$50$\:dB attenuation]{\label{fig:heatmap-scores-50db}    \includegraphics[width=0.49\linewidth]{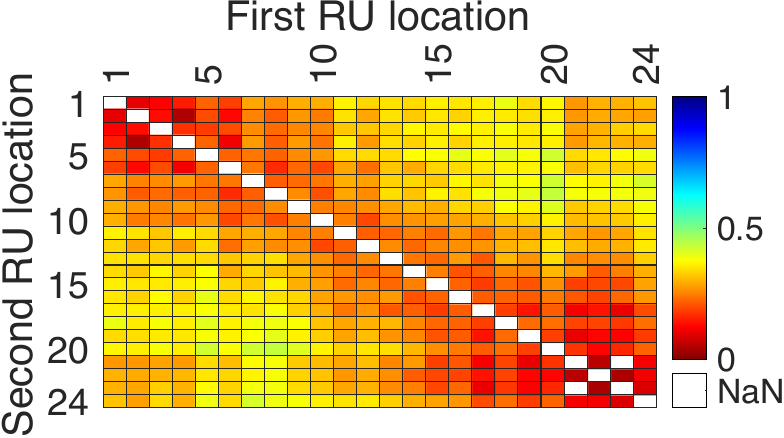}}
  \caption{Heatmap results of the normalized average \gls{sinr} $\blue{\Phi}(\Gamma)$ with 2 \glspl{ru}.}
  \label{fig:heatmap-scores}
\end{figure}
Figure~\ref{fig:heatmap-scores} visualizes the normalized values of \blue{the score $\Phi(\Gamma)$} for all possible combinations of \gls{ru} pairs and different attenuation values. Additionally, Table~\ref{table:score-results} provides the best \gls{ru} locations including the minimum and maximum values of $\blue{\Phi}(\Gamma)$ for the corresponding combinations.
As expected, locations with further \glspl{ru} exhibit higher average \gls{sinr} values, as they are less affected by interference. However, it is important to note that the score also considers coverage, \blue{as it is computed based on the \gls{sinr}}. Consequently, the optimal combination of locations identifies \glspl{ru} that are further apart but not necessarily the furthermost pair.
Considering these results, for the experiments in Section~\ref{sec:exp-results}, we select a TX attenuation of $20$\:dB, which exhibits a good trade-off between coverage and average \gls{sinr} values.
Moreover, during our real-world experiments, we observed that a $20$\:dB attenuation leads to increased system stability and reduced degradation compared to lower attenuation values, resulting in improved overall performance, as it reduces the likelihood of saturation at the \gls{ue} antenna.
Therefore, we select locations [6,23] for our \glspl{ru} deployment.

\begin{table}[t]
    \centering
    \footnotesize
    \caption{Best \glspl{ru} and average \gls{sinr} $\blue{\Phi}(\Gamma)$ range values.}
    \label{table:score-results}
    \begin{tabularx}{0.9\columnwidth}{
        >{\raggedright\arraybackslash\hsize=0.2\hsize}X
        >{\raggedright\arraybackslash\hsize=0.35\hsize}X
        >{\raggedright\arraybackslash\hsize=0.45\hsize}X }
        \toprule
        $A_{RU}$ [dB] & \gls{ru} locations with best \gls{sinr} & [Min, Max] $\blue{\blue{\Phi}}(\Gamma)$ [dB] \\
        \midrule
        0 & [8, 23] & [6.08, 23.33] \\
        10 & [6, 23] & [5.71, 22.66] \\
        20 & [6, 23] & [5.00, 21.03] \\
        30 & [8, 23] & [3.58, 17.82] \\
        40 & [7, 24] & [0.19, 12.94] \\
        50 & [8, 20] & [-6.23, 6.63] \\
        \bottomrule
    \end{tabularx}
\end{table}

\section{Experiment Results}
\label{sec:exp-results}



In this section, we describe the design and execution of a comprehensive set of experiments that illustrate the capabilities of the \testbed infrastructure in a variety of operational scenarios.
We assess the adaptability of the testbed through rigorous testing, utilizing iPerf to measure network throughput and MPEG-DASH to gauge video streaming quality. 
The experiments are \blue{mainly} conducted in the same indoor laboratory area modeled in Section~\ref{sec:ray-tracing}. They include static configurations with a single \gls{ue} as well as more complex setups with multiple \glspl{ue} and \glspl{ru}, \blue{leveraging the Keysight RuSIM emulator}, and scenarios with \gls{ue} mobility.

\subsection{Setup Overview}
\label{sec:exp-setup-overview}
%
\blue{We consider two different setups: (i)~Gigabyte \gls{ran} servers with a 2x2~\gls{mimo} configuration ($L_{DL}, L_{UL}$), 2~layers \gls{dl}, 1~layer \gls{ul}, a DDDSU \gls{tdd} pattern, and a modulation order ($Q_{m}$) up to 64-\gls{qam} (results for this setup are shown in Sections~\ref{sec:exp-static}, \ref{sec:exp-mobile}, and \ref{sec:exp-video}); and (ii)~\gls{gh} \gls{ran} servers with a 4x4~\gls{mimo} configuration, 4~layers \gls{dl}, 1~layer \gls{ul}, a DDDDDDSUUU \gls{tdd} pattern, and a $Q_{m}$ up to 256-\gls{qam} (Sections~\ref{sec:exp-peak} and \ref{sec:exp-long}).
All experiments utilize a carrier frequency of $3.75$\:GHz with a bandwidth ($\beta$) of $100$\:MHz.
}

%
%

\begin{figure}[t]
    \centering
    \includegraphics[width=1\linewidth]{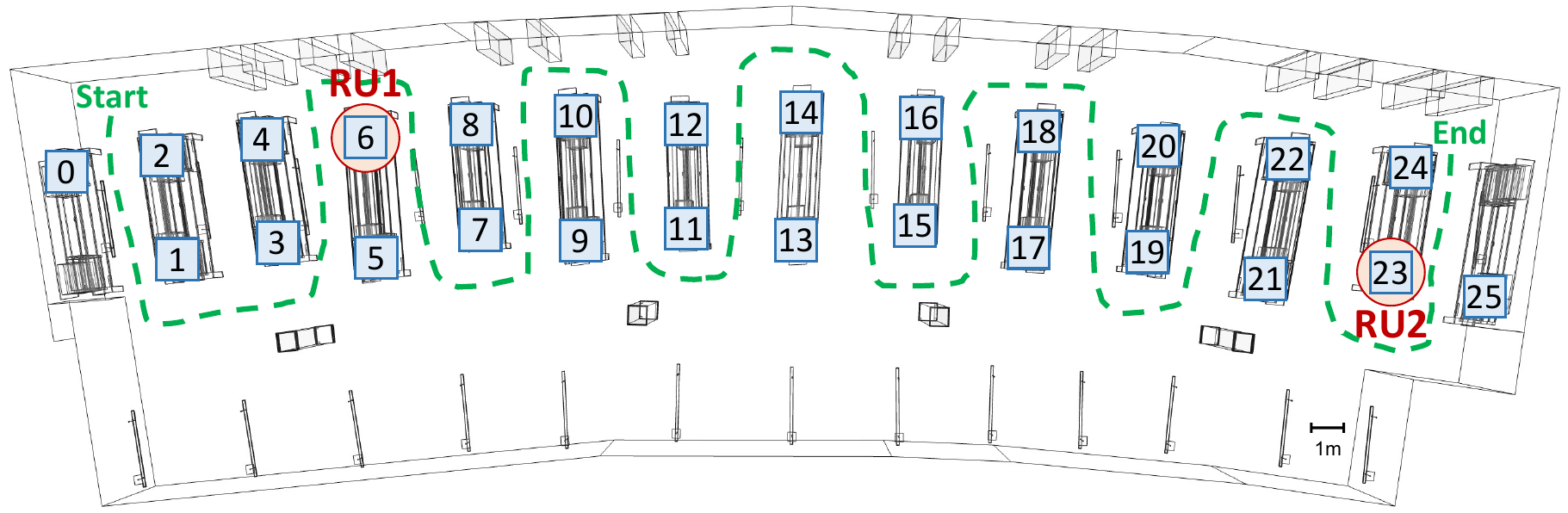}
    \setlength\belowcaptionskip{-.5cm}
    \caption{Node locations considered in our experiments: \glspl{ru} (red circles in 6 and 23); possible static \glspl{ue} (blue squares); and mobile \glspl{ue} (green dashed line).}
    \label{fig:node-locations}
\end{figure}

The experiments are \blue{mainly} conducted in the laboratory area shown in Figure~\ref{fig:node-locations}, which highlights the \gls{ru} locations (outcome of the ray-tracing study discussed in Section~\ref{sec:ray-tracing}), as well as the \gls{ue} locations and the mobility pattern for the non-static experiments. All tests involving a single \gls{ru} are conducted at location~6, as illustrated in Figure~\ref{fig:node-locations}.
%
%
%
%
%
An edge server, configured to support the iPerf and MPEG-DASH applications, is deployed within the campus network to ensure minimal latency, ranging from $1$ to $2$\:ms. During static throughput tests, \gls{tcp} backlogged traffic is transmitted first in the downlink and then in the uplink directions for $40$\:seconds each across different \gls{ue} configurations.
For video streaming, the server employs FFmpeg~\cite{ffmpeg} to deliver five distinct profiles simultaneously at various resolutions—ranging from 1080P at $250$~Mbps to 540P at $10$~Mbps—to the \glspl{ue}. On the device side, we leverage a pre-compiled iPerf3 binary for Android to generate \gls{tcp} traffic, and Google’s ExoPlayer for client-side video playback. 
Each set of experiments is replicated five times to ensure data reliability, with results including mean values and 95\% confidence intervals of the metrics plotted. These metrics encompass application layer measurements such as throughput, bitrate, and rebuffer ratio, alongside \gls{mac} layer metrics like \gls{sinr}, \gls{rsrp}, and \gls{mcs}, collected at the \gls{oai} \gls{gnb} level.



\begin{figure}[t]
\centering
    \subfloat[Throughput]{
    \label{fig:1ru1ue_static_iperf_th}
    \centering
    \setlength\fwidth{\linewidth}
    \setlength\fheight{.35\linewidth}
\begin{tikzpicture}
\pgfplotsset{every tick label/.append style={font=\scriptsize}}

\definecolor{darkgray176}{RGB}{176,176,176}
\definecolor{darkorange25512714}{RGB}{255,127,14}
\definecolor{darkorange2309111}{RGB}{230,91,11}
\definecolor{lightgray204}{RGB}{204,204,204}
\definecolor{steelblue31119180}{RGB}{31,119,180}


\begin{axis}[
width=0.951\fwidth,
height=\fheight,
at={(0\fwidth,0\fheight)},
legend cell align={left},
legend style={fill opacity=0.8, draw opacity=1, text opacity=1, draw=lightgray204, font=\footnotesize},
legend columns=2,
x grid style={darkgray176},
xmajorticks=false,
xmin=-0.69, xmax=10.09,
xtick={0.2,1.2,2.2,3.2,4.2,5.2,6.2,7.2,8.2,9.2},
xticklabel style={rotate=45.0},
y grid style={darkgray176},
ylabel=\textcolor{steelblue31119180}{DL TH [Mbps]},
ylabel style={font=\scriptsize},
xlabel style={font=\scriptsize},
ymin=0, ymax=350,
ytick pos=left,
ytick style={color=steelblue31119180},
yticklabel style={color=steelblue31119180},
xmajorgrids,
ymajorgrids
]


\addlegendimage{ybar,ybar legend,draw=black,fill=steelblue31119180,postaction={pattern=north east lines,pattern color=black}}
\addlegendentry{DL}
\addlegendimage{ybar,ybar legend,draw=black,fill=darkorange25512714,postaction={pattern=north west lines,pattern color=black}}
\addlegendentry{UL}


\draw[draw=black,fill=steelblue31119180,postaction={pattern=north east lines,pattern color=black}] (axis cs:-0.2,0) rectangle (axis cs:0.2,297.779273813333);
\draw[draw=black,fill=steelblue31119180,postaction={pattern=north east lines,pattern color=black}] (axis cs:0.8,0) rectangle (axis cs:1.2,278.951601066667);
\draw[draw=black,fill=steelblue31119180,postaction={pattern=north east lines,pattern color=black}] (axis cs:1.8,0) rectangle (axis cs:2.2,291.50394624);
\draw[draw=black,fill=steelblue31119180,postaction={pattern=north east lines,pattern color=black}] (axis cs:2.8,0) rectangle (axis cs:3.2,281.087304533333);
\draw[draw=black,fill=steelblue31119180,postaction={pattern=north east lines,pattern color=black}] (axis cs:3.8,0) rectangle (axis cs:4.2,286.19221888);
\draw[draw=black,fill=steelblue31119180,postaction={pattern=north east lines,pattern color=black}] (axis cs:4.8,0) rectangle (axis cs:5.2,293.477207893333);
\draw[draw=black,fill=steelblue31119180,postaction={pattern=north east lines,pattern color=black}] (axis cs:5.8,0) rectangle (axis cs:6.2,281.056259413333);
\draw[draw=black,fill=steelblue31119180,postaction={pattern=north east lines,pattern color=black}] (axis cs:6.8,0) rectangle (axis cs:7.2,282.361967786667);
\draw[draw=black,fill=steelblue31119180,postaction={pattern=north east lines,pattern color=black}] (axis cs:7.8,0) rectangle (axis cs:8.2,226.071218773333);
\draw[draw=black,fill=steelblue31119180,postaction={pattern=north east lines,pattern color=black}] (axis cs:8.8,0) rectangle (axis cs:9.2,177.708098986667);


\path [draw=black, line width=1pt]
(axis cs:0,293.561317740977)
--(axis cs:0,301.997229885689);
\path [draw=black, line width=1pt]
(axis cs:1,274.913928006811)
--(axis cs:1,282.989274126523);
\path [draw=black, line width=1pt]
(axis cs:2,285.413594787131)
--(axis cs:2,297.594297692869);
\path [draw=black, line width=1pt]
(axis cs:3,274.57133522018)
--(axis cs:3,287.603273846487);
\path [draw=black, line width=1pt]
(axis cs:4,280.463374578279)
--(axis cs:4,291.921063181721);
\path [draw=black, line width=1pt]
(axis cs:5,289.103429270324)
--(axis cs:5,297.850986516343);
\path [draw=black, line width=1pt]
(axis cs:6,276.853204662378)
--(axis cs:6,285.259314164289);
\path [draw=black, line width=1pt]
(axis cs:7,278.164506604369)
--(axis cs:7,286.559428968964);
\path [draw=black, line width=1pt]
(axis cs:8,215.497162859975)
--(axis cs:8,236.645274686691);
\path [draw=black, line width=1pt]
(axis cs:9,163.498710494908)
--(axis cs:9,191.917487478425);


\addplot [semithick, black, mark=-, mark size=1.5, mark options={solid}, only marks]
table {%
0 293.561317740977
1 274.913928006811
2 285.413594787131
3 274.57133522018
4 280.463374578279
5 289.103429270324
6 276.853204662378
7 278.164506604369
8 215.497162859975
9 163.498710494908
};


\addplot [semithick, black, mark=-, mark size=1.5, mark options={solid}, only marks]
table {%
0 301.997229885689
1 282.989274126523
2 297.594297692869
3 287.603273846487
4 291.921063181721
5 297.850986516343
6 285.259314164289
7 286.559428968964
8 236.645274686691
9 191.917487478425
};
\end{axis}


\begin{axis}[
width=0.951\fwidth,
height=\fheight,
at={(0\fwidth,0\fheight)},
axis y line*=right,
legend cell align={left},
legend style={fill opacity=0.8, draw opacity=1, text opacity=1, draw=lightgray204, font=\footnotesize},
legend columns=2,
x grid style={darkgray176},
xmin=-0.69, xmax=10.09,
xtick pos=left,
ytick style={color=darkorange2309111},
xtick={0.2,1.2,2.2,3.2,4.2,5.2,6.2,7.2,8.2,9.2},
xticklabels={0,2,4,6,8,10,12,14,16,18},
y grid style={darkgray176},
ylabel=\textcolor{darkorange2309111}{UL TH [Mbps]},
ylabel style={font=\scriptsize},
xlabel style={font=\scriptsize},
xlabel={UE Location},
ymin=0, ymax=50,
ytick pos=right,
ytick style={color=darkorange2309111},
yticklabel style={anchor=west, color=darkorange2309111},
ylabel shift=-5pt
]


\draw[draw=black,fill=darkorange25512714,postaction={pattern=north west lines,pattern color=black}] (axis cs:0.2,0) rectangle (axis cs:0.6,29.4579950933333);
\draw[draw=black,fill=darkorange25512714,postaction={pattern=north west lines,pattern color=black}] (axis cs:1.2,0) rectangle (axis cs:1.6,34.2534826666667);
\draw[draw=black,fill=darkorange25512714,postaction={pattern=north west lines,pattern color=black}] (axis cs:2.2,0) rectangle (axis cs:2.6,32.65265664);
\draw[draw=black,fill=darkorange25512714,postaction={pattern=north west lines,pattern color=black}] (axis cs:3.2,0) rectangle (axis cs:3.6,36.9588087466667);
\draw[draw=black,fill=darkorange25512714,postaction={pattern=north west lines,pattern color=black}] (axis cs:4.2,0) rectangle (axis cs:4.6,34.6589320533333);
\draw[draw=black,fill=darkorange25512714,postaction={pattern=north west lines,pattern color=black}] (axis cs:5.2,0) rectangle (axis cs:5.6,25.6132164266667);
\draw[draw=black,fill=darkorange25512714,postaction={pattern=north west lines,pattern color=black}] (axis cs:6.2,0) rectangle (axis cs:6.6,19.6153617066667);
\draw[draw=black,fill=darkorange25512714,postaction={pattern=north west lines,pattern color=black}] (axis cs:7.2,0) rectangle (axis cs:7.6,16.84013056);
\draw[draw=black,fill=darkorange25512714,postaction={pattern=north west lines,pattern color=black}] (axis cs:8.2,0) rectangle (axis cs:8.6,2.56551594666667);
\draw[draw=black,fill=darkorange25512714,postaction={pattern=north west lines,pattern color=black}] (axis cs:9.2,0) rectangle (axis cs:9.6,1.18139562666667);


\path [draw=black, line width=1pt]
(axis cs:0.4,28.0220236375514)
--(axis cs:0.4,30.8939665491152);
\path [draw=black, line width=1pt]
(axis cs:1.4,33.0880112122231)
--(axis cs:1.4,35.4189541211102);
\path [draw=black, line width=1pt]
(axis cs:2.4,31.3074846346825)
--(axis cs:2.4,33.9978286453175);
\path [draw=black, line width=1pt]
(axis cs:3.4,36.5276271183574)
--(axis cs:3.4,37.3899903749759);
\path [draw=black, line width=1pt]
(axis cs:4.4,33.8209615410315)
--(axis cs:4.4,35.4969025656351);
\path [draw=black, line width=1pt]
(axis cs:5.4,24.0539830782761)
--(axis cs:5.4,27.1724497750572);
\path [draw=black, line width=1pt]
(axis cs:6.4,18.2651274938699)
--(axis cs:6.4,20.9655959194635);
\path [draw=black, line width=1pt]
(axis cs:7.4,15.5330246472272)
--(axis cs:7.4,18.1472364727728);
\path [draw=black, line width=1pt]
(axis cs:8.4,2.16213861293845)
--(axis cs:8.4,2.96889328039488);
\path [draw=black, line width=1pt]
(axis cs:9.4,0.995143241485241)
--(axis cs:9.4,1.36764801184809);


\addplot [semithick, black, mark=-, mark size=1.5, mark options={solid}, only marks]
table {%
0.4 28.0220236375514
1.4 33.0880112122231
2.4 31.3074846346825
3.4 36.5276271183574
4.4 33.8209615410315
5.4 24.0539830782761
6.4 18.2651274938699
7.4 15.5330246472272
8.4 2.16213861293845
9.4 0.995143241485241
};


\addplot [semithick, black, mark=-, mark size=1.5, mark options={solid}, only marks]
table {%
0.4 30.8939665491152
1.4 35.4189541211102
2.4 33.9978286453175
3.4 37.3899903749759
4.4 35.4969025656351
5.4 27.1724497750572
6.4 20.9655959194635
7.4 18.1472364727728
8.4 2.96889328039488
9.4 1.36764801184809
};

\end{axis}

\end{tikzpicture}}
    
    \hfill
    
    \subfloat[\gls{rsrp}]{
    \label{fig:1ru1ue_static_iperf_rsrp}
    \centering
    \setlength\fwidth{\linewidth}
    \setlength\fheight{.35\linewidth}
\begin{tikzpicture}
\pgfplotsset{every tick label/.append style={font=\scriptsize}}

\definecolor{darkgray176}{RGB}{176,176,176}
\definecolor{darkorange25512714}{RGB}{255,127,14}
\definecolor{lightgray204}{RGB}{204,204,204}
\definecolor{steelblue31119180}{RGB}{31,119,180}


\begin{axis}[
width=0.951\fwidth,
height=\fheight,
at={(0\fwidth,0\fheight)},
legend cell align={left},
legend style={fill opacity=0.8, draw opacity=1, text opacity=1, draw=lightgray204, font=\footnotesize},
legend columns=2,
x grid style={darkgray176},
xmin=-0.69, xmax=10.09,
xtick style={color=black},
xtick={0.2,1.2,2.2,3.2,4.2,5.2,6.2,7.2,8.2,9.2},
xticklabels={0,2,4,6,8,10,12,14,16,18},
y grid style={darkgray176},
ylabel={RSRP [dBm]},
xlabel={UE Location},
ylabel style={font=\scriptsize},
xlabel style={font=\scriptsize},
ymin=-110, ymax=-75,
ytick pos=left,
xmajorgrids,
ymajorgrids
]


\addlegendimage{ybar,ybar legend,draw=black,fill=steelblue31119180,postaction={pattern=north east lines,pattern color=black}}
\addlegendentry{DL}


\draw[draw=black,fill=steelblue31119180,postaction={pattern=north east lines,pattern color=black}] (axis cs:-0.2,-110) rectangle (axis cs:0.2,-87.704347826087);
\draw[draw=black,fill=steelblue31119180,postaction={pattern=north east lines,pattern color=black}] (axis cs:0.8,-110) rectangle (axis cs:1.2,-82.6869565217391);
\draw[draw=black,fill=steelblue31119180,postaction={pattern=north east lines,pattern color=black}] (axis cs:1.8,-110) rectangle (axis cs:2.2,-81.2521739130435);
\draw[draw=black,fill=steelblue31119180,postaction={pattern=north east lines,pattern color=black}] (axis cs:2.8,-110) rectangle (axis cs:3.2,-81.2608695652174);
\draw[draw=black,fill=steelblue31119180,postaction={pattern=north east lines,pattern color=black}] (axis cs:3.8,-110) rectangle (axis cs:4.2,-83.9478260869565);
\draw[draw=black,fill=steelblue31119180,postaction={pattern=north east lines,pattern color=black}] (axis cs:4.8,-110) rectangle (axis cs:5.2,-87.4);
\draw[draw=black,fill=steelblue31119180,postaction={pattern=north east lines,pattern color=black}] (axis cs:5.8,-110) rectangle (axis cs:6.2,-93.2);
\draw[draw=black,fill=steelblue31119180,postaction={pattern=north east lines,pattern color=black}] (axis cs:6.8,-110) rectangle (axis cs:7.2,-94.8869565217391);
\draw[draw=black,fill=steelblue31119180,postaction={pattern=north east lines,pattern color=black}] (axis cs:7.8,-110) rectangle (axis cs:8.2,-98.704347826087);
\draw[draw=black,fill=steelblue31119180,postaction={pattern=north east lines,pattern color=black}] (axis cs:8.8,-110) rectangle (axis cs:9.2,-100.017391304348);


\addlegendimage{ybar,ybar legend,draw=black,fill=darkorange25512714,postaction={pattern=north west lines,pattern color=black}}
\addlegendentry{UL}


\draw[draw=black,fill=darkorange25512714,postaction={pattern=north west lines,pattern color=black}] (axis cs:0.2,-110) rectangle (axis cs:0.6,-87.9913043478261);
\draw[draw=black,fill=darkorange25512714,postaction={pattern=north west lines,pattern color=black}] (axis cs:1.2,-110) rectangle (axis cs:1.6,-83.3391304347826);
\draw[draw=black,fill=darkorange25512714,postaction={pattern=north west lines,pattern color=black}] (axis cs:2.2,-110) rectangle (axis cs:2.6,-82.4086956521739);
\draw[draw=black,fill=darkorange25512714,postaction={pattern=north west lines,pattern color=black}] (axis cs:3.2,-110) rectangle (axis cs:3.6,-81.0869565217391);
\draw[draw=black,fill=darkorange25512714,postaction={pattern=north west lines,pattern color=black}] (axis cs:4.2,-110) rectangle (axis cs:4.6,-83.3217391304348);
\draw[draw=black,fill=darkorange25512714,postaction={pattern=north west lines,pattern color=black}] (axis cs:5.2,-110) rectangle (axis cs:5.6,-87);
\draw[draw=black,fill=darkorange25512714,postaction={pattern=north west lines,pattern color=black}] (axis cs:6.2,-110) rectangle (axis cs:6.6,-93.6782608695652);
\draw[draw=black,fill=darkorange25512714,postaction={pattern=north west lines,pattern color=black}] (axis cs:7.2,-110) rectangle (axis cs:7.6,-94.904347826087);
\draw[draw=black,fill=darkorange25512714,postaction={pattern=north west lines,pattern color=black}] (axis cs:8.2,-110) rectangle (axis cs:8.6,-99.5391304347826);
\draw[draw=black,fill=darkorange25512714,postaction={pattern=north west lines,pattern color=black}] (axis cs:9.2,-110) rectangle (axis cs:9.6,-99.8550724637681);


\path [draw=black, line width=1pt]
(axis cs:0,-88.5521924932933)
--(axis cs:0,-86.8565031588807);
\path [draw=black, line width=1pt]
(axis cs:1,-83.5179907261917)
--(axis cs:1,-81.8559223172865);
\path [draw=black, line width=1pt]
(axis cs:2,-82.4865036072704)
--(axis cs:2,-80.0178442188166);
\path [draw=black, line width=1pt]
(axis cs:3,-82.1502509441677)
--(axis cs:3,-80.3714881862671);
\path [draw=black, line width=1pt]
(axis cs:4,-84.794320180374)
--(axis cs:4,-83.1013319935389);
\path [draw=black, line width=1pt]
(axis cs:5,-88.085693399849)
--(axis cs:5,-86.714306600151);
\path [draw=black, line width=1pt]
(axis cs:6,-94.1290629121523)
--(axis cs:6,-92.2709370878477);
\path [draw=black, line width=1pt]
(axis cs:7,-95.2560525581009)
--(axis cs:7,-94.5178604853773);
\path [draw=black, line width=1pt]
(axis cs:8,-99.3109476832644)
--(axis cs:8,-98.0976480689096);
\path [draw=black, line width=1pt]
(axis cs:9,-100.811918734186)
--(axis cs:9,-99.22286387451);


\addplot [semithick, black, mark=-, mark size=1.5, mark options={solid}, only marks]
table {%
0 -88.5521924932933
1 -83.5179907261917
2 -82.4865036072704
3 -82.15025094416779
4 -84.794320180374
5 -88.085693399849
6 -94.1290629121523
7 -95.2560525581009
8 -99.3109476832644
9 -100.811918734186
};


\addplot [semithick, black, mark=-, mark size=1.5, mark options={solid}, only marks]
table {%
0 -86.8565031588807
1 -81.8559223172865
2 -80.0178442188166
3 -80.3714881862671
4 -83.1013319935389
5 -86.714306600151
6 -92.2709370878477
7 -94.5178604853773
8 -98.0976480689096
9 -99.22286387451
};


\path [draw=black, line width=1pt]
(axis cs:0.4,-89.1190691933993)
--(axis cs:0.4,-86.8635395022529);
\path [draw=black, line width=1pt]
(axis cs:1.4,-84.1435803015243)
--(axis cs:1.4,-82.5346805670409);
\path [draw=black, line width=1pt]
(axis cs:2.4,-83.2022624633202)
--(axis cs:2.4,-81.6151288400276);
\path [draw=black, line width=1pt]
(axis cs:3.4,-82.2684306039206)
--(axis cs:3.4,-79.9054824395579);
\path [draw=black, line width=1pt]
(axis cs:4.4,-84.0417272649247)
--(axis cs:4.4,-82.6017509959449);
\path [draw=black, line width=1pt]
(axis cs:5.4,-87.5461186812728)
--(axis cs:5.4,-86.4538813187272);
\path [draw=black, line width=1pt]
(axis cs:6.4,-94.6292329123125)
--(axis cs:6.4,-92.7272888268179);
\path [draw=black, line width=1pt]
(axis cs:7.4,-95.7205640969140)
--(axis cs:7.4,-94.0881315552600);
\path [draw=black, line width=1pt]
(axis cs:8.4,-100.303317383006)
--(axis cs:8.4,-98.7749434865592);
\path [draw=black, line width=1pt]
(axis cs:9.4,-100.919087203785)
--(axis cs:9.4,-98.7910577237512);


\addplot [semithick, black, mark=-, mark size=1.5, mark options={solid}, only marks]
table {%
0.4 -89.1190691933993
1.4 -84.1435803015243
2.4 -83.2022624633202
3.4 -82.2684306039206
4.4 -84.0417272649247
5.4 -87.5461186812728
6.4 -94.6292329123125
7.4 -95.7205640969140
8.4 -100.303317383006
9.4 -100.919087203785
};


\addplot [semithick, black, mark=-, mark size=1.5, mark options={solid}, only marks]
table {%
0.4 -86.8635395022529
1.4 -82.5346805670409
2.4 -81.6151288400276
3.4 -79.9054824395579
4.4 -82.6017509959449
5.4 -86.4538813187272
6.4 -92.7272888268179
7.4 -94.0881315552600
8.4 -98.7749434865592
9.4 -98.7910577237512
};

\end{axis}

\end{tikzpicture}}

    \hfill
    
    \subfloat[\gls{mcs}]{
    \label{fig:1ru1ue_static_iperf_mcs}
    \centering
    \setlength\fwidth{\linewidth}
    \setlength\fheight{.35\linewidth}
\begin{tikzpicture}
\pgfplotsset{every tick label/.append style={font=\scriptsize}}

\definecolor{darkgray176}{RGB}{176,176,176}
\definecolor{darkorange25512714}{RGB}{255,127,14}
\definecolor{darkorange2309111}{RGB}{230,91,11}
\definecolor{lightgray204}{RGB}{204,204,204}
\definecolor{steelblue31119180}{RGB}{31,119,180}


\begin{axis}[
width=0.951\fwidth,
height=\fheight,
at={(0\fwidth,0\fheight)},
legend cell align={left},
legend style={fill opacity=0.8, draw opacity=1, text opacity=1, draw=lightgray204, font=\footnotesize},
legend columns=2,
x grid style={darkgray176},
xmajorticks=false,
xmin=-0.69, xmax=10.09,
xtick={0.2,1.2,2.2,3.2,4.2,5.2,6.2,7.2,8.2,9.2},
xticklabel style={rotate=45.0},
y grid style={darkgray176},
ylabel=\textcolor{steelblue31119180}{DL MCS},
ylabel style={font=\scriptsize},
xlabel style={font=\scriptsize},
ymin=0, ymax=30,
ytick pos=left,
ytick style={color=steelblue31119180},
yticklabel style={color=steelblue31119180},
xmajorgrids,
ymajorgrids
]


\addlegendimage{ybar,ybar legend,draw=black,fill=steelblue31119180,postaction={pattern=north east lines,pattern color=black}}
\addlegendentry{DL}
\addlegendimage{ybar,ybar legend,draw=black,fill=darkorange25512714,postaction={pattern=north west lines,pattern color=black}}
\addlegendentry{UL}


\draw[draw=black,fill=steelblue31119180,postaction={pattern=north east lines,pattern color=black}] (axis cs:-0.2,0) rectangle (axis cs:0.2,22.1130434782609);
\draw[draw=black,fill=steelblue31119180,postaction={pattern=north east lines,pattern color=black}] (axis cs:0.8,0) rectangle (axis cs:1.2,22.4434782608696);
\draw[draw=black,fill=steelblue31119180,postaction={pattern=north east lines,pattern color=black}] (axis cs:1.8,0) rectangle (axis cs:2.2,22.4260869565217);
\draw[draw=black,fill=steelblue31119180,postaction={pattern=north east lines,pattern color=black}] (axis cs:2.8,0) rectangle (axis cs:3.2,22.0782608695652);
\draw[draw=black,fill=steelblue31119180,postaction={pattern=north east lines,pattern color=black}] (axis cs:3.8,0) rectangle (axis cs:4.2,22.1826086956522);
\draw[draw=black,fill=steelblue31119180,postaction={pattern=north east lines,pattern color=black}] (axis cs:4.8,0) rectangle (axis cs:5.2,22.5565217391304);
\draw[draw=black,fill=steelblue31119180,postaction={pattern=north east lines,pattern color=black}] (axis cs:5.8,0) rectangle (axis cs:6.2,21.2260869565217);
\draw[draw=black,fill=steelblue31119180,postaction={pattern=north east lines,pattern color=black}] (axis cs:6.8,0) rectangle (axis cs:7.2,20.5391304347826);
\draw[draw=black,fill=steelblue31119180,postaction={pattern=north east lines,pattern color=black}] (axis cs:7.8,0) rectangle (axis cs:8.2,17.3304347826087);
\draw[draw=black,fill=steelblue31119180,postaction={pattern=north east lines,pattern color=black}] (axis cs:8.8,0) rectangle (axis cs:9.2,15.2956521739130);


\path [draw=black, line width=1pt]
(axis cs:0,20.5029589354697)
--(axis cs:0,23.7231280210521);
\path [draw=black, line width=1pt]
(axis cs:1,20.96147331055099)
--(axis cs:1,23.92548321118821);
\path [draw=black, line width=1pt]
(axis cs:2,21.04920470693957)
--(axis cs:2,23.80296920610383);
\path [draw=black, line width=1pt]
(axis cs:3,20.41257084297458)
--(axis cs:3,23.74395089615582);
\path [draw=black, line width=1pt]
(axis cs:4,20.75861253525148)
--(axis cs:4,23.60660485605292);
\path [draw=black, line width=1pt]
(axis cs:5,21.11654547015059)
--(axis cs:5,23.99649800811021);
\path [draw=black, line width=1pt]
(axis cs:6,19.64738623843593)
--(axis cs:6,22.80478767460747);
\path [draw=black, line width=1pt]
(axis cs:7,19.20610224696953)
--(axis cs:7,21.87215862259567);
\path [draw=black, line width=1pt]
(axis cs:8,13.53357149010824)
--(axis cs:8,21.12729807510916);
\path [draw=black, line width=1pt]
(axis cs:9,12.26084268238217)
--(axis cs:9,18.33046166544383);


\addplot [semithick, black, mark=-, mark size=1.5, mark options={solid}, only marks]
table {%
0 20.5029589354697
1 20.96147331055099
2 21.04920470693957
3 20.41257084297458
4 20.75861253525148
5 21.11654547015059
6 19.64738623843593
7 19.20610224696953
8 13.53357149010824
9 12.26084268238217
};


\addplot [semithick, black, mark=-, mark size=1.5, mark options={solid}, only marks]
table {%
0 23.7231280210521
1 23.92548321118821
2 23.80296920610383
3 23.74395089615582
4 23.60660485605292
5 23.99649800811021
6 22.80478767460747
7 21.87215862259567
8 21.12729807510916
9 18.33046166544383
};
\end{axis}


\begin{axis}[
width=0.951\fwidth,
height=\fheight,
at={(0\fwidth,0\fheight)},
axis y line*=right,
legend cell align={left},
legend style={fill opacity=0.8, draw opacity=1, text opacity=1, draw=lightgray204, font=\footnotesize},
legend columns=2,
x grid style={darkgray176},
xmin=-0.69, xmax=10.09,
xtick pos=left,
ytick style={color=darkorange2309111},
xtick={0.2,1.2,2.2,3.2,4.2,5.2,6.2,7.2,8.2,9.2},
xticklabels={0,2,4,6,8,10,12,14,16,18},
y grid style={darkgray176},
ylabel=\textcolor{darkorange2309111}{UL MCS},
ylabel style={font=\scriptsize},
xlabel style={font=\scriptsize},
xlabel={UE Location},
ymin=0, ymax=30,
ytick pos=right,
ytick style={color=darkorange2309111},
yticklabel style={anchor=west, color=darkorange2309111},
ylabel shift=-5pt
]


\draw[draw=black,fill=darkorange25512714,postaction={pattern=north west lines,pattern color=black}] (axis cs:0.2,0) rectangle (axis cs:0.6,15.6956521739130);
\draw[draw=black,fill=darkorange25512714,postaction={pattern=north west lines,pattern color=black}] (axis cs:1.2,0) rectangle (axis cs:1.6,15.2000000000000);
\draw[draw=black,fill=darkorange25512714,postaction={pattern=north west lines,pattern color=black}] (axis cs:2.2,0) rectangle (axis cs:2.6,15.1391304347826);
\draw[draw=black,fill=darkorange25512714,postaction={pattern=north west lines,pattern color=black}] (axis cs:3.2,0) rectangle (axis cs:3.6,14.6434782608696);
\draw[draw=black,fill=darkorange25512714,postaction={pattern=north west lines,pattern color=black}] (axis cs:4.2,0) rectangle (axis cs:4.6,14.7913043478261);
\draw[draw=black,fill=darkorange25512714,postaction={pattern=north west lines,pattern color=black}] (axis cs:5.2,0) rectangle (axis cs:5.6,15.8173913043478);
\draw[draw=black,fill=darkorange25512714,postaction={pattern=north west lines,pattern color=black}] (axis cs:6.2,0) rectangle (axis cs:6.6,15.8434782608696);
\draw[draw=black,fill=darkorange25512714,postaction={pattern=north west lines,pattern color=black}] (axis cs:7.2,0) rectangle (axis cs:7.6,16.4086956521739);
\draw[draw=black,fill=darkorange25512714,postaction={pattern=north west lines,pattern color=black}] (axis cs:8.2,0) rectangle (axis cs:8.6,16.6173913043478);
\draw[draw=black,fill=darkorange25512714,postaction={pattern=north west lines,pattern color=black}] (axis cs:9.2,0) rectangle (axis cs:9.6,9.89855072463768);


\path [draw=black, line width=1pt]
(axis cs:0.4,14.1844431015435)
--(axis cs:0.4,17.2068612462825);
\path [draw=black, line width=1pt]
(axis cs:1.4,13.79699570469953)
--(axis cs:1.4,16.60300429530047);
\path [draw=black, line width=1pt]
(axis cs:2.4,13.71326070006579)
--(axis cs:2.4,16.56500016949941);
\path [draw=black, line width=1pt]
(axis cs:3.4,13.47224930406393)
--(axis cs:3.4,15.81470721767527);
\path [draw=black, line width=1pt]
(axis cs:4.4,13.67920460966746)
--(axis cs:4.4,15.90340408698474);
\path [draw=black, line width=1pt]
(axis cs:5.4,13.92288118993618)
--(axis cs:5.4,17.71190141875942);
\path [draw=black, line width=1pt]
(axis cs:6.4,13.49419205218562)
--(axis cs:6.4,18.19276446955358);
\path [draw=black, line width=1pt]
(axis cs:7.4,14.21262399125452)
--(axis cs:7.4,18.60476731309328);
\path [draw=black, line width=1pt]
(axis cs:8.4,13.98299464351815)
--(axis cs:8.4,19.25178796517745);
\path [draw=black, line width=1pt]
(axis cs:9.4,3.45249171805185)
--(axis cs:9.4,16.34460973122351);


\addplot [semithick, black, mark=-, mark size=1.5, mark options={solid}, only marks]
table {%
0.4 14.1844431015435
1.4 13.79699570469953
2.4 13.71326070006579
3.4 13.47224930406393
4.4 13.67920460966746
5.4 13.92288118993618
6.4 13.49419205218562
7.4 14.21262399125452
8.4 13.98299464351815
9.4 3.45249171805185
};


\addplot [semithick, black, mark=-, mark size=1.5, mark options={solid}, only marks]
table {%
0.4 17.2068612462825
1.4 16.60300429530047
2.4 16.56500016949941
3.4 15.81470721767527
4.4 15.90340408698474
5.4 17.71190141875942
6.4 18.19276446955358
7.4 18.60476731309328
8.4 19.25178796517745
9.4 16.34460973122351
};

\end{axis}

\end{tikzpicture}}

    \hfill
    
    \subfloat[\gls{cqi}]{
    \label{fig:1ru1ue_static_iperf_cqi}
    \centering
    \setlength\fwidth{\linewidth}
    \setlength\fheight{.35\linewidth}
\begin{tikzpicture}
\pgfplotsset{every tick label/.append style={font=\scriptsize}}

\definecolor{darkgray176}{RGB}{176,176,176}
\definecolor{darkorange25512714}{RGB}{255,127,14}
\definecolor{lightgray204}{RGB}{204,204,204}
\definecolor{steelblue31119180}{RGB}{31,119,180}


\begin{axis}[
width=0.951\fwidth,
height=\fheight,
at={(0\fwidth,0\fheight)},
legend cell align={left},
legend style={fill opacity=0.8, draw opacity=1, text opacity=1, draw=lightgray204, font=\footnotesize},
legend columns=2,
x grid style={darkgray176},
xmin=-0.69, xmax=10.09,
xtick style={color=black},
xtick={0.2,1.2,2.2,3.2,4.2,5.2,6.2,7.2,8.2,9.2},
xticklabels={0,2,4,6,8,10,12,14,16,18},
y grid style={darkgray176},
ylabel={CQI},
xlabel={UE Location},
ylabel style={font=\scriptsize},
xlabel style={font=\scriptsize},
ymin=9, ymax=16,
ytick pos=left,
xmajorgrids,
ymajorgrids
]


\addlegendimage{ybar,ybar legend,draw=black,fill=steelblue31119180,postaction={pattern=north east lines,pattern color=black}}
\addlegendentry{DL}


\draw[draw=black,fill=steelblue31119180,postaction={pattern=north east lines,pattern color=black}] (axis cs:-0.2,9) rectangle (axis cs:0.2,14.0173913043478);
\draw[draw=black,fill=steelblue31119180,postaction={pattern=north east lines,pattern color=black}] (axis cs:0.8,9) rectangle (axis cs:1.2,15.0000000000000);
\draw[draw=black,fill=steelblue31119180,postaction={pattern=north east lines,pattern color=black}] (axis cs:1.8,9) rectangle (axis cs:2.2,14.8347826086957);
\draw[draw=black,fill=steelblue31119180,postaction={pattern=north east lines,pattern color=black}] (axis cs:2.8,9) rectangle (axis cs:3.2,14.5217391304348);
\draw[draw=black,fill=steelblue31119180,postaction={pattern=north east lines,pattern color=black}] (axis cs:3.8,9) rectangle (axis cs:4.2,14.9304347826087);
\draw[draw=black,fill=steelblue31119180,postaction={pattern=north east lines,pattern color=black}] (axis cs:4.8,9) rectangle (axis cs:5.2,14.6782608695652);
\draw[draw=black,fill=steelblue31119180,postaction={pattern=north east lines,pattern color=black}] (axis cs:5.8,9) rectangle (axis cs:6.2,12.8695652173913);
\draw[draw=black,fill=steelblue31119180,postaction={pattern=north east lines,pattern color=black}] (axis cs:6.8,9) rectangle (axis cs:7.2,12.3391304347826);
\draw[draw=black,fill=steelblue31119180,postaction={pattern=north east lines,pattern color=black}] (axis cs:7.8,9) rectangle (axis cs:8.2,10.9739130434783);
\draw[draw=black,fill=steelblue31119180,postaction={pattern=north east lines,pattern color=black}] (axis cs:8.8,9) rectangle (axis cs:9.2,9.93043478260870);


\addlegendimage{ybar,ybar legend,draw=black,fill=darkorange25512714,postaction={pattern=north west lines,pattern color=black}}
\addlegendentry{UL}


\draw[draw=black,fill=darkorange25512714,postaction={pattern=north west lines,pattern color=black}] (axis cs:0.2,9) rectangle (axis cs:0.6,14.1826086956522);
\draw[draw=black,fill=darkorange25512714,postaction={pattern=north west lines,pattern color=black}] (axis cs:1.2,9) rectangle (axis cs:1.6,14.9304347826087);
\draw[draw=black,fill=darkorange25512714,postaction={pattern=north west lines,pattern color=black}] (axis cs:2.2,9) rectangle (axis cs:2.6,14.6434782608696);
\draw[draw=black,fill=darkorange25512714,postaction={pattern=north west lines,pattern color=black}] (axis cs:3.2,9) rectangle (axis cs:3.6,14.6869565217391);
\draw[draw=black,fill=darkorange25512714,postaction={pattern=north west lines,pattern color=black}] (axis cs:4.2,9) rectangle (axis cs:4.6,14.9739130434783);
\draw[draw=black,fill=darkorange25512714,postaction={pattern=north west lines,pattern color=black}] (axis cs:5.2,9) rectangle (axis cs:5.6,14.9130434782609);
\draw[draw=black,fill=darkorange25512714,postaction={pattern=north west lines,pattern color=black}] (axis cs:6.2,9) rectangle (axis cs:6.6,13.0521739130435);
\draw[draw=black,fill=darkorange25512714,postaction={pattern=north west lines,pattern color=black}] (axis cs:7.2,9) rectangle (axis cs:7.6,12.3478260869565);
\draw[draw=black,fill=darkorange25512714,postaction={pattern=north west lines,pattern color=black}] (axis cs:8.2,9) rectangle (axis cs:8.6,10.6782608695652);
\draw[draw=black,fill=darkorange25512714,postaction={pattern=north west lines,pattern color=black}] (axis cs:9.2,9) rectangle (axis cs:9.6,9.94202898550725);


\path [draw=black, line width=1pt]
(axis cs:0,13.88609488761631)
--(axis cs:0,14.14868772107929);
\path [draw=black, line width=1pt]
(axis cs:1,14.9999)
--(axis cs:1,15.0001);
\path [draw=black, line width=1pt]
(axis cs:2,14.461780685399575)
--(axis cs:2,15.207784532991825);
\path [draw=black, line width=1pt]
(axis cs:3,14.020025818732202)
--(axis cs:3,15.023452442137398);
\path [draw=black, line width=1pt]
(axis cs:4,14.674908523772164)
--(axis cs:4,15.185961041445236);
\path [draw=black, line width=1pt]
(axis cs:5,14.209072954363661)
--(axis cs:5,15.147448784766739);
\path [draw=black, line width=1pt]
(axis cs:6,12.53131017164543)
--(axis cs:6,13.20782026313717);
\path [draw=black, line width=1pt]
(axis cs:7,11.86364441340896)
--(axis cs:7,12.81461645615624);
\path [draw=black, line width=1pt]
(axis cs:8,10.813821536464814)
--(axis cs:8,11.134004550491786);
\path [draw=black, line width=1pt]
(axis cs:9,9.674908523772164)
--(axis cs:9,10.185961041445236);


\addplot [semithick, black, mark=-, mark size=1.5, mark options={solid}, only marks]
table {%
0 13.88609488761631
1 14.9999
2 14.461780685399575
3 14.020025818732202
4 14.674908523772164
5 14.209072954363661
6 12.53131017164543
7 11.86364441340896
8 10.813821536464814
9 9.674908523772164
};


\addplot [semithick, black, mark=-, mark size=1.5, mark options={solid}, only marks]
table {%
0 14.14868772107929
1 15.0001
2 15.207784532991825
3 15.023452442137398
4 15.185961041445236
5 15.147448784766739
6 13.20782026313717
7 12.81461645615624
8 11.134004550491786
9 10.185961041445236
};


\path [draw=black, line width=1pt]
(axis cs:0.4,13.751726111091036)
--(axis cs:0.4,14.613491280213364);
\path [draw=black, line width=1pt]
(axis cs:1.4,14.674908523772165)
--(axis cs:1.4,15.185961041445235);
\path [draw=black, line width=1pt]
(axis cs:2.4,14.162410290550307)
--(axis cs:2.4,15.124546231188893);
\path [draw=black, line width=1pt]
(axis cs:3.4,14.221195168207128)
--(axis cs:3.4,15.152717875271072);
\path [draw=black, line width=1pt]
(axis cs:4.4,14.813821536464814)
--(axis cs:4.4,15.134004550491786);
\path [draw=black, line width=1pt]
(axis cs:5.4,14.630039002711876)
--(axis cs:5.4,15.196047953809924);
\path [draw=black, line width=1pt]
(axis cs:6.4,12.828823104842533)
--(axis cs:6.4,13.275524721244467);
\path [draw=black, line width=1pt]
(axis cs:7.4,11.851462498853784)
--(axis cs:7.4,12.844189675059216);
\path [draw=black, line width=1pt]
(axis cs:8.4,10.209072954363661)
--(axis cs:8.4,11.147448784766739);
\path [draw=black, line width=1pt]
(axis cs:9.4,9.651846160125718)
--(axis cs:9.4,10.232211810888782);


\addplot [semithick, black, mark=-, mark size=1.5, mark options={solid}, only marks]
table {%
0.4 13.751726111091036
1.4 14.674908523772165
2.4 14.162410290550307
3.4 14.221195168207128
4.4 14.813821536464814
5.4 14.630039002711876
6.4 12.828823104842533
7.4 11.851462498853784
8.4 10.209072954363661
9.4 9.651846160125718
};


\addplot [semithick, black, mark=-, mark size=1.5, mark options={solid}, only marks]
table {%
0.4 14.613491280213364
1.4 15.185961041445235
2.4 15.124546231188893
3.4 15.152717875271072
4.4 15.134004550491786
5.4 15.196047953809924
6.4 13.275524721244467
7.4 12.844189675059216
8.4 11.147448784766739
9.4 10.232211810888782
};

\end{axis}

\end{tikzpicture}}

    \hfill
    
    \subfloat[\gls{ph}]{
    \label{fig:1ru1ue_static_iperf_ph}
    \centering
    \setlength\fwidth{\linewidth}
    \setlength\fheight{.35\linewidth}
\begin{tikzpicture}
\pgfplotsset{every tick label/.append style={font=\scriptsize}}

\definecolor{darkgray176}{RGB}{176,176,176}
\definecolor{darkorange25512714}{RGB}{255,127,14}
\definecolor{lightgray204}{RGB}{204,204,204}
\definecolor{steelblue31119180}{RGB}{31,119,180}


\begin{axis}[
width=0.951\fwidth,
height=\fheight,
at={(0\fwidth,0\fheight)},
legend cell align={left},
legend style={fill opacity=0.8, draw opacity=1, text opacity=1, draw=lightgray204, font=\footnotesize},
legend columns=2,
x grid style={darkgray176},
xmin=-0.69, xmax=10.09,
xtick style={color=black},
xtick={0.2,1.2,2.2,3.2,4.2,5.2,6.2,7.2,8.2,9.2},
xticklabels={0,2,4,6,8,10,12,14,16,18},
y grid style={darkgray176},
ylabel={PH [dB]},
xlabel={UE Location},
ylabel style={font=\scriptsize},
xlabel style={font=\scriptsize},
ymin=0, ymax=60,
ytick pos=left,
xmajorgrids,
ymajorgrids
]


\addlegendimage{ybar,ybar legend,draw=black,fill=steelblue31119180,postaction={pattern=north east lines,pattern color=black}}
\addlegendentry{DL}


\draw[draw=black,fill=steelblue31119180,postaction={pattern=north east lines,pattern color=black}] (axis cs:-0.2,0) rectangle (axis cs:0.2,44.8782608695652);
\draw[draw=black,fill=steelblue31119180,postaction={pattern=north east lines,pattern color=black}] (axis cs:0.8,0) rectangle (axis cs:1.2,44.2434782608696);
\draw[draw=black,fill=steelblue31119180,postaction={pattern=north east lines,pattern color=black}] (axis cs:1.8,0) rectangle (axis cs:2.2,48.5565217391304);
\draw[draw=black,fill=steelblue31119180,postaction={pattern=north east lines,pattern color=black}] (axis cs:2.8,0) rectangle (axis cs:3.2,43.9826086956522);
\draw[draw=black,fill=steelblue31119180,postaction={pattern=north east lines,pattern color=black}] (axis cs:3.8,0) rectangle (axis cs:4.2,44.6347826086957);
\draw[draw=black,fill=steelblue31119180,postaction={pattern=north east lines,pattern color=black}] (axis cs:4.8,0) rectangle (axis cs:5.2,41.1478260869565);
\draw[draw=black,fill=steelblue31119180,postaction={pattern=north east lines,pattern color=black}] (axis cs:5.8,0) rectangle (axis cs:6.2,34.8608695652174);
\draw[draw=black,fill=steelblue31119180,postaction={pattern=north east lines,pattern color=black}] (axis cs:6.8,0) rectangle (axis cs:7.2,33.4521739130435);
\draw[draw=black,fill=steelblue31119180,postaction={pattern=north east lines,pattern color=black}] (axis cs:7.8,0) rectangle (axis cs:8.2,18.2695652173913);
\draw[draw=black,fill=steelblue31119180,postaction={pattern=north east lines,pattern color=black}] (axis cs:8.8,0) rectangle (axis cs:9.2,15.4434782608696);


\addlegendimage{ybar,ybar legend,draw=black,fill=darkorange25512714,postaction={pattern=north west lines,pattern color=black}}
\addlegendentry{UL}


\draw[draw=black,fill=darkorange25512714,postaction={pattern=north west lines,pattern color=black}] (axis cs:0.2,0) rectangle (axis cs:0.6,39.4869565217391);
\draw[draw=black,fill=darkorange25512714,postaction={pattern=north west lines,pattern color=black}] (axis cs:1.2,0) rectangle (axis cs:1.6,40.3130434782609);
\draw[draw=black,fill=darkorange25512714,postaction={pattern=north west lines,pattern color=black}] (axis cs:2.2,0) rectangle (axis cs:2.6,39.9565217391304);
\draw[draw=black,fill=darkorange25512714,postaction={pattern=north west lines,pattern color=black}] (axis cs:3.2,0) rectangle (axis cs:3.6,41.0347826086957);
\draw[draw=black,fill=darkorange25512714,postaction={pattern=north west lines,pattern color=black}] (axis cs:4.2,0) rectangle (axis cs:4.6,40.8000000000000);
\draw[draw=black,fill=darkorange25512714,postaction={pattern=north west lines,pattern color=black}] (axis cs:5.2,0) rectangle (axis cs:5.6,38.6434782608696);
\draw[draw=black,fill=darkorange25512714,postaction={pattern=north west lines,pattern color=black}] (axis cs:6.2,0) rectangle (axis cs:6.6,37.4521739130435);
\draw[draw=black,fill=darkorange25512714,postaction={pattern=north west lines,pattern color=black}] (axis cs:7.2,0) rectangle (axis cs:7.6,37.1739130434783);
\draw[draw=black,fill=darkorange25512714,postaction={pattern=north west lines,pattern color=black}] (axis cs:8.2,0) rectangle (axis cs:8.6,28.4347826086957);
\draw[draw=black,fill=darkorange25512714,postaction={pattern=north west lines,pattern color=black}] (axis cs:9.2,0) rectangle (axis cs:9.6,17.5797101449275);


\path [draw=black, line width=1pt]
(axis cs:0,42.036999023357)
--(axis cs:0,47.7195227157734);
\path [draw=black, line width=1pt]
(axis cs:1,41.520633694535)
--(axis cs:1,46.9663228272042);
\path [draw=black, line width=1pt]
(axis cs:2,44.7126583126817)
--(axis cs:2,52.4003851655791);
\path [draw=black, line width=1pt]
(axis cs:3,37.2966944518586)
--(axis cs:3,50.6685239394458);
\path [draw=black, line width=1pt]
(axis cs:4,40.7951681480093)
--(axis cs:4,48.4743970693821);
\path [draw=black, line width=1pt]
(axis cs:5,38.2361331500274)
--(axis cs:5,44.0595190238856);
\path [draw=black, line width=1pt]
(axis cs:6,31.2531610624788)
--(axis cs:6,38.468578067956);
\path [draw=black, line width=1pt]
(axis cs:7,30.2315074290579)
--(axis cs:7,36.6728403970291);
\path [draw=black, line width=1pt]
(axis cs:8,16.4327022397091)
--(axis cs:8,20.1064281950735);
\path [draw=black, line width=1pt]
(axis cs:9,8.65572961678848)
--(axis cs:9,22.2312269049507);


\addplot [semithick, black, mark=-, mark size=1.5, mark options={solid}, only marks]
table {%
0 42.036999023357
1 41.520633694535
2 44.7126583126817
3 37.2966944518586
4 40.7951681480093
5 38.2361331500274
6 31.2531610624788
7 30.2315074290579
8 16.4327022397091
9 8.65572961678848
};


\addplot [semithick, black, mark=-, mark size=1.5, mark options={solid}, only marks]
table {%
0 47.7195227157734
1 46.9663228272042
2 52.4003851655791
3 50.6685239394458
4 48.4743970693821
5 44.0595190238856
6 38.468578067956
7 36.6728403970291
8 20.1064281950735
9 22.2312269049507
};


\path [draw=black, line width=1pt]
(axis cs:0.4,37.2970065207425)
--(axis cs:0.4,41.6769065227357);
\path [draw=black, line width=1pt]
(axis cs:1.4,38.461703435418)
--(axis cs:1.4,42.1643835211038);
\path [draw=black, line width=1pt]
(axis cs:2.4,38.2023722508727)
--(axis cs:2.4,41.7106712273881);
\path [draw=black, line width=1pt]
(axis cs:3.4,39.6909831067427)
--(axis cs:3.4,42.3785821106487);
\path [draw=black, line width=1pt]
(axis cs:4.4,38.5868019487549)
--(axis cs:4.4,43.0131980512451);
\path [draw=black, line width=1pt]
(axis cs:5.4,36.4462259721833)
--(axis cs:5.4,40.8407305495559);
\path [draw=black, line width=1pt]
(axis cs:6.4,35.7239587387127)
--(axis cs:6.4,39.1803890873743);
\path [draw=black, line width=1pt]
(axis cs:7.4,35.5076278087119)
--(axis cs:7.4,38.8401982782447);
\path [draw=black, line width=1pt]
(axis cs:8.4,24.4608976061157)
--(axis cs:8.4,32.4086676112757);
\path [draw=black, line width=1pt]
(axis cs:9.4,10.4461443816284)
--(axis cs:9.4,24.7132759082266);


\addplot [semithick, black, mark=-, mark size=1.5, mark options={solid}, only marks]
table {%
0.4 37.2970065207425
1.4 38.461703435418
2.4 38.2023722508727
3.4 39.6909831067427
4.4 38.5868019487549
5.4 36.4462259721833
6.4 35.7239587387127
7.4 35.5076278087119
8.4 24.4608976061157
9.4 10.4461443816284
};


\addplot [semithick, black, mark=-, mark size=1.5, mark options={solid}, only marks]
table {%
0.4 41.6769065227357
1.4 42.1643835211038
2.4 41.7106712273881
3.4 42.3785821106487
4.4 43.0131980512451
5.4 40.8407305495559
6.4 39.1803890873743
7.4 38.8401982782447
8.4 32.4086676112757
9.4 24.7132759082266
};

\end{axis}

\end{tikzpicture}}

    \hfill
    
    \subfloat[\gls{bler}]{
    \label{fig:1ru1ue_static_iperf_bler}
    \centering
    \setlength\fwidth{\linewidth}
    \setlength\fheight{.35\linewidth}
\begin{tikzpicture}
\pgfplotsset{every tick label/.append style={font=\scriptsize}}

\definecolor{darkgray176}{RGB}{176,176,176}
\definecolor{darkorange25512714}{RGB}{255,127,14}
\definecolor{darkorange2309111}{RGB}{230,91,11}
\definecolor{lightgray204}{RGB}{204,204,204}
\definecolor{steelblue31119180}{RGB}{31,119,180}


\begin{axis}[
width=0.951\fwidth,
height=\fheight,
at={(0\fwidth,0\fheight)},
legend cell align={left},
legend style={fill opacity=0.8, draw opacity=1, text opacity=1, draw=lightgray204, font=\footnotesize},
legend columns=2,
x grid style={darkgray176},
xmajorticks=false,
xmin=-0.69, xmax=10.09,
xtick={0.2,1.2,2.2,3.2,4.2,5.2,6.2,7.2,8.2,9.2},
xticklabel style={rotate=45.0},
y grid style={darkgray176},
ylabel=\textcolor{steelblue31119180}{DL BLER (\%)},
ylabel style={font=\scriptsize},
xlabel style={font=\scriptsize},
ymin=0, ymax=20,
ytick pos=left,
ytick style={color=steelblue31119180},
yticklabel style={color=steelblue31119180},
xmajorgrids,
ymajorgrids
]


\addlegendimage{ybar,ybar legend,draw=black,fill=steelblue31119180,postaction={pattern=north east lines,pattern color=black}}
\addlegendentry{DL}
\addlegendimage{ybar,ybar legend,draw=black,fill=darkorange25512714,postaction={pattern=north west lines,pattern color=black}}
\addlegendentry{UL}


\draw[draw=black,fill=steelblue31119180,postaction={pattern=north east lines,pattern color=black}] (axis cs:-0.2,0) rectangle (axis cs:0.2,9.65520869565217);
\draw[draw=black,fill=steelblue31119180,postaction={pattern=north east lines,pattern color=black}] (axis cs:0.8,0) rectangle (axis cs:1.2,10.0563826086957);
\draw[draw=black,fill=steelblue31119180,postaction={pattern=north east lines,pattern color=black}] (axis cs:1.8,0) rectangle (axis cs:2.2,9.74211304347826);
\draw[draw=black,fill=steelblue31119180,postaction={pattern=north east lines,pattern color=black}] (axis cs:2.8,0) rectangle (axis cs:3.2,10.0126173913043);
\draw[draw=black,fill=steelblue31119180,postaction={pattern=north east lines,pattern color=black}] (axis cs:3.8,0) rectangle (axis cs:4.2,10.1451913043478);
\draw[draw=black,fill=steelblue31119180,postaction={pattern=north east lines,pattern color=black}] (axis cs:4.8,0) rectangle (axis cs:5.2,10.1578521739130);
\draw[draw=black,fill=steelblue31119180,postaction={pattern=north east lines,pattern color=black}] (axis cs:5.8,0) rectangle (axis cs:6.2,9.64549565217391);
\draw[draw=black,fill=steelblue31119180,postaction={pattern=north east lines,pattern color=black}] (axis cs:6.8,0) rectangle (axis cs:7.2,6.26393043478261);
\draw[draw=black,fill=steelblue31119180,postaction={pattern=north east lines,pattern color=black}] (axis cs:7.8,0) rectangle (axis cs:8.2,6.18401739130435);
\draw[draw=black,fill=steelblue31119180,postaction={pattern=north east lines,pattern color=black}] (axis cs:8.8,0) rectangle (axis cs:9.2,4.79395652173913);


\path [draw=black, line width=1pt]
(axis cs:0,5.48903857823112)
--(axis cs:0,13.82137881307322);
\path [draw=black, line width=1pt]
(axis cs:1,4.94947400010116)
--(axis cs:1,15.16329121729024);
\path [draw=black, line width=1pt]
(axis cs:2,5.06530920953951)
--(axis cs:2,14.41891687741701);
\path [draw=black, line width=1pt]
(axis cs:3,5.29759739435575)
--(axis cs:3,14.72763738825285);
\path [draw=black, line width=1pt]
(axis cs:4,4.97515481206821)
--(axis cs:4,15.31522779662739);
\path [draw=black, line width=1pt]
(axis cs:5,5.37444307204324)
--(axis cs:5,14.94126127578276);
\path [draw=black, line width=1pt]
(axis cs:6,5.21040864502897)
--(axis cs:6,14.08058265931885);
\path [draw=black, line width=1pt]
(axis cs:7,0.71300871944064)
--(axis cs:7,11.81485215012458);
\path [draw=black, line width=1pt]
(axis cs:8,0.000000)
--(axis cs:8,14.19494521607783);
\path [draw=black, line width=1pt]
(axis cs:9,0.000000)
--(axis cs:9,9.86617351753883);


\addplot [semithick, black, mark=-, mark size=1.5, mark options={solid}, only marks]
table {%
0 5.48903857823112
1 4.94947400010116
2 5.06530920953951
3 5.29759739435575
4 4.97515481206821
5 5.37444307204324
6 5.21040864502897
7 0.71300871944064
8 -1.82691043346913
9 -0.27826047406054
};


\addplot [semithick, black, mark=-, mark size=1.5, mark options={solid}, only marks]
table {%
0 13.82137881307322
1 15.16329121729024
2 14.41891687741701
3 14.72763738825285
4 15.31522779662739
5 14.94126127578276
6 14.08058265931885
7 11.81485215012458
8 14.19494521607783
9 9.86617351753883
};
\end{axis}


\begin{axis}[
width=0.951\fwidth,
height=\fheight,
at={(0\fwidth,0\fheight)},
axis y line*=right,
legend cell align={left},
legend style={fill opacity=0.8, draw opacity=1, text opacity=1, draw=lightgray204, font=\footnotesize},
legend columns=2,
x grid style={darkgray176},
xmin=-0.69, xmax=10.09,
xtick pos=left,
ytick style={color=darkorange2309111},
xtick={0.2,1.2,2.2,3.2,4.2,5.2,6.2,7.2,8.2,9.2},
xticklabels={0,2,4,6,8,10,12,14,16,18},
y grid style={darkgray176},
ylabel=\textcolor{darkorange2309111}{UL BLER (\%)},
ylabel style={font=\scriptsize},
xlabel style={font=\scriptsize},
xlabel={UE Location},
ymin=0, ymax=20,
ytick pos=right,
ytick style={color=darkorange2309111},
yticklabel style={anchor=west, color=darkorange2309111},
ylabel shift=-5pt
]


\draw[draw=black,fill=darkorange25512714,postaction={pattern=north west lines,pattern color=black}] (axis cs:0.2,0) rectangle (axis cs:0.6,9.56812173913044);
\draw[draw=black,fill=darkorange25512714,postaction={pattern=north west lines,pattern color=black}] (axis cs:1.2,0) rectangle (axis cs:1.6,9.50712173913043);
\draw[draw=black,fill=darkorange25512714,postaction={pattern=north west lines,pattern color=black}] (axis cs:2.2,0) rectangle (axis cs:2.6,9.27707826086957);
\draw[draw=black,fill=darkorange25512714,postaction={pattern=north west lines,pattern color=black}] (axis cs:3.2,0) rectangle (axis cs:3.6,9.17420000000000);
\draw[draw=black,fill=darkorange25512714,postaction={pattern=north west lines,pattern color=black}] (axis cs:4.2,0) rectangle (axis cs:4.6,9.45190434782609);
\draw[draw=black,fill=darkorange25512714,postaction={pattern=north west lines,pattern color=black}] (axis cs:5.2,0) rectangle (axis cs:5.6,9.82486956521739);
\draw[draw=black,fill=darkorange25512714,postaction={pattern=north west lines,pattern color=black}] (axis cs:6.2,0) rectangle (axis cs:6.6,9.78168695652174);
\draw[draw=black,fill=darkorange25512714,postaction={pattern=north west lines,pattern color=black}] (axis cs:7.2,0) rectangle (axis cs:7.6,9.39971304347826);
\draw[draw=black,fill=darkorange25512714,postaction={pattern=north west lines,pattern color=black}] (axis cs:8.2,0) rectangle (axis cs:8.6,5.20550434782609);
\draw[draw=black,fill=darkorange25512714,postaction={pattern=north west lines,pattern color=black}] (axis cs:9.2,0) rectangle (axis cs:9.6,0.81201449275362);


\path [draw=black, line width=1pt]
(axis cs:0.4,6.68591234005179)
--(axis cs:0.4,12.45033113820909);
\path [draw=black, line width=1pt]
(axis cs:1.4,6.4711601927064)
--(axis cs:1.4,12.54308328555446);
\path [draw=black, line width=1pt]
(axis cs:2.4,6.18722643012966)
--(axis cs:2.4,12.36693009160948);
\path [draw=black, line width=1pt]
(axis cs:3.4,6.1539307335098)
--(axis cs:3.4,12.1944692664902);
\path [draw=black, line width=1pt]
(axis cs:4.4,6.37777071442081)
--(axis cs:4.4,12.52603798123137);
\path [draw=black, line width=1pt]
(axis cs:5.4,6.61743310313804)
--(axis cs:5.4,13.03230602729674);
\path [draw=black, line width=1pt]
(axis cs:6.4,5.5995890267589)
--(axis cs:6.4,13.96378488628458);
\path [draw=black, line width=1pt]
(axis cs:7.4,5.65376859843443)
--(axis cs:7.4,13.14565748852209);
\path [draw=black, line width=1pt]
(axis cs:8.4,0.32297303824742)
--(axis cs:8.4,10.08803565740476);
\path [draw=black, line width=1pt]
(axis cs:9.4,0.000000)
--(axis cs:9.4,3.43483478899158);


\addplot [semithick, black, mark=-, mark size=1.5, mark options={solid}, only marks]
table {%
0.4 6.68591234005179
1.4 6.4711601927064
2.4 6.18722643012966
3.4 6.1539307335098
4.4 6.37777071442081
5.4 6.61743310313804
6.4 5.5995890267589
7.4 5.65376859843443
8.4 0.32297303824742
9.4 0.000000
};


\addplot [semithick, black, mark=-, mark size=1.5, mark options={solid}, only marks]
table {%
0.4 12.45033113820909
1.4 12.54308328555446
2.4 12.36693009160948
3.4 12.1944692664902
4.4 12.52603798123137
5.4 13.03230602729674
6.4 13.96378488628458
7.4 13.14565748852209
8.4 10.08803565740476
9.4 3.43483478899158
};

\end{axis}

\end{tikzpicture}}
    
\caption{Performance profiling with one \gls{ue} and single \gls{ru} for the static iPerf use case during \gls{dl} (blue bars) and \gls{ul} (orange bars) data transmissions.}
\label{fig:1ru1ue_static_iperf}
\vspace{-10pt}
\end{figure}
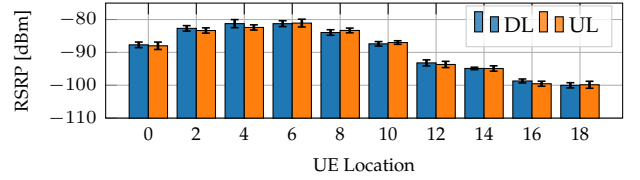
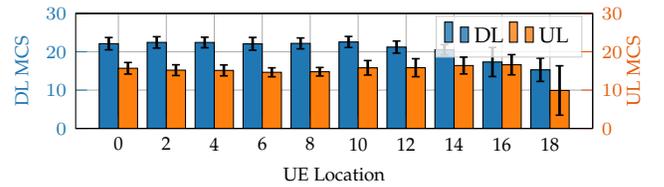
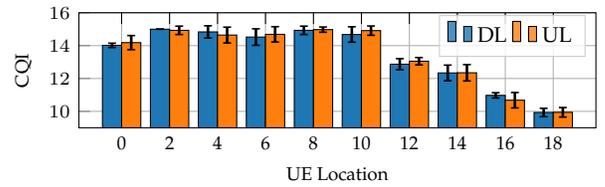
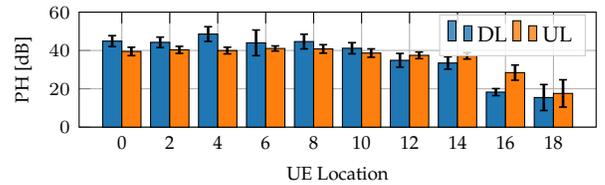
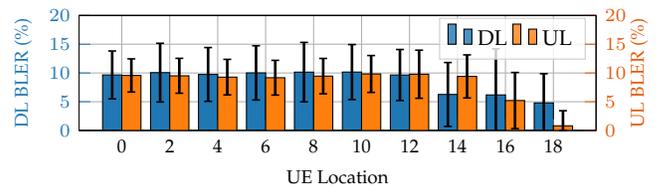

\subsection{Static Experiments}
\label{sec:exp-static}
\textbf{1 \gls{ue}, static, iPerf.} In the initial series of tests, we analyze the performance of a single \gls{ue} in ten static locations at varying distances from the \gls{ru}, as
shown in Figure~\ref{fig:1ru1ue_static_iperf}, \blue{using the first configuration setup of 2x2 \gls{mimo}, 2~layers \gls{dl}, 1~layer \gls{ul}, a DDDSU \gls{tdd} pattern, and a $Q_{m}$ up to 64-\gls{qam}}.

The iPerf throughput results in Figure~\ref{fig:1ru1ue_static_iperf_th} highlight the upper layer's responsiveness, showing a significant reduction from an average downlink throughput of $300$~Mbps and an uplink one of $38$~Mbps at locations near the \gls{ru}, to significantly lower rates of $177$~Mbps in downlink and $1.5$~Mbps in uplink at the most remote point, i.e., location 18. This high-level data throughput behavior is supported by corresponding shifts in the lower layers.

For example, this trend is clearly noticeable from the results of Figure~\ref{fig:1ru1ue_static_iperf_rsrp}, which shows the \gls{rsrp} values reported by the \gls{ue} to the \gls{gnb} during \gls{dl} (blue bars) and \gls{ul} (orange bars) data transmissions. As the distance from the \gls{ru} initially decreases starting from location~0 to location~6, the \gls{rsrp} values peak at around $-80$\:dBm. Subsequently, as the distance starts to increase again, moving from location~6 towards location~18, the \gls{rsrp} values begin to decrease, reaching as low as $-100$\:dBm.

In the same way, the \gls{mcs} values for both downlink and uplink initially mirror each other, then begin to diverge from around location~10, as shown in Figure~\ref{fig:1ru1ue_static_iperf_mcs}. In uplink, the \gls{mcs} values tend to remain stable or slightly increase, whereas downlink \gls{mcs} values experience a slight decline possibly due to power control strategies that better favor the uplink at more distant locations.

Similarly, the \gls{cqi} (Figure~\ref{fig:1ru1ue_static_iperf_cqi}) for both downlink and uplink starts closely matched but begins to show variation past the midpoint of the locations. The downlink \gls{cqi} experiences a modest decline, while the uplink \gls{cqi} sustains higher values. This suggests better channel conditions or more effective adaptation mechanisms in the uplink, due to adaptive power adjustments in uplink transmissions that maintain signal quality over distance.

\gls{ph} metrics, shown in Figure~\ref{fig:1ru1ue_static_iperf_ph}, reveal that downlink power headroom remains relatively stable across all locations, indicating a consistent application of power levels for downlink transmissions. In contrast, the uplink displays greater variability and generally higher values in distant locations to compensate for potential path loss and ensure that the transmit power remains adequate to maintain quality of service as the \gls{ue} moves further from the \gls{ru}.

Finally, the \gls{bler} (Figure~\ref{fig:1ru1ue_static_iperf_bler}) for downlink remains below 10\% for most locations, pointing to good reliability and effective adaptation of the \gls{mcs}. However, while the uplink \gls{bler} is generally low, it exhibits some peaks, especially around mid-range locations which may be related to specific lab obstacles or multipath effects and slow adaptation loops. 

\begin{figure}[t]
\centering
    \subfloat[Downlink]{
    \label{fig:dl_1ru2ue}
    \centering
    \setlength\fwidth{\linewidth}
    \setlength\fheight{.35\linewidth}
\begin{tikzpicture}
\pgfplotsset{every tick label/.append style={font=\scriptsize}}

\definecolor{darkgray176}{RGB}{176,176,176}
\definecolor{darkorange25512714}{RGB}{255,127,14}
\definecolor{lightgray204}{RGB}{204,204,204}
\definecolor{steelblue31119180}{RGB}{31,119,180}


\begin{axis}[
width=0.951\fwidth,
height=\fheight,
at={(0\fwidth,0\fheight)},
legend cell align={left},
legend columns=2,
legend style={fill opacity=0.8, draw opacity=1, text opacity=1, draw=lightgray204, at={(0.5,1.02)}, anchor=south, font =\footnotesize},
x grid style={darkgray176},
xmin=-0.69, xmax=10.09,
xtick style={color=black},
xtick={0.2,1.2,2.2,3.2,4.2,5.2,6.2,7.2,8.2,9.2},
xticklabels={0,2,4,6,8,10,12,14,16,18},
y grid style={darkgray176},
ylabel={DL Throughput [Mbps]},
xlabel={Location},
ylabel style={font=\scriptsize},
xlabel style={font=\scriptsize},
ymin=0, ymax=250,
ytick pos=left,
xmajorgrids,
ymajorgrids
]


\addlegendimage{ybar,ybar legend,draw=black,fill=steelblue31119180,postaction={pattern=north east lines,pattern color=black}}
\addlegendentry{UE1}


\addlegendimage{ybar,ybar legend,draw=black,fill=darkorange25512714,postaction={pattern=north west lines,pattern color=black}}
\addlegendentry{UE2}

\draw[draw=black,fill=steelblue31119180,postaction={pattern=north east lines,pattern color=black}] (axis cs:-0.2,0) rectangle (axis cs:0.2,209.31375104);
\draw[draw=black,fill=steelblue31119180,postaction={pattern=north east lines,pattern color=black}] (axis cs:0.8,0) rectangle (axis cs:1.2,186.617992838095);
\draw[draw=black,fill=steelblue31119180,postaction={pattern=north east lines,pattern color=black}] (axis cs:1.8,0) rectangle (axis cs:2.2,219.202990933333);
\draw[draw=black,fill=steelblue31119180,postaction={pattern=north east lines,pattern color=black}] (axis cs:2.8,0) rectangle (axis cs:3.2,217.906626986667);
\draw[draw=black,fill=steelblue31119180,postaction={pattern=north east lines,pattern color=black}] (axis cs:3.8,0) rectangle (axis cs:4.2,199.6595072);
\draw[draw=black,fill=steelblue31119180,postaction={pattern=north east lines,pattern color=black}] (axis cs:4.8,0) rectangle (axis cs:5.2,191.50603008);
\draw[draw=black,fill=steelblue31119180,postaction={pattern=north east lines,pattern color=black}] (axis cs:5.8,0) rectangle (axis cs:6.2,188.67632768);
\draw[draw=black,fill=steelblue31119180,postaction={pattern=north east lines,pattern color=black}] (axis cs:6.8,0) rectangle (axis cs:7.2,214.866522026667);
\draw[draw=black,fill=steelblue31119180,postaction={pattern=north east lines,pattern color=black}] (axis cs:7.8,0) rectangle (axis cs:8.2,202.916541866667);
\draw[draw=black,fill=steelblue31119180,postaction={pattern=north east lines,pattern color=black}] (axis cs:8.8,0) rectangle (axis cs:9.2,175.75314304);


\draw[draw=black,fill=darkorange25512714,postaction={pattern=north west lines,pattern color=black}] (axis cs:0.2,0) rectangle (axis cs:0.6,197.091720533333);
\draw[draw=black,fill=darkorange25512714,postaction={pattern=north west lines,pattern color=black}] (axis cs:1.2,0) rectangle (axis cs:1.6,178.269318095238);
\draw[draw=black,fill=darkorange25512714,postaction={pattern=north west lines,pattern color=black}] (axis cs:2.2,0) rectangle (axis cs:2.6,167.423474773333);
\draw[draw=black,fill=darkorange25512714,postaction={pattern=north west lines,pattern color=black}] (axis cs:3.2,0) rectangle (axis cs:3.6,178.305252693333);
\draw[draw=black,fill=darkorange25512714,postaction={pattern=north west lines,pattern color=black}] (axis cs:4.2,0) rectangle (axis cs:4.6,197.397266346667);
\draw[draw=black,fill=darkorange25512714,postaction={pattern=north west lines,pattern color=black}] (axis cs:5.2,0) rectangle (axis cs:5.6,206.263659946667);
\draw[draw=black,fill=darkorange25512714,postaction={pattern=north west lines,pattern color=black}] (axis cs:6.2,0) rectangle (axis cs:6.6,201.951863466667);
\draw[draw=black,fill=darkorange25512714,postaction={pattern=north west lines,pattern color=black}] (axis cs:7.2,0) rectangle (axis cs:7.6,201.720223573333);
\draw[draw=black,fill=darkorange25512714,postaction={pattern=north west lines,pattern color=black}] (axis cs:8.2,0) rectangle (axis cs:8.6,200.045983573333);
\draw[draw=black,fill=darkorange25512714,postaction={pattern=north west lines,pattern color=black}] (axis cs:9.2,0) rectangle (axis cs:9.6,187.889971626667);


\path [draw=black, line width=1pt]
(axis cs:0,201.847342649942)
--(axis cs:0,216.780159430058);
\path [draw=black, line width=1pt]
(axis cs:1,177.31164661874)
--(axis cs:1,195.924339057451);
\path [draw=black, line width=1pt]
(axis cs:2,214.074751221685)
--(axis cs:2,224.331230644982);
\path [draw=black, line width=1pt]
(axis cs:3,213.073034039615)
--(axis cs:3,222.740219933718);
\path [draw=black, line width=1pt]
(axis cs:4,189.806104456157)
--(axis cs:4,209.512909943843);
\path [draw=black, line width=1pt]
(axis cs:5,186.048853581522)
--(axis cs:5,196.963206578478);
\path [draw=black, line width=1pt]
(axis cs:6,180.538676605231)
--(axis cs:6,196.813978754769);
\path [draw=black, line width=1pt]
(axis cs:7,210.616087254379)
--(axis cs:7,219.116956798954);
\path [draw=black, line width=1pt]
(axis cs:8,200.64366464943)
--(axis cs:8,205.189419083904);
\path [draw=black, line width=1pt]
(axis cs:9,172.587760013351)
--(axis cs:9,178.918526066649);


\addplot [semithick, black, mark=-, mark size=1.5, mark options={solid}, only marks]
table {%
0 201.847342649942
1 177.31164661874
2 214.074751221685
3 213.073034039615
4 189.806104456157
5 186.048853581522
6 180.538676605231
7 210.616087254379
8 200.64366464943
9 172.587760013351
};


\addplot [semithick, black, mark=-, mark size=1.5, mark options={solid}, only marks]
table {%
0 216.780159430058
1 195.924339057451
2 224.331230644982
3 222.740219933718
4 209.512909943843
5 196.963206578478
6 196.813978754769
7 219.116956798954
8 205.189419083904
9 178.918526066649
};


\path [draw=black, line width=1pt]
(axis cs:0.4,189.979224182537)
--(axis cs:0.4,204.20421688413);
\path [draw=black, line width=1pt]
(axis cs:1.4,168.810219824794)
--(axis cs:1.4,187.728416365682);
\path [draw=black, line width=1pt]
(axis cs:2.4,160.967087065603)
--(axis cs:2.4,173.879862481064);
\path [draw=black, line width=1pt]
(axis cs:3.4,172.736335356394)
--(axis cs:3.4,183.874170030273);
\path [draw=black, line width=1pt]
(axis cs:4.4,190.338614817949)
--(axis cs:4.4,204.455917875384);
\path [draw=black, line width=1pt]
(axis cs:5.4,201.296433481144)
--(axis cs:5.4,211.23088641219);
\path [draw=black, line width=1pt]
(axis cs:6.4,197.585972034881)
--(axis cs:6.4,206.317754898452);
\path [draw=black, line width=1pt]
(axis cs:7.4,195.473528009698)
--(axis cs:7.4,207.966919136969);
\path [draw=black, line width=1pt]
(axis cs:8.4,196.563119925935)
--(axis cs:8.4,203.528847220731);
\path [draw=black, line width=1pt]
(axis cs:9.4,183.098790508473)
--(axis cs:9.4,192.681152744861);


\addplot [semithick, black, mark=-, mark size=1.5, mark options={solid}, only marks]
table {%
0.4 189.979224182537
1.4 168.810219824794
2.4 160.967087065603
3.4 172.736335356394
4.4 190.338614817949
5.4 201.296433481144
6.4 197.585972034881
7.4 195.473528009698
8.4 196.563119925935
9.4 183.098790508473
};


\addplot [semithick, black, mark=-, mark size=1.5, mark options={solid}, only marks]
table {%
0.4 204.20421688413
1.4 187.728416365682
2.4 173.879862481064
3.4 183.874170030273
4.4 204.455917875384
5.4 211.23088641219
6.4 206.317754898452
7.4 207.966919136969
8.4 203.528847220731
9.4 192.681152744861
};

\end{axis}

\end{tikzpicture}
    \setlength\abovecaptionskip{.05cm}}
    \hfill    
    \subfloat[Uplink]{
    \label{fig:ul_1ru2ue}
    \centering
        \setlength\fwidth{\linewidth}
        \setlength\fheight{.35\linewidth}
\begin{tikzpicture}
\pgfplotsset{every tick label/.append style={font=\scriptsize}}

\definecolor{darkgray176}{RGB}{176,176,176}
\definecolor{darkorange25512714}{RGB}{255,127,14}
\definecolor{lightgray204}{RGB}{204,204,204}
\definecolor{steelblue31119180}{RGB}{31,119,180}

\begin{axis}[
width=0.951\fwidth,
height=\fheight,
at={(0\fwidth,0\fheight)},
legend cell align={left},
legend columns=2,
legend style={fill opacity=0.8, draw opacity=1, text opacity=1, draw=lightgray204, at={(0.5,1.02)}, anchor=south, font =\footnotesize},
x grid style={darkgray176},
xmin=-0.69, xmax=10.09,
xtick style={color=black},
xtick={0.2,1.2,2.2,3.2,4.2,5.2,6.2,7.2,8.2,9.2},
xticklabels={0,2,4,6,8,10,12,14,16,18},
y grid style={darkgray176},
ylabel={UL Throughput [Mbps]},
xlabel={Location},
ylabel style={font=\scriptsize},
xlabel style={font=\scriptsize},
ymin=0, ymax=30,
ytick pos=left,
xmajorgrids,
ymajorgrids
]


\addlegendimage{ybar,ybar legend,draw=black,fill=steelblue31119180,postaction={pattern=north east lines,pattern color=black}}
\addlegendentry{UE1}


\addlegendimage{ybar,ybar legend,draw=black,fill=darkorange25512714,postaction={pattern=north west lines,pattern color=black}}
\addlegendentry{UE2}

\draw[draw=black,fill=steelblue31119180,postaction={pattern=north east lines,pattern color=black}] (axis cs:-0.2,0) rectangle (axis cs:0.2,15.9243741866667);
\draw[draw=black,fill=steelblue31119180,postaction={pattern=north east lines,pattern color=black}] (axis cs:0.8,0) rectangle (axis cs:1.2,16.58847232);
\draw[draw=black,fill=steelblue31119180,postaction={pattern=north east lines,pattern color=black}] (axis cs:1.8,0) rectangle (axis cs:2.2,22.79604224);
\draw[draw=black,fill=steelblue31119180,postaction={pattern=north east lines,pattern color=black}] (axis cs:2.8,0) rectangle (axis cs:3.2,21.8173713066667);
\draw[draw=black,fill=steelblue31119180,postaction={pattern=north east lines,pattern color=black}] (axis cs:3.8,0) rectangle (axis cs:4.2,20.69889024);
\draw[draw=black,fill=steelblue31119180,postaction={pattern=north east lines,pattern color=black}] (axis cs:4.8,0) rectangle (axis cs:5.2,20.3074218666667);
\draw[draw=black,fill=steelblue31119180,postaction={pattern=north east lines,pattern color=black}] (axis cs:5.8,0) rectangle (axis cs:6.2,15.2113425066667);
\draw[draw=black,fill=steelblue31119180,postaction={pattern=north east lines,pattern color=black}] (axis cs:6.8,0) rectangle (axis cs:7.2,9.48611754666667);
\draw[draw=black,fill=steelblue31119180,postaction={pattern=north east lines,pattern color=black}] (axis cs:7.8,0) rectangle (axis cs:8.2,11.6042410666667);
\draw[draw=black,fill=steelblue31119180,postaction={pattern=north east lines,pattern color=black}] (axis cs:8.8,0) rectangle (axis cs:9.2,4.04051285333333);
\draw[draw=black,fill=darkorange25512714,postaction={pattern=north west lines,pattern color=black}] (axis cs:0.2,0) rectangle (axis cs:0.6,24.30599168);

\draw[draw=black,fill=darkorange25512714,postaction={pattern=north west lines,pattern color=black}] (axis cs:1.2,0) rectangle (axis cs:1.6,23.6838365866667);
\draw[draw=black,fill=darkorange25512714,postaction={pattern=north west lines,pattern color=black}] (axis cs:2.2,0) rectangle (axis cs:2.6,21.3559978666667);
\draw[draw=black,fill=darkorange25512714,postaction={pattern=north west lines,pattern color=black}] (axis cs:3.2,0) rectangle (axis cs:3.6,21.1952162133333);
\draw[draw=black,fill=darkorange25512714,postaction={pattern=north west lines,pattern color=black}] (axis cs:4.2,0) rectangle (axis cs:4.6,20.5381085866667);
\draw[draw=black,fill=darkorange25512714,postaction={pattern=north west lines,pattern color=black}] (axis cs:5.2,0) rectangle (axis cs:5.6,20.1606212266667);
\draw[draw=black,fill=darkorange25512714,postaction={pattern=north west lines,pattern color=black}] (axis cs:6.2,0) rectangle (axis cs:6.6,18.1473553066667);
\draw[draw=black,fill=darkorange25512714,postaction={pattern=north west lines,pattern color=black}] (axis cs:7.2,0) rectangle (axis cs:7.6,10.0313770666667);
\draw[draw=black,fill=darkorange25512714,postaction={pattern=north west lines,pattern color=black}] (axis cs:8.2,0) rectangle (axis cs:8.6,5.48055722666667);
\draw[draw=black,fill=darkorange25512714,postaction={pattern=north west lines,pattern color=black}] (axis cs:9.2,0) rectangle (axis cs:9.6,3.08980394666667);
\path [draw=black, line width=1pt]
(axis cs:0,15.1854121927106)
--(axis cs:0,16.6633361806227);

\path [draw=black, line width=1pt]
(axis cs:1,15.9339611695867)
--(axis cs:1,17.2429834704133);

\path [draw=black, line width=1pt]
(axis cs:2,22.4809501886116)
--(axis cs:2,23.1111342913884);

\path [draw=black, line width=1pt]
(axis cs:3,21.4878175036607)
--(axis cs:3,22.1469251096727);

\path [draw=black, line width=1pt]
(axis cs:4,20.4020033048218)
--(axis cs:4,20.9957771751782);

\path [draw=black, line width=1pt]
(axis cs:5,19.6595932286659)
--(axis cs:5,20.9552505046674);

\path [draw=black, line width=1pt]
(axis cs:6,14.1801751599062)
--(axis cs:6,16.2425098534271);

\path [draw=black, line width=1pt]
(axis cs:7,8.36420179649132)
--(axis cs:7,10.608033296842);

\path [draw=black, line width=1pt]
(axis cs:8,10.3518558129289)
--(axis cs:8,12.8566263204044);

\path [draw=black, line width=1pt]
(axis cs:9,3.34300492484811)
--(axis cs:9,4.73802078181856);

\addplot [semithick, black, mark=-, mark size=1.5, mark options={solid}, only marks]
table {%
0 15.1854121927106
1 15.9339611695867
2 22.4809501886116
3 21.4878175036607
4 20.4020033048218
5 19.6595932286659
6 14.1801751599062
7 8.36420179649132
8 10.3518558129289
9 3.34300492484811
};
\addplot [semithick, black, mark=-, mark size=1.5, mark options={solid}, only marks]
table {%
0 16.6633361806227
1 17.2429834704133
2 23.1111342913884
3 22.1469251096727
4 20.9957771751782
5 20.9552505046674
6 16.2425098534271
7 10.608033296842
8 12.8566263204044
9 4.73802078181856
};
\path [draw=black, line width=1pt]
(axis cs:0.4,23.6666234045624)
--(axis cs:0.4,24.9453599554376);

\path [draw=black, line width=1pt]
(axis cs:1.4,23.0062202721671)
--(axis cs:1.4,24.3614529011663);

\path [draw=black, line width=1pt]
(axis cs:2.4,21.0196821373176)
--(axis cs:2.4,21.6923135960157);

\path [draw=black, line width=1pt]
(axis cs:3.4,20.8733946143571)
--(axis cs:3.4,21.5170378123096);

\path [draw=black, line width=1pt]
(axis cs:4.4,20.2846734632244)
--(axis cs:4.4,20.7915437101089);

\path [draw=black, line width=1pt]
(axis cs:5.4,19.5017001671464)
--(axis cs:5.4,20.8195422861869);

\path [draw=black, line width=1pt]
(axis cs:6.4,17.1807637278958)
--(axis cs:6.4,19.1139468854375);

\path [draw=black, line width=1pt]
(axis cs:7.4,8.89414285024027)
--(axis cs:7.4,11.1686112830931);

\path [draw=black, line width=1pt]
(axis cs:8.4,4.66292328647226)
--(axis cs:8.4,6.29819116686107);

\path [draw=black, line width=1pt]
(axis cs:9.4,2.62564703322214)
--(axis cs:9.4,3.55396086011119);

\addplot [semithick, black, mark=-, mark size=1.5, mark options={solid}, only marks]
table {%
0.4 23.6666234045624
1.4 23.0062202721671
2.4 21.0196821373176
3.4 20.8733946143571
4.4 20.2846734632244
5.4 19.5017001671464
6.4 17.1807637278958
7.4 8.89414285024027
8.4 4.66292328647226
9.4 2.62564703322214
};
\addplot [semithick, black, mark=-, mark size=1.5, mark options={solid}, only marks]
table {%
0.4 24.9453599554376
1.4 24.3614529011663
2.4 21.6923135960157
3.4 21.5170378123096
4.4 20.7915437101089
5.4 20.8195422861869
6.4 19.1139468854375
7.4 11.1686112830931
8.4 6.29819116686107
9.4 3.55396086011119
};
\end{axis}

\end{tikzpicture}
        \setlength\abovecaptionskip{.05cm}}
\caption{Performance profiling for one \gls{ru} and two static \glspl{ue} for the static iPerf use case.}
\label{fig:1RU2UE_Static}
\end{figure}
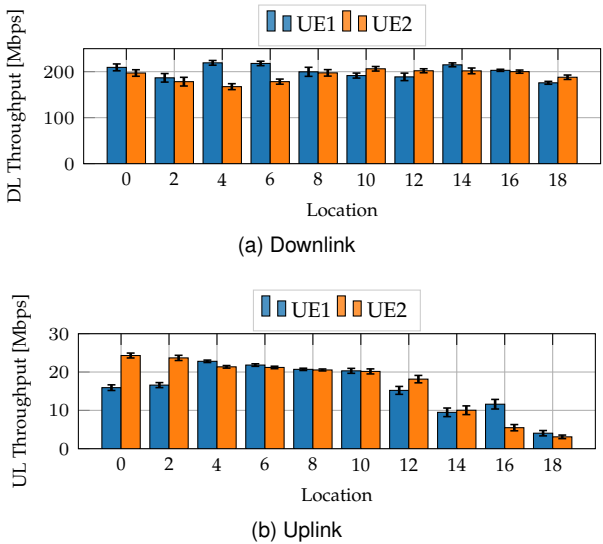

\textbf{2 \glspl{ue}, static, iPerf.} We test the performance of 2 \glspl{ue} for a single \gls{ru}. We position the \glspl{ue} at the same static locations as in the previous single \gls{ue} static case. The results are plotted in Figure~\ref{fig:1RU2UE_Static} for both downlink and uplink transmissions. We observe that, in most cases, the \glspl{ue} are able to share bandwidth fairly. The best achievable average aggregate throughput from both \glspl{ue} is around $400$\:Mbps in \gls{dl} and $44$\:Mbps in \gls{ul}. This shows that the total cell throughput can be higher than the single \gls{ue} throughput. As discussed in Table~\ref{table:testbeds-features}, this is due to a limitation in the number of transport blocks that can be acknowledged in a single slot for a single \gls{ue} \blue{in the case of the DDDSU \gls{tdd} pattern used in this first set of experiments}. Therefore, scheduling multiple \glspl{ue} improves the resource utilization of the system.


\textbf{1-4 \glspl{ue}, static, iPerf.}
We further extend our evaluation to include additional tests with multiple \glspl{ue}. At fixed location~4, we compare system performance with varying numbers of \glspl{ue} (from 1 to 4) connected to our network. The average throughput and $95$\% confidence intervals are plotted in Figure~\ref{fig:4ue_static}. We observe that the \glspl{ue} achieve steady throughput in all the cases, as indicated by the small confidence interval values. Additionally, the combined throughput increases with the number of \glspl{ue} connected: with four \glspl{ue}, the aggregate throughput reaches $512$\:Mbps in \gls{dl} and $46$\:Mbps in \gls{ul}.
This scenario highlights the maximum throughput performance that \testbed is able to achieve with the current \blue{2x2 \gls{mimo} configuration, featuring 2~layers in \gls{dl} and a DDDSU \gls{tdd} pattern, ensuring a fair distribution of resources among all \glspl{ue}.} It is worth noting that \blue{the peak performance of \testbed is detailed in Section~\ref{sec:exp-peak}}.

\begin{figure}[t]
\centering
    \setlength\fwidth{0.98\linewidth}
    \setlength\fheight{.3\linewidth}
    \input{figures/fig_tex/1ru4ue_static_iperf_th}
\vspace{-0.10in}
\caption{Performance profiling with multiple \glspl{ue} at fixed location~4 for the static iPerf use case \blue{using a DDDSU \gls{tdd} pattern, a 2x2 MIMO configuration, 2~layers DL and 1~layer UL}.}
\label{fig:4ue_static}
\end{figure}



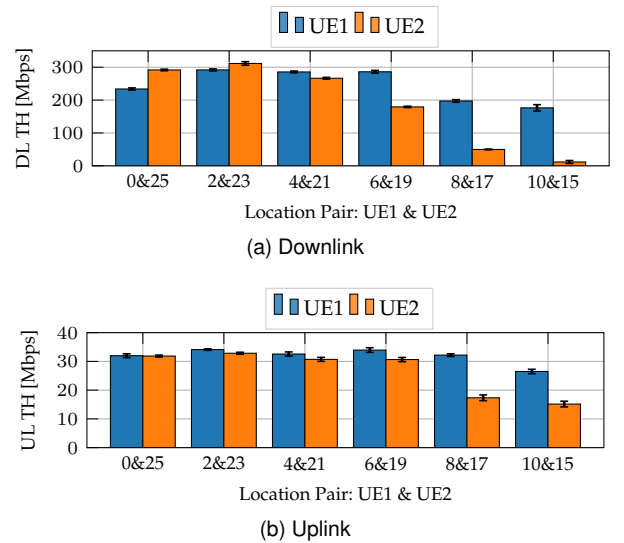
\begin{figure}[b]
\centering
    \subfloat[Downlink]{
    \label{fig:dl_2ru2ue_new}
    \centering
    \setlength\fwidth{\linewidth}
    \setlength\fheight{.35\linewidth}
\begin{tikzpicture}
\pgfplotsset{every tick label/.append style={font=\scriptsize}}

\definecolor{darkgray176}{RGB}{176,176,176}
\definecolor{darkorange25512714}{RGB}{255,127,14}
\definecolor{lightgray204}{RGB}{204,204,204}
\definecolor{steelblue31119180}{RGB}{31,119,180}

\begin{axis}[
width=0.951\fwidth,
height=\fheight,
at={(0\fwidth,0\fheight)},
legend cell align={left},
legend columns=2,
legend style={fill opacity=0.8, 
draw opacity=1, 
text opacity=1, 
draw=lightgray204, 
font=\footnotesize,
at={(0.67, 1.39)}},
x grid style={darkgray176},
xmin=-0.49, xmax=5.89,
xtick style={color=black},
xtick={0.2,1.2,2.2,3.2,4.2,5.2},
xticklabels={0\&25,2\&23,4\&21,6\&19,8\&17,10\&15},
y grid style={darkgray176},
ylabel={DL TH [Mbps]},
xlabel={Location Pair: UE1 \& UE2},
ylabel style={font=\scriptsize},
xlabel style={font=\scriptsize},
ymin=0, ymax=350,
ytick pos=left,
ytick style={color=black},
xmajorgrids,
ymajorgrids
]

\addlegendimage{ybar,ybar legend,draw=black,fill=steelblue31119180,postaction={pattern=north east lines,pattern color=black}}
\addlegendentry{UE1}

\draw[draw=black,fill=steelblue31119180,postaction={pattern=north east lines,pattern color=black}] (axis cs:-0.2,0) rectangle (axis cs:0.2,233.86456832);
\draw[draw=black,fill=steelblue31119180,postaction={pattern=north east lines,pattern color=black}] (axis cs:0.8,0) rectangle (axis cs:1.2,291.89891328);
\draw[draw=black,fill=steelblue31119180,postaction={pattern=north east lines,pattern color=black}] (axis cs:1.8,0) rectangle (axis cs:2.2,285.713666133333);
\draw[draw=black,fill=steelblue31119180,postaction={pattern=north east lines,pattern color=black}] (axis cs:2.8,0) rectangle (axis cs:3.2,286.180349866667);
\draw[draw=black,fill=steelblue31119180,postaction={pattern=north east lines,pattern color=black}] (axis cs:3.8,0) rectangle (axis cs:4.2,197.37538432);
\draw[draw=black,fill=steelblue31119180,postaction={pattern=north east lines,pattern color=black}] (axis cs:4.8,0) rectangle (axis cs:5.2,176.3935744);

\addlegendimage{ybar,ybar legend,draw=black,fill=darkorange25512714,postaction={pattern=north west lines,pattern color=black}}
\addlegendentry{UE2}

\draw[draw=black,fill=darkorange25512714,postaction={pattern=north west lines,pattern color=black}] (axis cs:0.2,0) rectangle (axis cs:0.6,291.741805226667);
\draw[draw=black,fill=darkorange25512714,postaction={pattern=north west lines,pattern color=black}] (axis cs:1.2,0) rectangle (axis cs:1.6,311.55290112);
\draw[draw=black,fill=darkorange25512714,postaction={pattern=north west lines,pattern color=black}] (axis cs:2.2,0) rectangle (axis cs:2.6,266.541028693333);
\draw[draw=black,fill=darkorange25512714,postaction={pattern=north west lines,pattern color=black}] (axis cs:3.2,0) rectangle (axis cs:3.6,179.292514986667);
\draw[draw=black,fill=darkorange25512714,postaction={pattern=north west lines,pattern color=black}] (axis cs:4.2,0) rectangle (axis cs:4.6,49.80736);
\draw[draw=black,fill=darkorange25512714,postaction={pattern=north west lines,pattern color=black}] (axis cs:5.2,0) rectangle (axis cs:5.6,11.7059211636364);

\path [draw=black, line width=1pt]
(axis cs:0,230.547928827315)
--(axis cs:0,237.181207812685);

\path [draw=black, line width=1pt]
(axis cs:1,288.8207988582)
--(axis cs:1,294.9770277018);

\path [draw=black, line width=1pt]
(axis cs:2,282.98840323712)
--(axis cs:2,288.438929029546);

\path [draw=black, line width=1pt]
(axis cs:3,282.045078669539)
--(axis cs:3,290.315621063794);

\path [draw=black, line width=1pt]
(axis cs:4,193.352877110513)
--(axis cs:4,201.397891529487);

\path [draw=black, line width=1pt]
(axis cs:5,166.642034928531)
--(axis cs:5,186.145113871469);

\addplot [semithick, black, mark=-, mark size=1.5, mark options={solid}, only marks]
table {%
0 230.547928827315
1 288.8207988582
2 282.98840323712
3 282.045078669539
4 193.352877110513
5 166.642034928531
};
\addplot [semithick, black, mark=-, mark size=1.5, mark options={solid}, only marks]
table {%
0 237.181207812685
1 294.9770277018
2 288.438929029546
3 290.315621063794
4 201.397891529487
5 186.145113871469
};

\path [draw=black, line width=1pt]
(axis cs:0.4,289.430103281427)
--(axis cs:0.4,294.053507171906);

\path [draw=black, line width=1pt]
(axis cs:1.4,306.152562756234)
--(axis cs:1.4,316.953239483766);

\path [draw=black, line width=1pt]
(axis cs:2.4,263.856146802523)
--(axis cs:2.4,269.225910584144);

\path [draw=black, line width=1pt]
(axis cs:3.4,177.173672478772)
--(axis cs:3.4,181.411357494562);

\path [draw=black, line width=1pt]
(axis cs:4.4,48.6648455686547)
--(axis cs:4.4,50.9498744313453);

\path [draw=black, line width=1pt]
(axis cs:5.4,7.21392651935996)
--(axis cs:5.4,16.1979158079128);

\addplot [semithick, black, mark=-, mark size=1.5, mark options={solid}, only marks]
table {%
0.4 289.430103281427
1.4 306.152562756234
2.4 263.856146802523
3.4 177.173672478772
4.4 48.6648455686547
5.4 7.21392651935996
};
\addplot [semithick, black, mark=-, mark size=1.5, mark options={solid}, only marks]
table {%
0.4 294.053507171906
1.4 316.953239483766
2.4 269.225910584144
3.4 181.411357494562
4.4 50.9498744313453
5.4 16.1979158079128
};

\end{axis}

\end{tikzpicture}
    \setlength\abovecaptionskip{.05cm}}
    \hfill    
    \subfloat[Uplink]{
    \label{fig:ul_2ru2ue_new}
    \centering
        \setlength\fwidth{\linewidth}
        \setlength\fheight{.35\linewidth}
\begin{tikzpicture}
\pgfplotsset{every tick label/.append style={font=\scriptsize}}

\definecolor{darkgray176}{RGB}{176,176,176}
\definecolor{darkorange25512714}{RGB}{255,127,14}
\definecolor{lightgray204}{RGB}{204,204,204}
\definecolor{steelblue31119180}{RGB}{31,119,180}

\begin{axis}[
width=0.951\fwidth,
height=\fheight,
at={(0\fwidth,0\fheight)},
legend cell align={left},
legend columns=2,
legend style={fill opacity=0.8, draw opacity=1, text opacity=1, draw=lightgray204, font=\footnotesize,at={(0.67, 1.39)}},
x grid style={darkgray176},
xmin=-0.49, xmax=5.89,
xtick style={color=black},
xtick={0.2,1.2,2.2,3.2,4.2,5.2},
xticklabels={0\&25,2\&23,4\&21,6\&19,8\&17,10\&15},
y grid style={darkgray176},
ylabel={UL TH [Mbps]},
xlabel={Location Pair: UE1 \& UE2},
ylabel style={font=\scriptsize},
xlabel style={font=\scriptsize},
ymin=0, ymax=40,
ytick pos=left,
ytick style={color=black},
xmajorgrids,
ymajorgrids
]

\addlegendimage{ybar,ybar legend,draw=black,fill=steelblue31119180,postaction={pattern=north east lines,pattern color=black}}
\addlegendentry{UE1}

\addlegendimage{ybar,ybar legend,draw=black,fill=darkorange25512714,postaction={pattern=north west lines,pattern color=black}}
\addlegendentry{UE2}

\draw[draw=black,fill=steelblue31119180,postaction={pattern=north east lines,pattern color=black}] (axis cs:-0.2,0) rectangle (axis cs:0.2,31.9885585066667);
\draw[draw=black,fill=steelblue31119180,postaction={pattern=north east lines,pattern color=black}] (axis cs:0.8,0) rectangle (axis cs:1.2,34.1066820266667);
\draw[draw=black,fill=steelblue31119180,postaction={pattern=north east lines,pattern color=black}] (axis cs:1.8,0) rectangle (axis cs:2.2,32.5617800533333);
\draw[draw=black,fill=steelblue31119180,postaction={pattern=north east lines,pattern color=black}] (axis cs:2.8,0) rectangle (axis cs:3.2,33.93191936);
\draw[draw=black,fill=steelblue31119180,postaction={pattern=north east lines,pattern color=black}] (axis cs:3.8,0) rectangle (axis cs:4.2,32.1842926933333);
\draw[draw=black,fill=steelblue31119180,postaction={pattern=north east lines,pattern color=black}] (axis cs:4.8,0) rectangle (axis cs:5.2,26.4940202666667);

\draw[draw=black,fill=darkorange25512714,postaction={pattern=north west lines,pattern color=black}] (axis cs:0.2,0) rectangle (axis cs:0.6,31.85573888);
\draw[draw=black,fill=darkorange25512714,postaction={pattern=north west lines,pattern color=black}] (axis cs:1.2,0) rectangle (axis cs:1.6,32.8344098133333);
\draw[draw=black,fill=darkorange25512714,postaction={pattern=north west lines,pattern color=black}] (axis cs:2.2,0) rectangle (axis cs:2.6,30.70230528);
\draw[draw=black,fill=darkorange25512714,postaction={pattern=north west lines,pattern color=black}] (axis cs:3.2,0) rectangle (axis cs:3.6,30.6463812266667);
\draw[draw=black,fill=darkorange25512714,postaction={pattern=north west lines,pattern color=black}] (axis cs:4.2,0) rectangle (axis cs:4.6,17.3364565333333);
\draw[draw=black,fill=darkorange25512714,postaction={pattern=north west lines,pattern color=black}] (axis cs:5.2,0) rectangle (axis cs:5.6,15.1554184533333);

\path [draw=black, line width=1pt]
(axis cs:0.4,31.513056012484)
--(axis cs:0.4,32.198421747516);

\path [draw=black, line width=1pt]
(axis cs:1.4,32.4809834366442)
--(axis cs:1.4,33.1878361900225);

\path [draw=black, line width=1pt]
(axis cs:2.4,29.9925850167168)
--(axis cs:2.4,31.4120255432832);

\path [draw=black, line width=1pt]
(axis cs:3.4,29.9176908246676)
--(axis cs:3.4,31.3750716286657);

\path [draw=black, line width=1pt]
(axis cs:4.4,16.3248632855358)
--(axis cs:4.4,18.3480497811309);

\path [draw=black, line width=1pt]
(axis cs:5.4,14.1746933344082)
--(axis cs:5.4,16.1361435722584);

\path [draw=black, line width=1pt]
(axis cs:0,31.3367893521494)
--(axis cs:0,32.6403276611839);

\path [draw=black, line width=1pt]
(axis cs:1,33.8468768830825)
--(axis cs:1,34.3664871702508);

\path [draw=black, line width=1pt]
(axis cs:2,31.8263617954621)
--(axis cs:2,33.2971983112046);

\path [draw=black, line width=1pt]
(axis cs:3,33.1077024863186)
--(axis cs:3,34.7561362336814);

\path [draw=black, line width=1pt]
(axis cs:4,31.7343581731514)
--(axis cs:4,32.6342272135153);

\path [draw=black, line width=1pt]
(axis cs:5,25.7266767289913)
--(axis cs:5,27.2613638043421);

\addplot [semithick, black, mark=-, mark size=1.5, mark options={solid}, only marks]
table {%
0.4 31.513056012484
1.4 32.4809834366442
2.4 29.9925850167168
3.4 29.9176908246676
4.4 16.3248632855358
5.4 14.1746933344082
};

\addplot [semithick, black, mark=-, mark size=1.5, mark options={solid}, only marks]
table {%
0.4 32.198421747516
1.4 33.1878361900225
2.4 31.4120255432832
3.4 31.3750716286657
4.4 18.3480497811309
5.4 16.1361435722584
};

\addplot [semithick, black, mark=-, mark size=1.5, mark options={solid}, only marks]
table {%
0 31.3367893521494
1 33.8468768830825
2 31.8263617954621
3 33.1077024863186
4 31.7343581731514
5 25.7266767289913
};

\addplot [semithick, black, mark=-, mark size=1.5, mark options={solid}, only marks]
table {%
0 32.6403276611839
1 34.3664871702508
2 33.2971983112046
3 34.7561362336814
4 32.6342272135153
5 27.2613638043421
};

\end{axis}

\end{tikzpicture}
        \setlength\abovecaptionskip{.05cm}}
\caption{Performance profiling for two \glspl{ru} in the static iPerf use case, each with one assigned \gls{ue}: \gls{ue}1 to \gls{ru}1 and \gls{ue}2 to \gls{ru}2.}
\label{fig:RF2RU2UE_Static_th_New}
\end{figure}

\begin{figure}[t]
\centering
    \subfloat[MCS]{
    \label{fig:mcs_2ru2ue_new}
    \centering
    \setlength\fwidth{.97\linewidth}
    \setlength\fheight{.35\linewidth}
\begin{tikzpicture}
\pgfplotsset{every tick label/.append style={font=\scriptsize}}

\definecolor{black}{RGB}{38,38,38}
\definecolor{lightgray204}{RGB}{204,204,204}
\definecolor{steelblue31119180}{RGB}{31,119,180}
\definecolor{darkorange25512714}{RGB}{255,127,14}

\begin{axis}[
width=1\fwidth, 
height=1.05\fheight, 
at={(0\fwidth,0\fheight)},
axis line style={color=black},
legend cell align={center},
legend style={
  fill opacity=0.8,
  draw opacity=1,
  text opacity=1,
  at={(0.5, 1.5)},
  anchor=south,
  draw=lightgray204,
  font=\scriptsize,
},
axis line style={color=black},
tick align=inside,
x dir=reverse,
x grid style={lightgray204},
xlabel=\textcolor{black}{UE2 Location},
label style={font=\scriptsize},
xmajorgrids,
xmin=15, xmax=25,
xtick={15, 17, 19, 21, 23, 25},
xtick pos=right,
xtick style={color=black},
y grid style={lightgray204},
ymajorgrids,
ymin=0, ymax=27,
ytick pos=left,
ytick style={color=black},
ylabel={MCS},
legend columns=4,
]

\path [draw=darkorange25512714, semithick]
(axis cs:15,5.16716845905602)
--(axis cs:15,7.22172042983287);
\path [draw=darkorange25512714, semithick]
(axis cs:17,6)
--(axis cs:17,6);
\path [draw=darkorange25512714, semithick]
(axis cs:19,12.1842084276582)
--(axis cs:19,14.3248824814328);
\path [draw=darkorange25512714, semithick]
(axis cs:21,17.4511202560732)
--(axis cs:21,20.730697925745);
\path [draw=darkorange25512714, semithick]
(axis cs:23,20.9435835065462)
--(axis cs:23,24.8745983116356);
\path [draw=darkorange25512714, semithick]
(axis cs:25,19.0582607645917)
--(axis cs:25,22.2144665081356);

\addplot [semithick, darkorange25512714, mark=*, mark size=1pt]
table {%
15 6.19444444444444
17 6
19 13.2545454545455
21 19.0909090909091
23 22.9090909090909
25 20.6363636363636
};

\path [draw=steelblue31119180, semithick]
(axis cs:0,14.8066916065866)
--(axis cs:0,18.629672029777);
\path [draw=steelblue31119180, semithick]
(axis cs:2,17.6831786740086)
--(axis cs:2,20.3713667805369);
\path [draw=steelblue31119180, semithick]
(axis cs:4,18.4606777382919)
--(axis cs:4,23.8847768071626);
\path [draw=steelblue31119180, semithick]
(axis cs:6,17.5306145275091)
--(axis cs:6,22.6875672906727);
\path [draw=steelblue31119180, semithick]
(axis cs:8,15.5148644570635)
--(axis cs:8,20.3942264520274);
\path [draw=steelblue31119180, semithick]
(axis cs:10,10.0883025514888)
--(axis cs:10,17.1227066228231);

\addplot [semithick, darkorange25512714, densely dashed, mark=*, mark size=1pt]
table {%
15 6
17 6.54545454545455
19 16.8545454545455
21 21.5272727272727
23 21.5909090909091
25 17.9636363636364
};

\end{axis}

\begin{axis}[
width=1\fwidth, 
height=1.05\fheight, 
axis line style={color=black},
at={(0\fwidth,0\fheight)},
legend cell align={center},
legend columns=4,
legend style={
  fill opacity=0.8,
  draw opacity=1,
  text opacity=1,
  at={(0.5, 1.5)},
  anchor=south,
  draw=lightgray204,
  font=\scriptsize,
},
tick align=inside,
tick pos=left,
x grid style={lightgray204},
xlabel=\textcolor{black}{UE1 Location},
label style={font=\scriptsize},
xmin=0, xmax=10,
xtick style={color=black},
y grid style={lightgray204},
ymin=0, ymax=27,
ytick style={color=black}, 
yticklabel style={text=black} 
]
\path [draw=steelblue31119180, semithick]
(axis cs:0,14.954290766315)
--(axis cs:0,18.6638910518668);
\path [draw=steelblue31119180, semithick]
(axis cs:2,19.2813613623151)
--(axis cs:2,22.5913659104122);
\path [draw=steelblue31119180, semithick]
(axis cs:4,18.6056533575758)
--(axis cs:4,21.8670739151515);
\path [draw=steelblue31119180, semithick]
(axis cs:6,19.2865882524875)
--(axis cs:6,22.7134117475125);
\path [draw=steelblue31119180, semithick]
(axis cs:8,13.6759077642524)
--(axis cs:8,17.663714877257);
\path [draw=steelblue31119180, semithick]
(axis cs:10,11.6354323899127)
--(axis cs:10,19.9533526568163);

\addplot [semithick, steelblue31119180, forget plot]
table {%
0 16.8090909090909
2 20.9363636363636
4 20.2363636363636
6 21
8 15.6698113207547
10 15.7943925233645
};

\path [draw=darkorange25512714, semithick]
(axis cs:15,6)
--(axis cs:15,6);
\path [draw=darkorange25512714, semithick]
(axis cs:17,5.16647497567179)
--(axis cs:17,7.9244341152373);
\path [draw=darkorange25512714, semithick]
(axis cs:19,14.8461394334378)
--(axis cs:19,18.8629514756531);
\path [draw=darkorange25512714, semithick]
(axis cs:21,20.0328851173779)
--(axis cs:21,23.0216603371676);
\path [draw=darkorange25512714, semithick]
(axis cs:23,19.8284282781021)
--(axis cs:23,23.3533899037161);
\path [draw=darkorange25512714, semithick]
(axis cs:25,15.2547408145402)
--(axis cs:25,20.6725319127326);

\addplot [semithick, steelblue31119180, densely dashed, forget plot]
table {%
0 16.7181818181818
2 19.0272727272727
4 21.1727272727273
6 20.1090909090909
8 17.9545454545455
10 13.605504587156
};

\addlegendimage{semithick, steelblue31119180}
\addlegendentry{UE1 DL}

\addlegendimage{semithick, steelblue31119180, densely dashed}
\addlegendentry{UE1 UL}

\addlegendimage{semithick, darkorange25512714, mark=*, mark size=1pt}
\addlegendentry{UE2 DL}

\addlegendimage{semithick, darkorange25512714, densely dashed, mark=*, mark size=1pt}
\addlegendentry{UE2 UL}

\end{axis}

\end{tikzpicture}
    \setlength\abovecaptionskip{.05cm}}
    \hfill    
    \subfloat[RSRP]{
    \label{fig:rsrp_2ru2ue_new}
    \centering
    \setlength\fwidth{.97\linewidth}
    \setlength\fheight{.35\linewidth}
\begin{tikzpicture}
\pgfplotsset{every tick label/.append style={font=\scriptsize}}

\definecolor{darkslategray38}{RGB}{38,38,38}
\definecolor{lightgray204}{RGB}{204,204,204}
\definecolor{steelblue31119180}{RGB}{31,119,180}
\definecolor{darkorange25512714}{RGB}{255,127,14}

\begin{axis}[
width=1\fwidth, 
height=1.05\fheight, 
axis line style={color=black},
legend cell align={center},
legend style={
  fill opacity=0.8,
  draw opacity=1,
  text opacity=1,
  at={(0.5, 1.5)},
  anchor=south,
  draw=lightgray204,
  font=\scriptsize,
},
tick align=inside,
tick pos=left,
x grid style={lightgray204},
xlabel=\textcolor{darkslategray38}{UE1 Location},
label style={font=\scriptsize},
xmajorgrids,
xmin=0, xmax=10,
xtick style={color=darkslategray38},
y grid style={lightgray204},
ymajorgrids,
ymin=-97, ymax=-78,
ytick style={color=darkslategray38},
ylabel={RSRP [dBm]},
legend columns=4,
]

\path [draw=steelblue31119180, semithick]
(axis cs:0,-91.1617796100565)
--(axis cs:0,-90.0382203899435);
\path [draw=steelblue31119180, semithick]
(axis cs:2,-85.9742678518038)
--(axis cs:2,-84.9711866936508);
\path [draw=steelblue31119180, semithick]
(axis cs:4,-84.4175242453835)
--(axis cs:4,-83.1279303000711);
\path [draw=steelblue31119180, semithick]
(axis cs:6,-82.4039581621078)
--(axis cs:6,-81.3778600197104);
\path [draw=steelblue31119180, semithick]
(axis cs:8,-84.1587910102666)
--(axis cs:8,-82.6525297444503);
\path [draw=steelblue31119180, semithick]
(axis cs:10,-90.373455643225)
--(axis cs:10,-88.822806039018);

\addplot [semithick, steelblue31119180]
table {%
0 -90.6
2 -85.4727272727273
4 -83.7727272727273
6 -81.8909090909091
8 -83.4056603773585
10 -89.5981308411215
};

\path [draw=steelblue31119180, semithick]
(axis cs:0,-90.9922381737809)
--(axis cs:0,-89.9532163716736);
\path [draw=steelblue31119180, semithick]
(axis cs:2,-85.9350707755739)
--(axis cs:2,-84.6103837698806);
\path [draw=steelblue31119180, semithick]
(axis cs:4,-85.8730828648639)
--(axis cs:4,-83.636008044227);
\path [draw=steelblue31119180, semithick]
(axis cs:6,-82.3494258368982)
--(axis cs:6,-81.3051196176472);
\path [draw=steelblue31119180, semithick]
(axis cs:8,-84.6933401012945)
--(axis cs:8,-82.2157508077964);
\path [draw=steelblue31119180, semithick]
(axis cs:10,-90.6362571091663)
--(axis cs:10,-89.3453942669805);

\addplot [semithick, steelblue31119180, densely dashed]
table {%
0 -90.4727272727273
2 -85.2727272727273
4 -84.7545454545455
6 -81.8272727272727
8 -83.4545454545455
10 -89.9908256880734
};

\end{axis}

\begin{axis}[
width=1\fwidth, 
height=1.05\fheight, 
axis line style={lightgray204},
axis x line=top,
legend cell align={left},
legend columns=4,
legend style={
  fill opacity=0.8,
  draw opacity=1,
  text opacity=1,
  at={(0.5, 1.5)},
  anchor=south,
  draw=lightgray204,
  font=\scriptsize,
},
tick align=inside,
x dir=reverse,
x grid style={lightgray204},
xlabel=\textcolor{darkslategray38}{UE2 Location},
label style={font=\scriptsize},
xmin=15, xmax=25,
xtick={15, 17, 19, 21, 23, 25},
xtick pos=right,
xtick style={color=darkslategray38},
y grid style={lightgray204},
ymin=-97, ymax=-78,
ytick pos=left,
]
\path [draw=darkorange25512714, semithick]
(axis cs:15,-95.6116938433622)
--(axis cs:15,-94.2031209714526);
\path [draw=darkorange25512714, semithick]
(axis cs:17,-95.2085797073974)
--(axis cs:17,-94.4823293835117);
\path [draw=darkorange25512714, semithick]
(axis cs:19,-88.3043882602975)
--(axis cs:19,-87.4592481033389);
\path [draw=darkorange25512714, semithick]
(axis cs:21,-85.0848253737061)
--(axis cs:21,-84.0969928081121);
\path [draw=darkorange25512714, semithick]
(axis cs:23,-81.2736036168899)
--(axis cs:23,-80.0536691103828);
\path [draw=darkorange25512714, semithick]
(axis cs:25,-82.7318807301365)
--(axis cs:25,-81.7044829062271);

\addplot [semithick, darkorange25512714, forget plot, mark=*, mark size=1pt]
table {%
15 -94.9074074074074
17 -94.8454545454545
19 -87.8818181818182
21 -84.5909090909091
23 -80.6636363636364
25 -82.2181818181818
};

\path [draw=darkorange25512714, semithick]
(axis cs:15,-95.6695166076239)
--(axis cs:15,-94.7850288469215);
\path [draw=darkorange25512714, semithick]
(axis cs:17,-95.2284101819335)
--(axis cs:17,-94.1534079998847);
\path [draw=darkorange25512714, semithick]
(axis cs:19,-88.7989408624991)
--(axis cs:19,-87.6374227738646);
\path [draw=darkorange25512714, semithick]
(axis cs:21,-86.1603548691006)
--(axis cs:21,-85.2396451308994);
\path [draw=darkorange25512714, semithick]
(axis cs:23,-81.1997146382569)
--(axis cs:23,-80.0548308162886);
\path [draw=darkorange25512714, semithick]
(axis cs:25,-84.3737428732838)
--(axis cs:25,-82.644438944898);

\addplot [semithick, darkorange25512714, densely dashed, forget plot, mark=*, mark size=1pt]
table {%
15 -95.2272727272727
17 -94.6909090909091
19 -88.2181818181818
21 -85.7
23 -80.6272727272727
25 -83.5090909090909
};

\addlegendimage{semithick, steelblue31119180}
\addlegendentry{UE1 DL}

\addlegendimage{semithick, steelblue31119180, densely dashed}
\addlegendentry{UE1 UL}

\addlegendimage{semithick, darkorange25512714, mark=*, mark size=1pt}
\addlegendentry{UE2 DL}

\addlegendimage{semithick, darkorange25512714, densely dashed, mark=*, mark size=1pt}
\addlegendentry{UE2 UL}

\end{axis}

\end{tikzpicture}
    \setlength\abovecaptionskip{.05cm}}
\caption{\gls{mac} \glspl{kpi} in the two \glspl{ru} iPerf use case, each with one static \gls{ue} (\gls{ue}1 assigned to \gls{ru}1 and \gls{ue}2 to \gls{ru}2): (a) averages and confidence intervals for \gls{dl} \gls{mcs} (solid lines) during \gls{dl} data transmissions, and \gls{ul} \gls{mcs} (dashed lines) during \gls{ul} transmissions, for \gls{ue}1 (blue) and \gls{ue}2 (orange); (b) averages and confidence intervals of \gls{rsrp} reported by \gls{ue}1 (blue) and \gls{ue}2 (orange) during \gls{dl} (solid lines) and \gls{ul} (dashed lines) transmissions.}
\label{fig:RF2RU2UE_Static_New}
\end{figure}
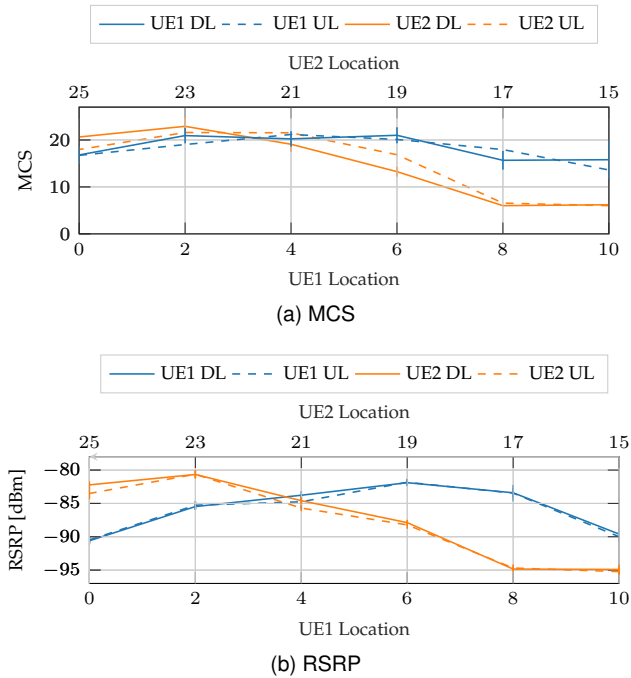

\textbf{2 \glspl{ru}, 1 \gls{ue} per \gls{ru}, static, iPerf.}
Finally, we evaluate \testbed performance with two \glspl{ru} by connecting \gls{ue}1 to \gls{ru}1 and \gls{ue}2 to \gls{ru}2. \gls{ru}1 is located at position~6, and \gls{ru}2 is at position~23. We select six pairs of locations---{(0,25), (2,23), (4,21), (6,19), (8,17), (10,15)}---for the \glspl{ue} to ensure different distances among them and the \gls{ru}.

From Figure~\ref{fig:dl_2ru2ue_new}, we observe that the \gls{dl} throughput is significantly impacted by interference, particularly at cell edge locations. The throughput for \gls{ue}1 shows a reduction of up to 90\% as the \glspl{ue} approach each other, while \gls{ue}2 throughput decreases by up to 50\%. These observations indicate that interference predominantly affects the \gls{dl} direction. Conversely, as depicted in Figure~\ref{fig:ul_2ru2ue_new}, \gls{ul} throughput remains relatively stable across different location pairs, suggesting that \gls{ul} is less susceptible to the types of interference affecting \gls{dl} throughput.

Figure~\ref{fig:RF2RU2UE_Static_New} further supports these observations by presenting additional \glspl{kpi} from the \gls{mac} layer. Figure~\ref{fig:mcs_2ru2ue_new} shows that the \gls{mcs} for both \glspl{ue} decreases as the distance between the \glspl{ue} diminishes, indicative of increasing interference levels. 
Figure~\ref{fig:rsrp_2ru2ue_new} illustrates the \gls{rsrp}, which varies in response to the \glspl{ue} locations. Notably, despite adequate \gls{rsrp} levels, the throughput remains low, highlighting the significant impact of interference, particularly in the \gls{dl} direction.

\subsection{Mobile Experiments}
\label{sec:exp-mobile}

\begin{figure}[t]
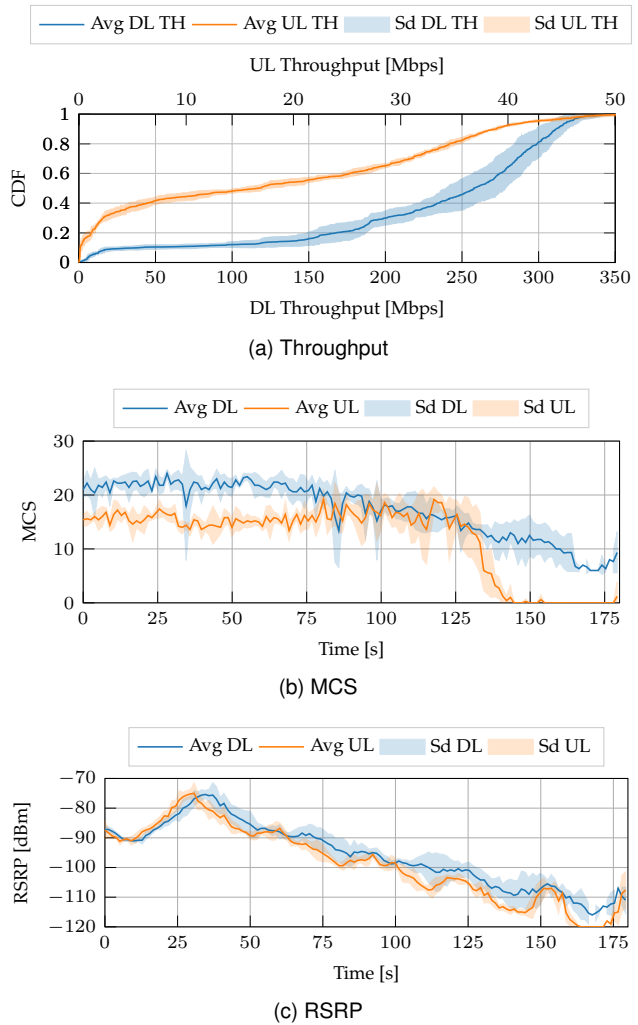

\centering
    \subfloat[Throughput]{
    \label{fig:1ru1ue_mobile_iperf_th}
    \centering
    \setlength\fwidth{.98\columnwidth}
    \setlength\fheight{.4\columnwidth}
    \input{figures/fig_tex/throughput_combined_cdf2}
    }

    \hfill
    
    \subfloat[\gls{mcs}]{
    \label{fig:1ru1ue_mobile_iperf_mcs}
    \centering
    \setlength\fwidth{.98\columnwidth}
    \setlength\fheight{.4\columnwidth}
    \input{figures/fig_tex/average_mcs_variability}
    }

    \hfill
    
    \subfloat[\gls{rsrp}]{
    \label{fig:1ru1ue_mobile_iperf_rsrp}
    \centering
    \setlength\fwidth{.96\columnwidth}
    \setlength\fheight{.4\columnwidth}
    \input{figures/fig_tex/average_rsrp_variability}
    }
\caption{Performance profiling with one \gls{ru} and one mobile \gls{ue} in the iPerf use case: (a)~\gls{cdf} of \gls{dl} and \gls{ul} throughputs with averages (solid lines) and \gls{sd} (shaded areas); (b)~averages (solid lines) and \gls{sd} (shaded areas) of the \gls{dl} \gls{mcs} during \gls{dl} transmissions (blue) and of the \gls{ul} \gls{mcs} during \gls{ul} transmissions (orange); (c)~averages (solid lines) and \gls{sd} (shaded areas) of the \gls{rsrp} reported by the \gls{ue} during \gls{dl} (blue) and \gls{ul} (orange) data transmissions.}
\label{fig:1ru1ue_mobile_iperf}
\vspace{-10pt}
\end{figure}

%
We assess the network performance by measuring throughput as the \gls{ue} follows the walking pattern around the laboratory space depicted by the dashed green line in Figure~\ref{fig:node-locations}. The entire walk from the start to the end point spans approximately $3$\:minutes at regular walking speed.
The mobile use case results are illustrated in Figure~\ref{fig:1ru1ue_mobile_iperf}. The application layer throughput is depicted by \gls{cdf} plots in Figure~\ref{fig:1ru1ue_mobile_iperf_th}, where solid lines represent the averaged curve for all runs, while the shaded areas around these lines illustrate the variation across different runs, indicating the range of values within one \gls{sd} above and below the mean.
The \gls{mcs} and \gls{rsrp} results at the \gls{mac} layer are shown in Figure~\ref{fig:1ru1ue_mobile_iperf_mcs} and Figure~\ref{fig:1ru1ue_mobile_iperf_rsrp}, respectively. Also here, the average values are depicted using solid lines, while the shaded areas indicate the \gls{sd}.

The throughput results of Figure~\ref{fig:1ru1ue_mobile_iperf_th} highlight notable variability in network quality influenced by mobility. Throughout the test, the \gls{ue} achieves peaks of up to $350$\:Mbps in \gls{dl} and $50$\:Mbps in \gls{ul}. However, significant fluctuations in performance are observed, particularly as the \gls{ue} moves further from the initial \gls{ru} position.
%
%
%
Figure~\ref{fig:1ru1ue_mobile_iperf_mcs} illustrates a significant drop in \gls{ul} \gls{mcs} values around the 100-second mark, where averages initially above 10 drop sharply, while \gls{dl} \gls{mcs} fluctuate more gradually until they fall below 10. This sudden decline in \gls{ul} \gls{mcs} at this specific time is likely due to deteriorating signal conditions, as corroborated by the corresponding \gls{rsrp} trends in Figure~\ref{fig:1ru1ue_mobile_iperf_rsrp}. This is most probably due to increased distance from the base station or physical obstructions, leading to a necessary reduction in \gls{mcs} to maintain connectivity under compromised signal strength.


The \gls{mcs} results of Figure~\ref{fig:1ru1ue_mobile_iperf_mcs} show that both \gls{dl} and \gls{ul} \gls{mcs} values start relatively high but decrease as the \gls{ue} moves further from the \gls{ru}. The \gls{dl} \gls{mcs} exhibits more variability and sharper declines compared to the \gls{ul}, which maintains a more stable profile until the final part of the walk. This suggests that the uplink benefits from more aggressive modulation and coding strategies due to \gls{5g} adaptive power control mechanisms that mitigate the impact of increasing distance and obstacles more effectively.
Additionally, the \gls{rsrp} data, shown in Figure~\ref{fig:1ru1ue_mobile_iperf_rsrp}, indicates a gradual decline in signal strength as the \gls{ue} moves along its trajectory. \gls{rsrp} values for both \gls{dl} (blue) and \gls{ul} (orange) cases decrease over time, with the most significant drops observed after $100$ seconds. This reduction in signal quality corresponds with declines in throughput and \gls{mcs}, highlighting the strong dependency of these metrics on signal strength.
%


\subsection{Video Streaming Experiments}
\label{sec:exp-video}

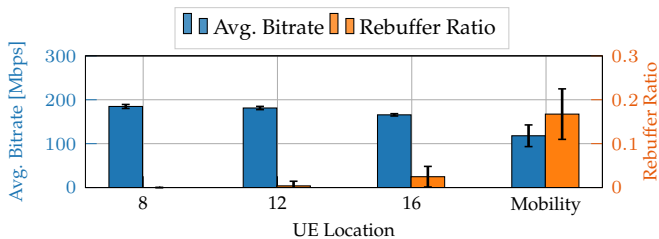
\begin{figure}[t]
\centering
    \setlength\fwidth{\linewidth}
    \setlength\fheight{.25\linewidth}
\begin{tikzpicture}
\pgfplotsset{every tick label/.append style={font=\scriptsize}}

\definecolor{darkgray176}{RGB}{176,176,176}
\definecolor{darkorange25512714}{RGB}{255,127,14}
\definecolor{darkorange2309111}{RGB}{230,91,11}
\definecolor{lightgray204}{RGB}{204,204,204}
\definecolor{steelblue31119180}{RGB}{31,119,180}

\begin{axis}[
width=0.951\fwidth,
height=1.5\fheight,
at={(0\fwidth,0\fheight)},
legend cell align={left},
legend columns=2,
legend style={fill opacity=0.8, 
draw opacity=1, 
text opacity=1, 
draw=lightgray204, 
font=\footnotesize,
at={(0.82, 1.37)}},
x grid style={darkgray176},
xmajorticks=false,
xmin=-0.3, xmax=3.55,
xtick style={color=steelblue31119180},
xtick={0.13,1.13,2.13,3.13},
xticklabel style={rotate=45.0},
xticklabels={8,12,16,Mobility},
y grid style={darkgray176},
ylabel=\textcolor{steelblue31119180}{Avg. Bitrate [Mbps]},
ylabel style={font=\scriptsize},
xlabel style={font=\scriptsize},
ymin=0, ymax=300,
ytick pos=left,
ytick style={color=steelblue31119180},
yticklabel style={color=steelblue31119180},
xmajorgrids,
ymajorgrids
]

\addlegendimage{ybar,ybar legend,draw=black,fill=steelblue31119180,postaction={pattern=north east lines,pattern color=black}}
\addlegendentry{Avg. Bitrate}
\addlegendimage{ybar,ybar legend,draw=black,fill=darkorange25512714,postaction={pattern=north west lines,pattern color=black}}
\addlegendentry{Rebuffer Ratio}

\draw[draw=black,fill=steelblue31119180,postaction={pattern=north east lines,pattern color=black}] (axis cs:-0.125,0) rectangle (axis cs:0.125,184.684677810809);

\draw[draw=black,fill=steelblue31119180,postaction={pattern=north east lines,pattern color=black}] (axis cs:0.875,0) rectangle (axis cs:1.125,181.192263115416);

\draw[draw=black,fill=steelblue31119180,postaction={pattern=north east lines,pattern color=black}] (axis cs:1.875,0) rectangle (axis cs:2.125,165.600715680784);

\draw[draw=black,fill=steelblue31119180,postaction={pattern=north east lines,pattern color=black}] (axis cs:2.875,0) rectangle (axis cs:3.125,117.942643215823);

\path [draw=black, line width=1pt]
(axis cs:0,180.108878846041)
--(axis cs:0,189.260476775578);

\addplot [semithick, black, mark=-, mark size=1.5, mark options={solid}, only marks]
table {%
0 180.108878846041
};
\addplot [semithick, black, mark=-, mark size=1.5, mark options={solid}, only marks]
table {%
0 189.260476775578
};
\path [draw=black, line width=1pt]
(axis cs:1,177.525601291505)
--(axis cs:1,184.858924939326);

\addplot [semithick, black, mark=-, mark size=1.5, mark options={solid}, only marks]
table {%
1 177.525601291505
};
\addplot [semithick, black, mark=-, mark size=1.5, mark options={solid}, only marks]
table {%
1 184.858924939326
};
\path [draw=black, line width=1pt]
(axis cs:2,162.956109049625)
--(axis cs:2,168.245322311942);

\addplot [semithick, black, mark=-, mark size=1.5, mark options={solid}, only marks]
table {%
2 162.956109049625
};
\addplot [semithick, black, mark=-, mark size=1.5, mark options={solid}, only marks]
table {%
2 168.245322311942
};
\path [draw=black, line width=1pt]
(axis cs:3,93.2293861692557)
--(axis cs:3,142.65590026239);

\addplot [semithick, black, mark=-, mark size=1.5, mark options={solid}, only marks]
table {%
3 93.2293861692557
};
\addplot [semithick, black, mark=-, mark size=1.5, mark options={solid}, only marks]
table {%
3 142.65590026239
};
\end{axis}

\begin{axis}[
width=0.951\fwidth,
height=1.5\fheight,
at={(0\fwidth,0\fheight)},
axis y line*=right,
legend cell align={left},
legend style={
  fill opacity=0.8,
  draw opacity=1,
  text opacity=1,
  font=\footnotesize,
  draw=lightgray204
},
x grid style={darkgray176},
xmin=-0.3, xmax=3.55,
xtick pos=left,
xtick style={color=black},
xtick={0.13,1.13,2.13,3.13},
xticklabels={8,12,16,Mobility},
x label style={at={(axis description cs:0.5,-0.20)},anchor=north},
ylabel style={font=\scriptsize},
xlabel style={font=\scriptsize},
xlabel={UE Location},
y grid style={darkgray176},
ylabel=\textcolor{darkorange2309111}{Rebuffer Ratio},
ymin=0, ymax=0.3,
ytick pos=right,
legend columns=4,
ytick style={color=darkorange2309111},
yticklabel style={anchor=west,color=darkorange2309111},
ylabel shift=-5pt
]
\draw[draw=black,fill=darkorange25512714,postaction={pattern=north west lines,pattern color=black}] (axis cs:0.125,0) rectangle (axis cs:0.375,0);


\draw[draw=black,fill=darkorange25512714,postaction={pattern=north west lines,pattern color=black}] (axis cs:1.125,0) rectangle (axis cs:1.375,0.0037973392);
\draw[draw=black,fill=darkorange25512714,postaction={pattern=north west lines,pattern color=black}] (axis cs:2.125,0) rectangle (axis cs:2.375,0.0246761052);
\draw[draw=black,fill=darkorange25512714,postaction={pattern=north west lines,pattern color=black}] (axis cs:3.125,0) rectangle (axis cs:3.375,0.167254825);
\path [draw=black, line width=1pt]
(axis cs:0.25,0)
--(axis cs:0.25,0);

\addplot [semithick, black, mark=-, mark size=1.5, mark options={solid}, only marks, forget plot]
table {%
0.25 0
};
\addplot [semithick, black, mark=-, mark size=1.5, mark options={solid}, only marks, forget plot]
table {%
0.25 0
};
\path [draw=black, line width=1pt]
(axis cs:1.25,-0.00674576463461573)
--(axis cs:1.25,0.0143404430346157);

\addplot [semithick, black, mark=-, mark size=1.5, mark options={solid}, only marks, forget plot]
table {%
1.25 -0.00674576463461573
};
\addplot [semithick, black, mark=-, mark size=1.5, mark options={solid}, only marks, forget plot]
table {%
1.25 0.0143404430346157
};
\path [draw=black, line width=1pt]
(axis cs:2.25,0.00120152041884096)
--(axis cs:2.25,0.048150689981159);

\addplot [semithick, black, mark=-, mark size=1.5, mark options={solid}, only marks, forget plot]
table {%
2.25 0.00120152041884096
};
\addplot [semithick, black, mark=-, mark size=1.5, mark options={solid}, only marks, forget plot]
table {%
2.25 0.048150689981159
};
\path [draw=black, line width=1pt]
(axis cs:3.25,0.1097362637768)
--(axis cs:3.25,0.2247733862232);

\addplot [semithick, black, mark=-, mark size=1.5, mark options={solid}, only marks, forget plot]
table {%
3.25 0.1097362637768
};
\addplot [semithick, black, mark=-, mark size=1.5, mark options={solid}, only marks, forget plot]
table {%
3.25 0.2247733862232
};
\end{axis}

\end{tikzpicture}
    \setlength\abovecaptionskip{-0.1cm}
    \label{fig:static_video_1ru1ue}
\caption{Video streaming performance with one \gls{ue} and single \gls{ru} across both static (8—close, 12—mid, 16—far) and mobile use cases.}
\label{fig:video_static}
\end{figure}

%
We place the \gls{ue} at three static locations at different distances from the \gls{ru}: location~8 (close); 12 (mid); and 16 (far). We run each video session for three minutes, streaming five distinct profiles simultaneously at various resolutions as described in Section~\ref{sec:exp-setup-overview}. We then plot the mean bitrate over five runs, as well as the rebuffer ratio, in Figure~\ref{fig:video_static}.
As expected, the average bitrate decreases and the rebuffer ratio increases as further distances between \gls{ue} and \gls{ru} are considered, transitioning from close to far static locations. We observe that the \gls{ue} can achieve a steady mean bitrate of around $180$\:Mbps in all static cases. Note that, unlike test results achieved through iPerf backlogged traffic, the mean bitrate for video streaming is lower.
The video client fetches segments in an intermittent fashion (causing flows to be short), which depends on parameters, e.g., video buffer and segment size. Because of this, throughput sometimes does not increase to the fullest during that short period of time, and the client algorithm \gls{abr} downgrades the bitrate based on the estimate it gets. This is due to a slow \gls{mcs} selection loop in the \gls{oai} L2, which will be improved as part of our future work. However, this shows that our setup is capable of supporting up to 8K \gls{hdr} videos that require $150-300$\:Mbps bitrates according to YouTube guidelines~\cite{Youtube}.
During mobility, the average bitrate is $120$\:Mbps, and the rebuffer ratio increases to $15$\%. This is once again because the \gls{ue} moves away from the \gls{ru}, gradually entering low-coverage regions and eventually disconnecting.


\subsection{\blue{Peak Performance Experiments}}
\label{sec:exp-peak}

\blue{In this second set of experiments, we expand our evaluation to stress-test the system and attain peak performance results. To achieve these compared to previous tests, we leverage a \gls{gh} \gls{ran} server with a DDDDDDSUUU \gls{tdd} pattern, a 4x4~\gls{mimo} configuration, 4~layers \gls{dl}, 1~layer \gls{ul} and a $Q_{m}$) up to 256-\gls{qam}. We compare system output with a single and double commercial \gls{ota} \glspl{ue} connected to our network at a fixed location using Open5GS as \gls{cn} and iPerf to generate traffic, as well as with the Keysight RuSIM emulator device, emulating both \gls{ru} and up to 25 \glspl{ue}, using Keysight CoreSIM to emulate the \gls{cn}. The average throughput and $95$\% confidence intervals are plotted in Figure~\ref{fig:2ue_static_rusim}.
We observe that in \gls{ota} at a fixed location, the \glspl{ue} achieve steady throughput in all cases, as indicated by the small confidence interval values, with a peak of up to $1.05$\:Gbps in \gls{dl} and $100$\:Mbps in \gls{ul} for a single \gls{ue} (\textit{1-ota}). Furthermore, the combined throughput increases with the number of connected \glspl{ue} (\textit{2-ota}), reaching a maximum of $1.2$\:Gbps in \gls{dl}, while remaining close to $100$\:Mbps in \gls{ul}.}

\blue{By using the Keysight RuSIM emulator with the same configuration as \gls{ota}, performance improves to over $1.26$\:Gbps with a single \gls{ue} (\textit{1-sim}) and $1.42$\:Gbps with two \glspl{ue} in \gls{dl} (\textit{2-sim}), and close to $110$\:Mbps in \gls{ul} for both one and two \glspl{ue}. This performance increase can be attributed to the more controlled environment provided by RuSIM, which eliminates external interference and impairments. In this case, an \textit{ExcellentRadioConditions} channel model---also used for \gls{bs} conformance testing as specified in the \gls{3gpp} specifications~\cite{3gpptesting}---is enabled to simulate ideal radio conditions.
To achieve the current peak cell throughput, we leverage a DDDDDDDSUU \gls{tdd} pattern in \gls{dl} and a DDDSU pattern in \gls{ul}, utilizing a reduced number of guard symbols (only one) enabled by RuSIM during two separate experiment runs with two emulated \glspl{ue}. This approach results in an aggregate throughput of $1.68$\:Gbps in \gls{dl} (\textit{2-simdl}) and $143$\:Mbps in \gls{ul} (\textit{2-simul}).
Moreover, we stress-test the system by simultaneously connecting up to 25 emulated \glspl{ue} while exchanging traffic, achieving similar performance (\textit{25-simdl, 25-simul}). This demonstrates that the network can reliably sustain multiple \glspl{ue} and reaches its peak with two \glspl{ue}, while fairly distributing resources when more devices are connected.
These results highlight the maximum performance currently achievable by \testbed, showcasing values comparable to those of production-level systems.}

\begin{figure}[t]
\centering
    \setlength\fwidth{0.95\linewidth}
    \setlength\fheight{.3\linewidth}
    \input{figures/fig_tex/1ru2ue_rusim_all}
\vspace{-15pt}
\caption{\blue{Performance profiling to achieve peak network throughput, leveraging: one (\textit{1-ota}) and two (\textit{2-ota}) \gls{ota} \glspl{ue} at a fixed location using iPerf and a DDDDDDSUUU \gls{tdd} pattern; one (\textit{1-sim}) and two (\textit{2-sim}) emulated \glspl{ue} using Keysight RuSIM and CoreSIM with a DDDDDDSUUU \gls{tdd} pattern; and two (\textit{2-simdl, 2-simul}) and twenty-five (\textit{25-simdl, 25-simul}) emulated \glspl{ue} with Keysight RuSIM and CoreSIM, a reduced number of guard symbols, a DDDDDDDSUU \gls{tdd} pattern for \gls{dl} cases, and a DDDSU \gls{tdd} pattern for \gls{ul} cases.}}
\label{fig:2ue_static_rusim}
\end{figure}

\subsection{\blue{Long-running Experiments}}
\label{sec:exp-long}

\begin{figure}[t]
    \centering
    \includegraphics[width=\linewidth]{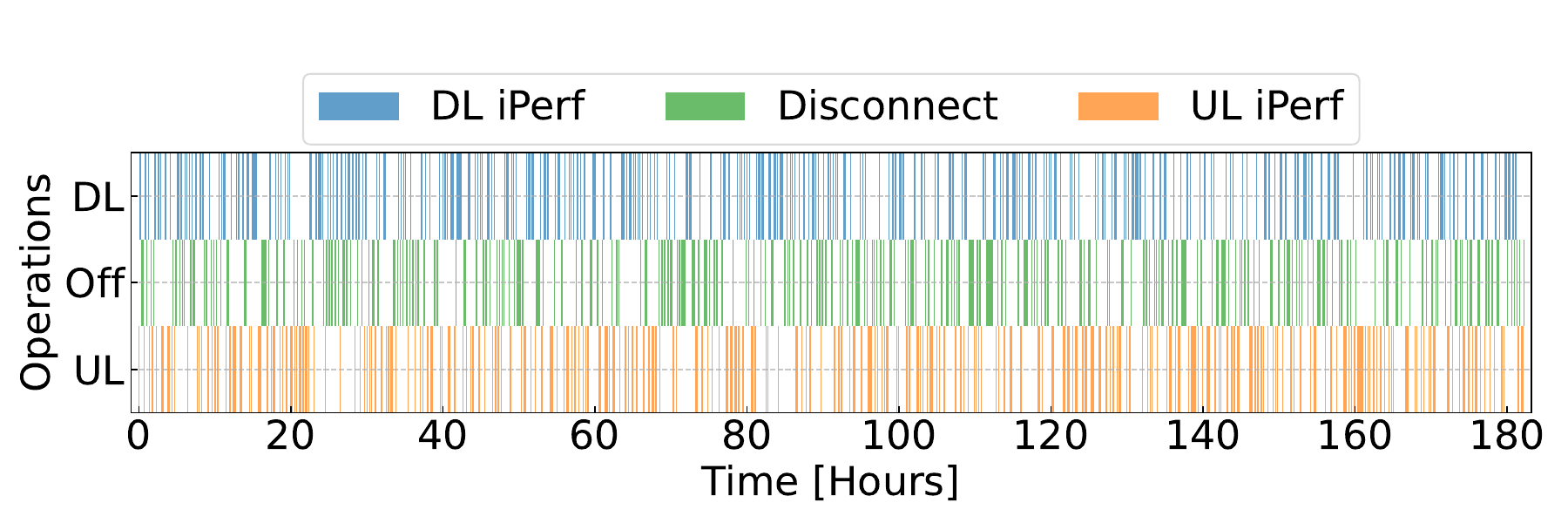}
    \caption{\blue{Long-running stability experiment involving one UE randomly cycling for over 180 hours among three operations, repeated every 10 minutes: (blue) DL iPerf for 1 minute; (orange) UL iPerf for 1 minute; and (green) disconnection from the network for the remainder of the 10-minute cycle window.}}
    \label{fig:ue-ops}
\end{figure}

\begin{figure}[t]
    \centering
    \includegraphics[width=\linewidth]{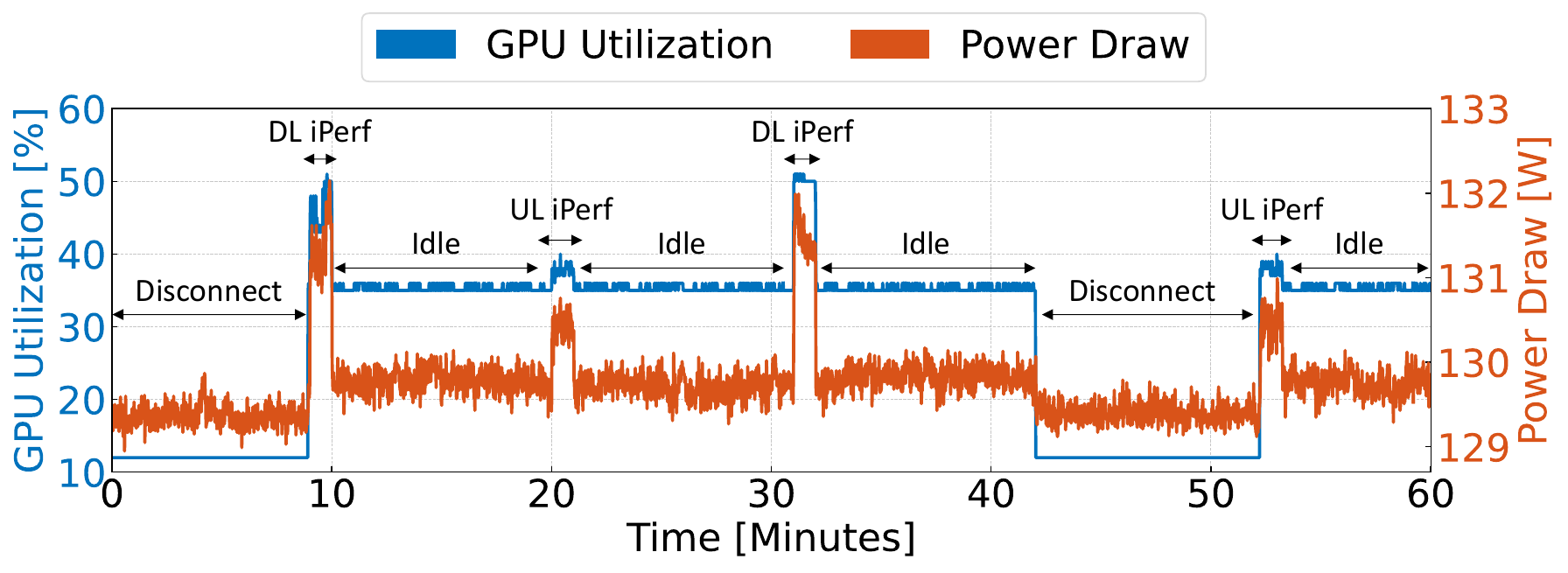}
    \caption{\blue{GPU utilization (blue) and power draw (orange) of the NVIDIA Grace Hopper server node during a one-hour window of the long-running stability experiment. The results show the behavior of the system when the \gls{ue} cycles through three operations: disconnecting for 10 minutes, performing a DL iPerf test for 1 minute, followed by 10 minutes of idling, and performing a UL iPerf test for 1 minute, followed by 10 minutes of idling.}}
    \label{fig:ue-ops-gpu}
    \vspace{-8pt}
\end{figure}

\blue{To validate stability and reliability, we evaluate \testbed through long-running experiments with a single \gls{ue} performing continuous operations. The cell configuration remains the same as in Section~\ref{sec:exp-peak}, utilizing the DDDDDDSUUU \gls{tdd} pattern, a 4x4~\gls{mimo} setup with 4~\gls{dl} layers and 1~\gls{ul} layer. The \gls{ue} is a Samsung S23 phone, which cycles randomly every $10$\:minutes between three different operations:
\begin{itemize}
    \item \gls{dl} test, a 1-minute \gls{udp} downlink iPerf data test targeting $50$\:Mbps.
    \item \gls{ul} test, a 1-minute \gls{udp} uplink iPerf data test targeting $10$\:Mbps.
    \item Disconnection, the \gls{ue} disconnects from the network, remains disconnected for the remaining $10$\:minutes, and then reconnects.
\end{itemize}
Figure~\ref{fig:ue-ops} shows the results of the long-running experiment, where the operations performed by the \gls{ue} every $10$\:minutes are represented with colored bars. The system can sustain indefinite uptime, as highlighted in the figure with over $180$\:hours of operation before the cell was manually shut down to vacate the spectrum for other planned experiments in the area.
Additionally, Figure~\ref{fig:ue-ops-gpu} presents some of the metrics available on the \gls{ran} server side, showing the resource utilization required to run the NVIDIA \gls{arc} \gls{gnb} with a single cell on a GH200 \gls{gpu}. Specifically, \gls{gpu} utilization and power draw are depicted for a \gls{gh} server during a one-hour window of the previous long-running stability experiment. We can see how the utilization (in blue) drops to nearly 10\% when no \gls{ue} is connected, and rises to approximately 50\% during \gls{dl} data traffic, reflecting the system's computation demand. On the other hand, during idle periods and \gls{ul} communication, \gls{gpu} utilization remains stable between 35\% and 40\%, respectively. The power draw (in orange) follows a similar trend, ranging from $129$ to $132$~W. It is important to note that these results apply to a single cell, but the resource requirements for multiple cells do not scale linearly. Each \gls{gh} server can support up to 20 cells~\cite{fujitsu1} while maintaining a high-level of energy efficiency for \gls{ran} communications~\cite{kundu2024energy}.
Overall, these results highlight the high reliability of \testbed in terms of both performance and stability, positioning it as a suitable candidate for \gls{p5g} deployments, as well as a valuable playground to develop, test, and evaluate novel \gls{ai}/\gls{ml} algorithms and solutions for the \gls{ran}.}

\section{Related Work}
\label{sec:related-work}

This section compares the features and capabilities of the \testbed testbed within the context of similar programmable open RAN and \gls{5g}, highlighting its unique features and contributions beyond \gls{5g} research and experimentation. Surveys of testbeds for open and programmable wireless networks can also be found in~\cite{bonati2020open,polese2023understanding}.

The \gls{pawr}~\cite{pawr} offers a set of geographically and technically diverse testbeds designed to enhance specific wireless communication areas. These include POWDER, AERPAW, COSMOS, ARA, and Colosseum, each equipped with specialized technologies to address varied research needs.

The POWDER facility, located at the University of Utah in Salt Lake City, UT, supports a wide spectrum of research areas, including next-generation wireless networks and dynamic spectrum access~\cite{breen2020powder}. Its \gls{5g} stack is based primarily on a combination of open-source stacks, combined with \glspl{sdr} or \glspl{ru} but not accelerated at the physical layer, and on a commercial Mavenir system, which does not support access to the source code from the \gls{phy} to the core network, differently from the \testbed stack. 

Similarly, AERPAW, deployed on the campus of North Carolina State University in Raleigh, NC, focuses on aerial and drone communications, diverging from our emphasis on private \gls{5g} network configurations~\cite{panicker2021aerpaw}. The AERPAW facility hosts an Ericsson \gls{5g} deployment with similar limitations with respect to stack programmability for research use cases.

The COSMOS project~\cite{chen2023open-access} leverages an array of programmable and software-defined radios, including USRP and Xilinx RFSoC boards, to facilitate mmWave communication experiments across a city-scale environment. The outdoor facilities of COSMOS are deployed in the Harlem area, in New York City, while its indoor wireless facilities are on the Rutgers campus in North Brunswick, NJ. Unlike \testbed, COSMOS is designed for broad academic and industry use and is more focused on mmWave deployments enabling diverse external contributions to its development without specific emphasis on any single network architecture.

The ARA testbed \cite{zhang2021ara}, deployed across Iowa State University (ISU), in the city of Ames, and surrounding rural areas in central Iowa, serves as a large-scale platform for advanced wireless research tailored to rural settings. ARA includes diverse wireless platforms ranging from low-UHF massive \gls{mimo} to mmWave access, long-distance backhaul, free-space optical, and \gls{leo} satellite communications, utilizing both \gls{sdr} and programmable \gls{cots} platforms and leveraging open-source software like \gls{oai}, srsRAN, and SD-RAN~\cite{ARASDR}. However, unlike the \testbed testbed,
ARA focuses primarily on rural connectivity without focusing on specialized hardware for \gls{phy} layer optimization or digital twin frameworks for \gls{rf} planning.

Colosseum is the world's largest Open RAN digital twin~\cite{villa2024dt, polese2024colosseum}. This testbed allows users to quickly instantiate softwarized cellular protocol stacks, e.g., the \gls{oai} one, on its 128~compute nodes. These nodes control 128~\glspl{sdr} that are used as \gls{rf} front-ends and are connected to a massive channel emulator, which enables experimentation in a variety of emulated \gls{rf} environments. However, the Colosseum servers are not equipped to offload lower-layer cellular operations on \glspl{gpu}, and the available \glspl{sdr} are USRP~X310 from NI, instead of commercial \glspl{ru}. 

The \gls{osc} is also involved in the creation of laboratory facilities~\cite{SoAOSC} that comply with O-RAN standards and support the testing and integration of O-RAN-compliant components. These testing facilities, distributed across multiple laboratories, foster a diverse ecosystem through their commitment to open standards and collaborative development. However, unlike \testbed, they do not explicitly focus on the deployment complexities of private networks, nor do they provide any \gls{phy} layer acceleration technology or utilize a digital twin for \gls{rf} planning. Instead, they aim to promote multi-vendor interoperability within an open collaborative framework.


6G-SANDBOX~\cite{6GSANDBOX} is a versatile facility that includes four geographically displaced platforms in Europe, each equipped to support a variety of advanced wireless technologies and experimental setups. It uses a mix of commercial solutions (for example, Nokia microcells, Ericsson \gls{bbu}, and the Amarisoft stack) and open source solutions (for example, \gls{oai} and srsRAN) in diverse environments ranging from urban to rural settings. Unlike \testbed, 6G-SANDBOX primarily facilitates wide-ranging 6G research through its extensive, multi-location infrastructure. Its predecessor, 5GENESIS~\cite{5Genesis}, featured a modular and flexible experimentation methodology, supporting both per-component and \gls{e2e} validation of \gls{5g} technologies and \gls{kpi} across five European locations. This testbed emphasizes a comprehensive approach to \gls{5g} performance assessment, integrating diverse technologies such as \gls{sdn}, \gls{nfv}, and network slicing to enable rigorous testing of vertical applications but not including O-RAN architectures.


The \gls{oaic} testbed~\cite{OAIC}, developed at Virginia Tech, is an open-source \gls{5g} O-RAN-based platform designed to facilitate AI-based \gls{ran} management algorithms. It includes the \gls{oaic}-Control framework for designing AI-based \gls{ran} controllers and the \gls{oaic}-Testing framework for automated testing of these controllers. The \gls{oaic} testbed introduces a new real-time \gls{ric}, zApps, and a Z1 interface to support use cases requiring latency under $10$\:ms, integrated with the CORNET infrastructure for remote accessibility.


The CCI xG Testbed provides a comprehensive platform for advanced wireless research, particularly in the realm of \gls{5g} and beyond. It features a disaggregated architecture with multiple servers distributed across geographically disparate cloud sites, leveraging a combination of central and edge cloud infrastructures to optimize resource allocation and latency. The testbed includes several \gls{sdr}-based \gls{cbrs} Base Station Device (CBSD) integrated with an open-source \gls{sas} for dynamic spectrum sharing in the \gls{cbrs} band~\cite{CCI1}. Additionally, the testbed supports a full O-RAN stack using srsRAN and Open5GS and features both non-RT \gls{ric} and near-RT \gls{ric} for real-time and non-real-time radio resource management~\cite{CCI2,CCI3}.

The testbed in~\cite{NEC} provides a prototypical environment designed to experiment with vRAN deployments and evaluate resource allocation and orchestration algorithms. It focuses on the decoupling of radio software components from hardware to facilitate efficient and cost-effective \gls{ran} deployments. This testbed includes datasets that characterize computing usage, energy consumption, and application performance, which are made publicly available to foster further research. Unlike the \testbed testbed, the O-RAN platform primarily addresses the flexibility and cost efficiency of virtualized RANs without incorporating specialized hardware for PHY layer tasks.

\blue{The disaggregated 5G testbed for live audio production~\cite{ReviewSoA1} emphasizes ultra-reliable low-latency communication for media applications. Its scope is narrower than \testbed, which supports a broader range of experimental scenarios and computationally intensive network configurations.}

\blue{The data usage control framework~\cite{ReviewSoA2} addresses privacy challenges in hybrid private-public 5G networks. While it highlights the importance of secure orchestration and policy management, its focus on analytics differs from \testbed capabilities in physical layer acceleration and network performance experimentation.}

\blue{Finally, the Microsoft enterprise-scale Open RAN testbed~\cite{reviewerref9} highlights the potential of virtualized RAN functions on commodity servers, employing disaggregated architectures to demonstrate scalability and flexibility. By integrating Kubernetes for dynamic orchestration and using Intel FlexRAN with ACC100 accelerators for \gls{ldpc} look-aside offloading, this testbed achieves functional disaggregation of RAN workloads.}

\blue{While state-of-the-art software stacks such as srsRAN already offer similar performance in terms of core \gls{5g} functionalities, including handovers, \testbed distinguishes itself through its integration of \gls{gpu} acceleration, enabling enhanced flexibility and computational power for future innovations. Unlike traditional platforms that rely on \gls{cpu}-based architectures, which achieve performance parity for standard \gls{ran} tasks, it leverages \glspl{gpu} not only for optimized \gls{phy} processing but also as a unified platform for \gls{ai}/\gls{ml} workloads. Indeed, the \gls{gpu} architecture of \testbed supports the development and deployment of dApps~\cite{lacava2025dApps} that utilize \gls{ai}/\gls{ml} models for real-time network optimization. This capability aligns directly with the vision outlined by the AI-RAN Alliance~\cite{airanwhite}, which emphasizes the integration of \gls{ai}-driven decision-making processes across three key development areas: (i) AI-for-RAN, (ii) AI-and-RAN, and (iii) AI-on-RAN, making our platform an ideal candidate for advancing these areas. Moreover, the modular design of \testbed guarantees compatibility with both open-source and commercial cores, facilitating future experiments with advanced technologies like massive \gls{mimo}, mmWave, and beamforming that are currently under development.}

\section{Conclusions and Future Work}
\label{sec:conclusions}

We introduced \testbed, an open, programmable, and multi-vendor private 5G O-RAN testbed deployed at Northeastern University in Boston, MA.
We demonstrated the integration of NVIDIA Aerial, a \gls{phy} layer implementation on \glspl{gpu}, with higher layers based on \gls{oai}, resulting in the creation of the NVIDIA \gls{arc} platform.
We provided an overview of the \gls{arc} software and hardware implementations, designed for a \blue{multiple}-node deployment, including a \blue{Red Hat} OpenShift cluster for the \gls{osc} \gls{ric} deployment, \blue{as well as examples of a \gls{kpm} xApp and a slicing xApp}.
Additionally, we conducted a ray-tracing study using our digital twin framework to determine the optimal placement of \testbed \glspl{ru}.
Finally, we discussed platform performance with varying numbers of \gls{cots} \blue{and emulated} \glspl{ue} and applications, such as iPerf and video streaming, \blue{as well as through long-running and stress-test experiments to evaluate its stability}.

Next, we plan to \blue{continue} the deployment of \testbed \glspl{gnb} comprising a mix of indoor and outdoor locations for more realistic experiments and comprehensive development of \gls{ue} handover procedures.
We are targeting the integration of \glspl{ru} from different vendors and supporting bands for \gls{5g} \gls{nr} \gls{fr2}.
We will also develop pipelines for the automatic deployment\blue{, testing}, and management of workloads leveraging the Red Hat OpenShift cluster already in use for the \gls{osc} \gls{ric} integration.
Our aims include \blue{(i) deploying a fully functional and reliable private \gls{5g} network that remains continuously up and running, providing an infrastructure for users to operate on and for researchers to collect realistic datasets, and (ii) enabling full \gls{ran} control to facilitate dynamic changes in network behavior by enhancing the capabilities of \testbed, thereby} offering the research community an end-to-end \blue{open and programmable} platform for the development and testing of next-generation wireless networks and algorithms.



\footnotesize  
\bibliographystyle{IEEEtran}
\bibliography{biblio}
\balance

\vskip -1\baselineskip plus -1fil

\begin{IEEEbiography}[{\includegraphics[width=1in,height=1.25in,clip,keepaspectratio]{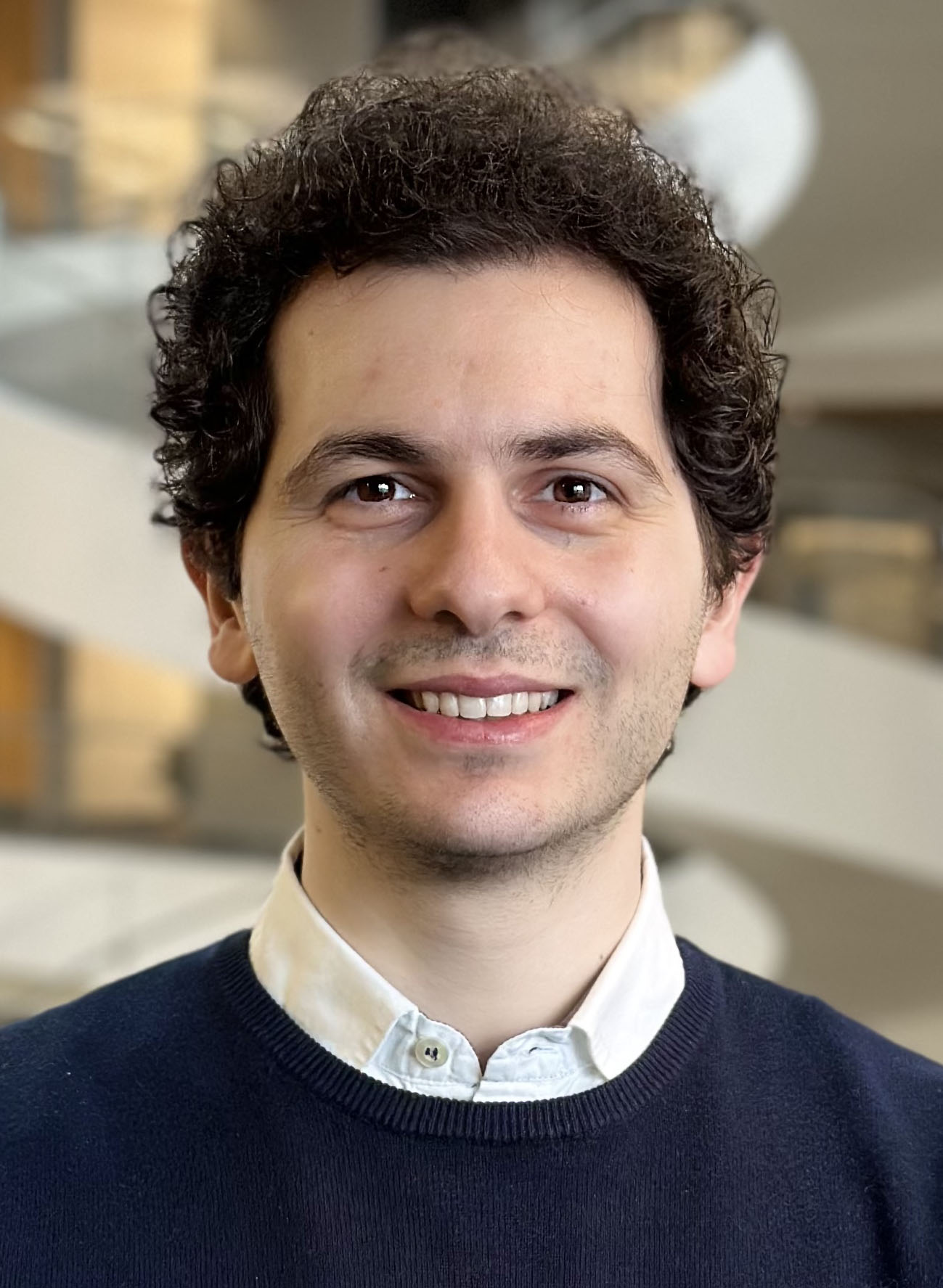}}]{Davide Villa} received his B.S. in Computer Engineering from the University of Pisa, Italy, in 2015, and his M.S. in Embedded Computing Systems from Sant’Anna School of Advanced Studies and the University of Pisa, Italy, in 2018. He worked as a Research Scientist in the Embedded Systems and Network Group at United Technologies Research Center in Cork, Ireland, from 2018 to 2020. He is currently pursuing a Ph.D. in Computer Engineering at the Institute for the Wireless Internet of Things at Northeastern University in Boston, USA. His research interests include 5G and beyond cellular networks, channel characterization for wireless systems, O-RAN, and software-defined networking for experimental wireless testbeds.
\end{IEEEbiography}

\vskip -3\baselineskip plus -1fil

\begin{IEEEbiography}[{\includegraphics[width=1in,height=1.25in,clip,keepaspectratio]{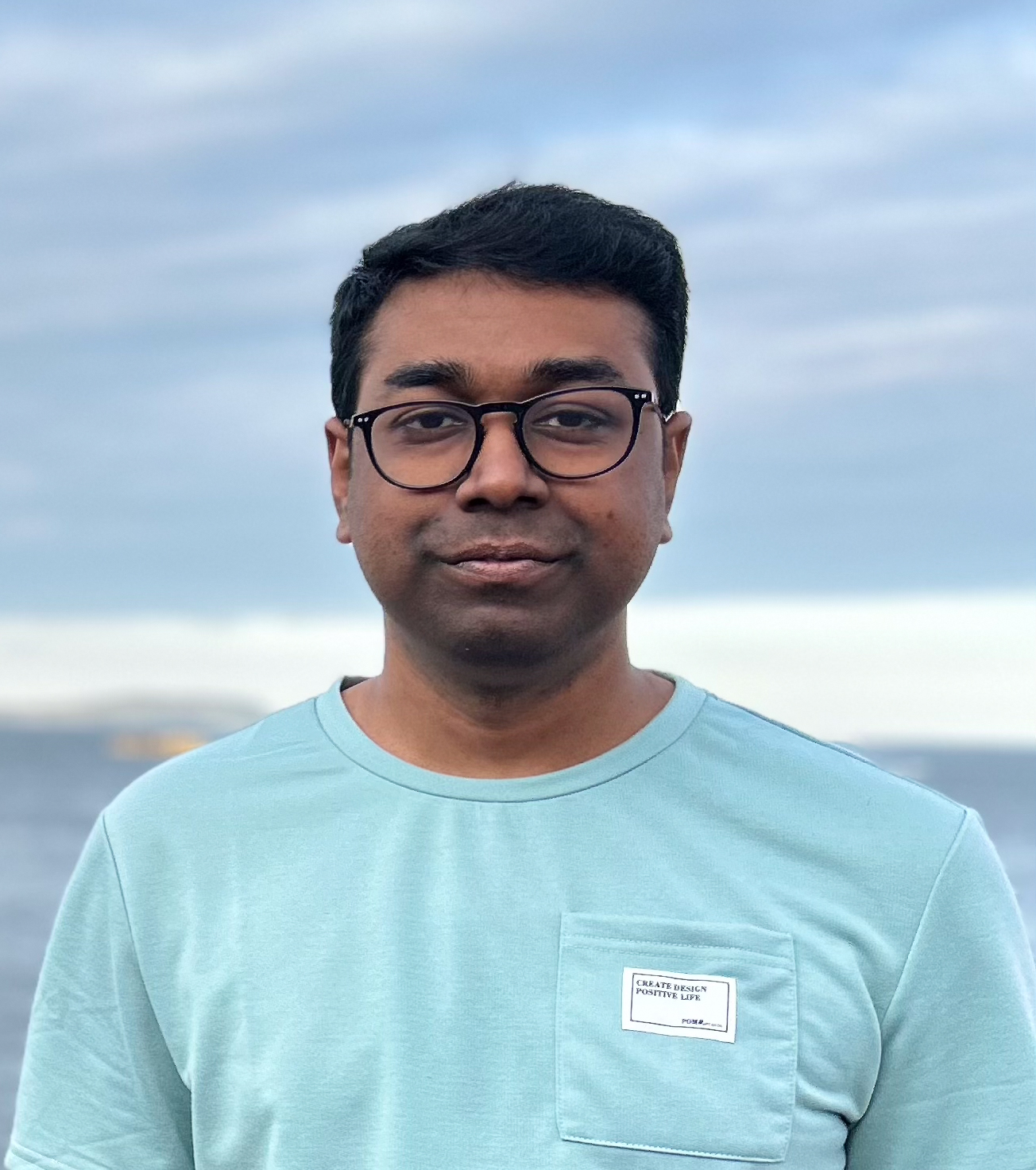}}]{Imran Khan} received his B.S. in Electrical Engineering from Bangladesh University of Engineering and Technology, in 2014, and his M.S. in Computer Engineering from Southern Illinois University of Carbondale, USA, in 2020. He is currently pursuing a Ph.D. in Computer Engineering at Northeastern University in Boston, USA. His research interest revolves around ensuring comprehensive performance, seamless mobility, and dependable reliability in 5G/6G networks.
\end{IEEEbiography}

\vskip -3\baselineskip plus -1fil

\begin{IEEEbiography}[{\includegraphics[width=1in,height=1.25in,clip,keepaspectratio]{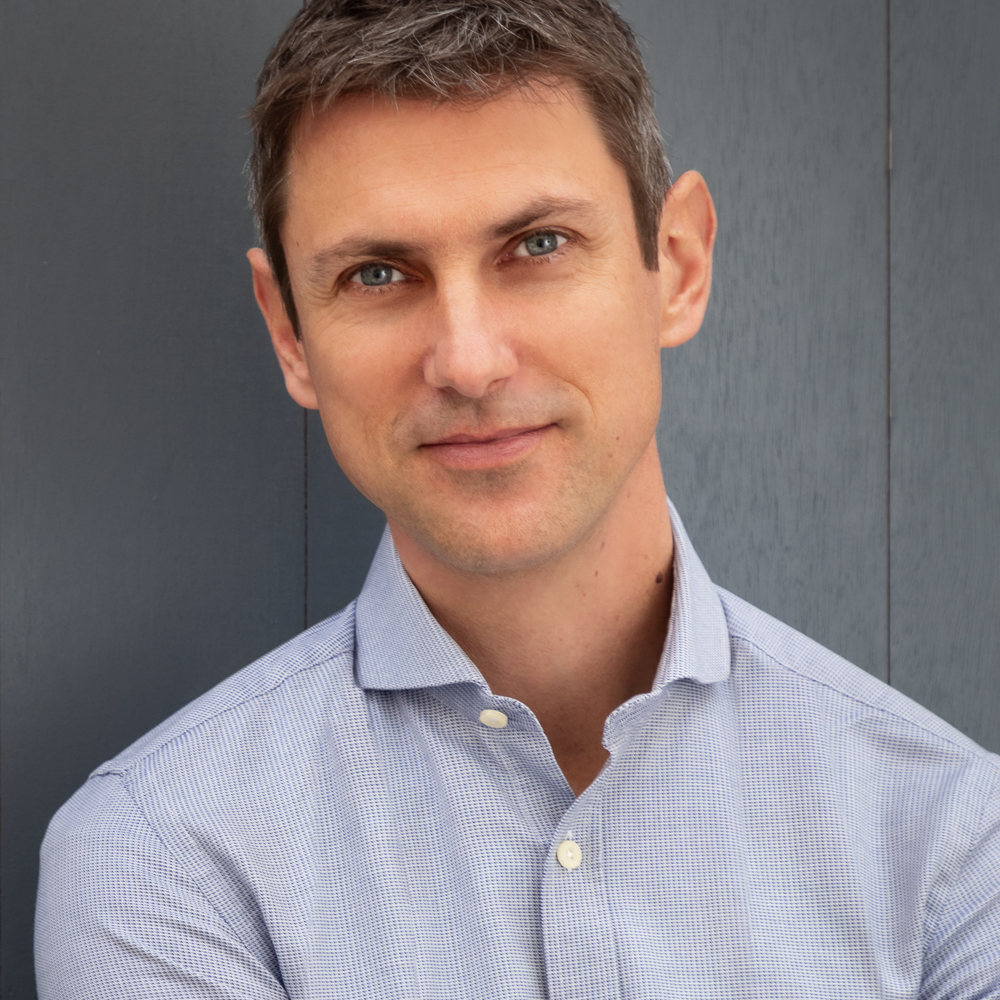}}]{Florian Kaltenberger}  is an Associate Professor in the Communication Systems department at EURECOM (France). He received his Diploma degree (Dipl.-Ing.) and his Ph.D. both in Technical Mathematics from the Vienna University of Technology in 2002 and 2007 respectively. 
He is part of the management team for the real-time open-source 5G platform OpenAirInterface.org where he is coordinating the developments of the OAI radio access network project group, which delivered support for 5G non-standalone access in 2020 and for 5G standalone access in 2021. 
Florian is currently on sabbatical at Northeastern University, where he is working on bringing different open-source communities around 5G and open RAN together to build an end-to-end reference architecture for 6G research.
 
\end{IEEEbiography}

\vskip -3\baselineskip plus -1fil


\begin{IEEEbiographynophoto}{Nicholas Hedberg} is a Senior Engineer in the Public Sector at NVIDIA in Zurich, Switzerland, having joined in October 2021. He brings extensive experience from his previous tenure at Viasat Inc., where he was a System Engineering Team Lead in Lausanne, focusing on cutting-edge phased array antennas for satellite communications. Prior roles at Viasat in Carlsbad involved significant contributions to mobile terminal Verilog modules and ASIC development. Nicholas holds a BS in Physics and a BA in Economics from UC San Diego (2003-2007).
\end{IEEEbiographynophoto}

\vskip -2\baselineskip plus -1fil

\begin{IEEEbiography}[{\includegraphics[width=1in,height=1.25in,clip,keepaspectratio]{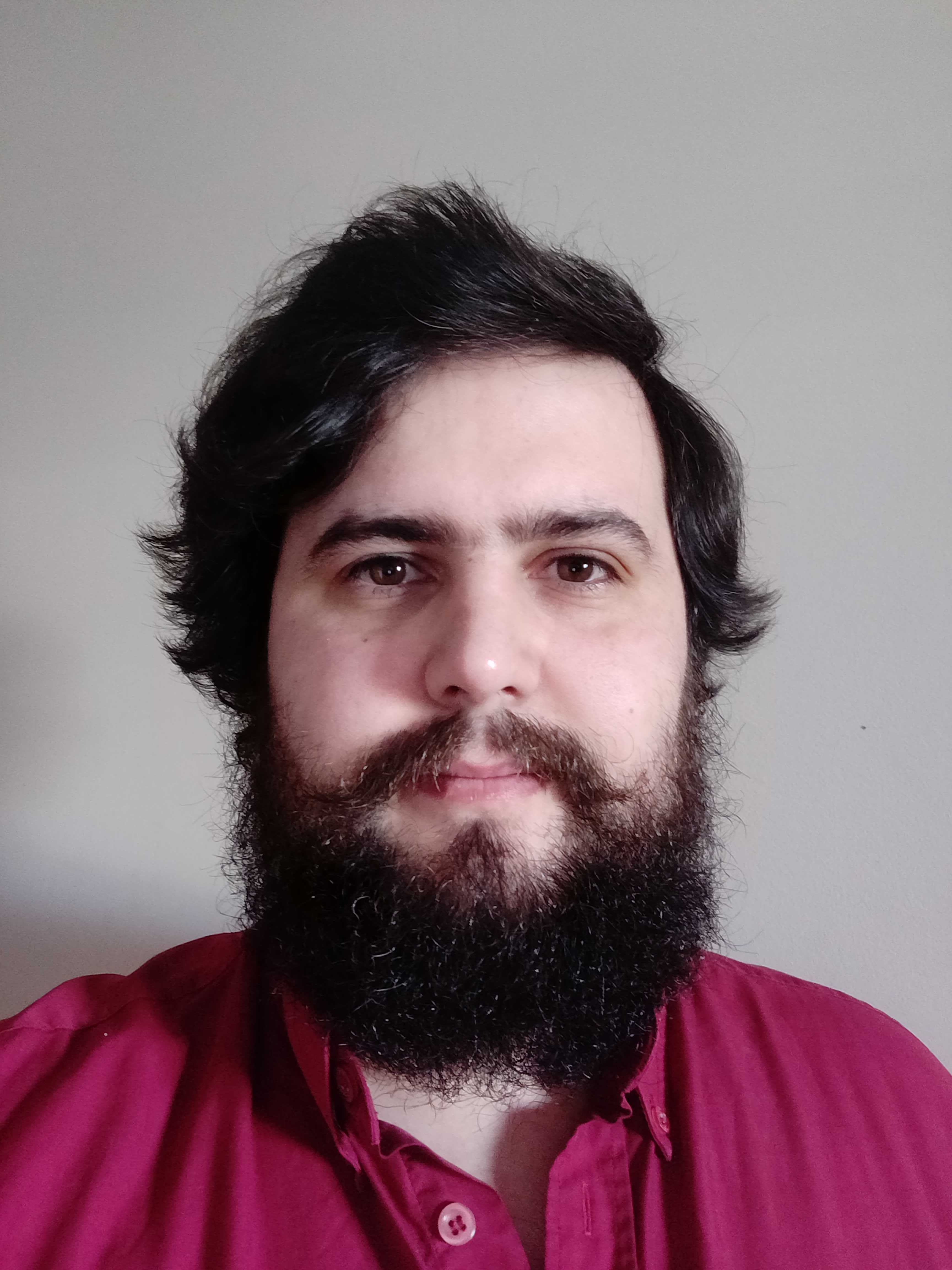}}]{R\'{u}ben Soares da Silva} received his degree in Computer Engineering from Castelo Branco Polytechnic Institute (IPCB) in 2018. Having joined Allbesmart in 2021, began work for OpenAirInterface Software Alliance in integrating the OAI L2 with the NVIDIA Aerial L1 following Small Cell Forums' FAPI standard, and since continued development in support of the FAPI split, as well as continued support for the deployments using this L2/L1 integration.

\end{IEEEbiography}

\vskip -3\baselineskip plus -1fil

\begin{IEEEbiography}[{\includegraphics[width=1in,height=1.25in,clip,keepaspectratio]{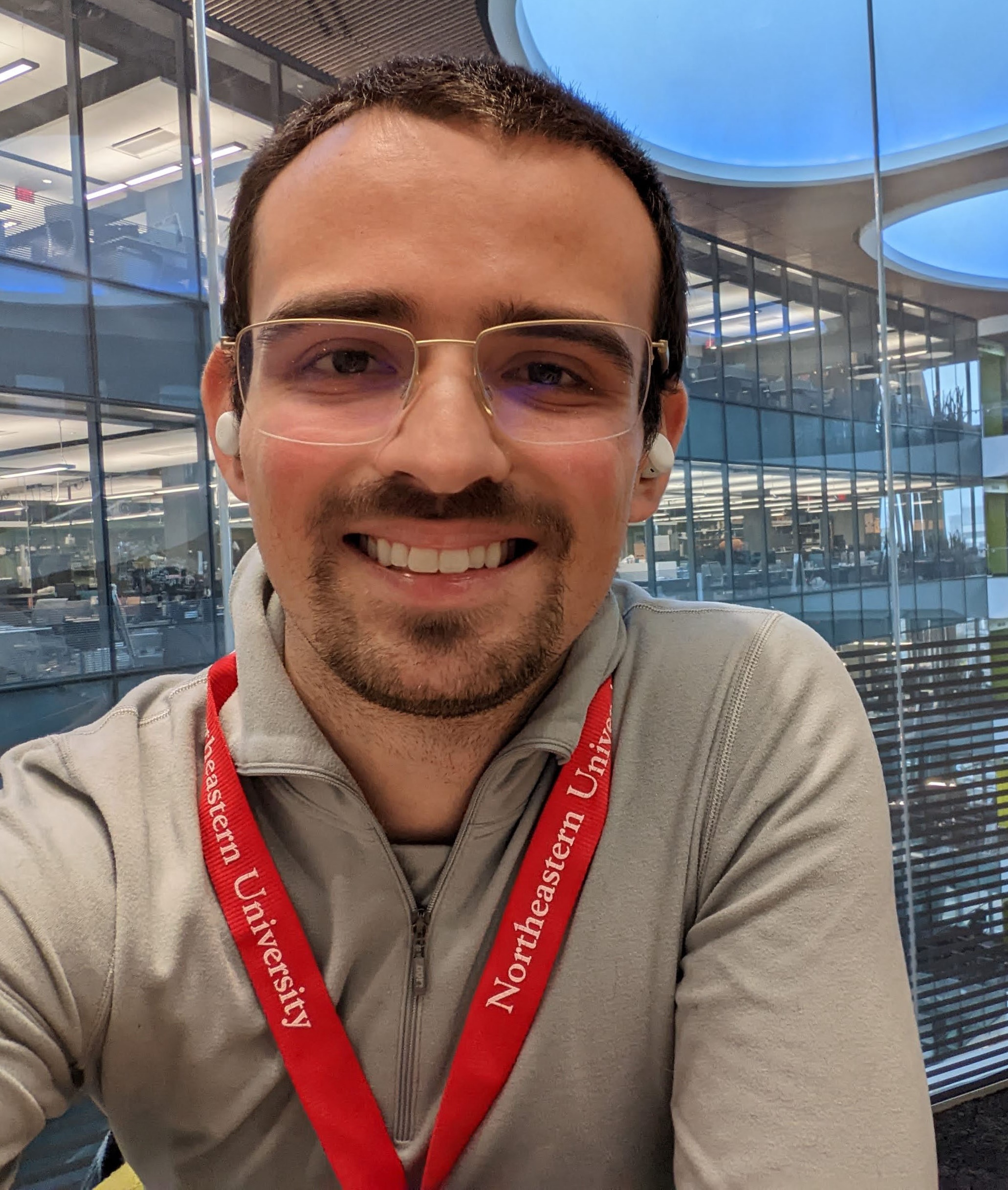}}]{Stefano Maxenti} is a Ph.D. Candidate in Computer Engineering at the Institute for the Wireless Internet of Things (WIoT) at Northeastern University, under Prof. Tommaso Melodia. He received a B.Sc. in Engineering of Computing Systems in 2020 and a M.Sc. in Telecommunication Engineering in 2023 from Politecnico di Milano, Italy. His research is linked with AI applications for wireless communications and orchestration, integration, and automation of O-RAN networks
\end{IEEEbiography}

\vskip -3\baselineskip plus -1fil

\begin{IEEEbiography}[{\includegraphics[width=1in,height=1.25in,clip,keepaspectratio]{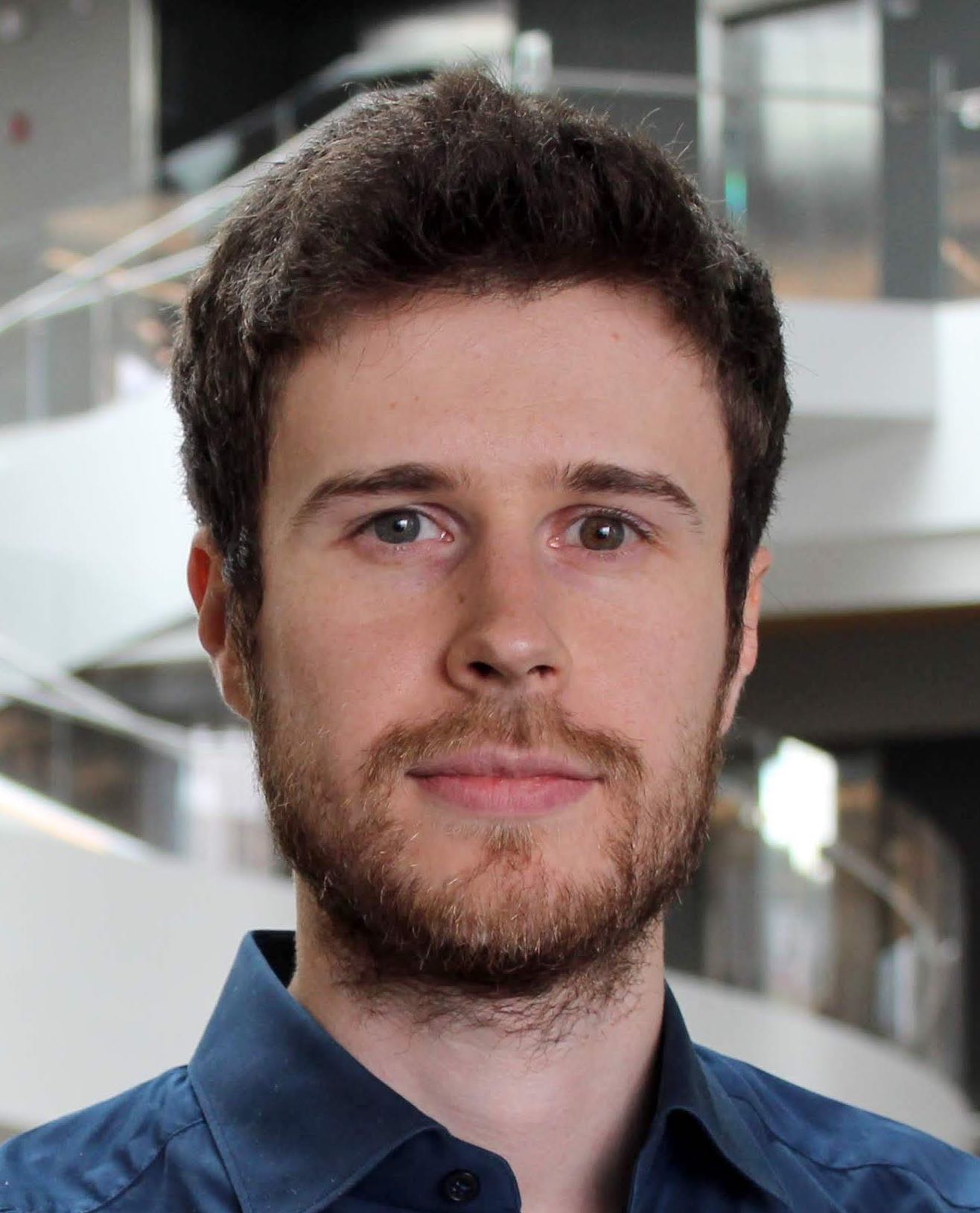}}]{Leonardo Bonati} is an Associate Research Scientist at the Institute for the Wireless Internet of Things, Northeastern University, Boston, MA. He received a Ph.D. degree in Computer Engineering from Northeastern University in 2022. His main research focuses on softwarized approaches for the Open Radio Access Network (RAN) of the next generation of cellular networks, on O-RAN-managed networks, and on network automation and orchestration. He served as guest editor of the special issue of Elsevier's Computer Networks Journal on Advances in Experimental Wireless Platforms and Systems.
\end{IEEEbiography}

\vskip -3\baselineskip plus -1fil

\begin{IEEEbiography}[{\includegraphics[width=1in,height=1.25in,clip,keepaspectratio]{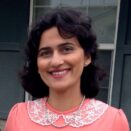}}]{Anupa Kelkar} is the product manager for NVIDIA 5G Aerial converged radio access and edge compute platform. She has over 20 years of telecommunications and networking industry experience in software engineering and product management leadership spanning wireline, wireless, and satellite networks. Before NVIDIA, Anupa worked at Apple in incubation for AR/VR low-latency on-device, cloud, and edge-use cases, at Qualcomm in the connected cars (CV2X) ecosystem. And if you ever fly JetBlue, United, Continental, or Quantas, she was instrumental in the inflight broadband connectivity over satellite networks. Anupa graduated from University of California, Berkeley in electrical engineering and computer science.
\end{IEEEbiography}

\vskip -3\baselineskip plus -1fil

\begin{IEEEbiography}[{\includegraphics[width=1in,height=1.25in,clip,keepaspectratio]{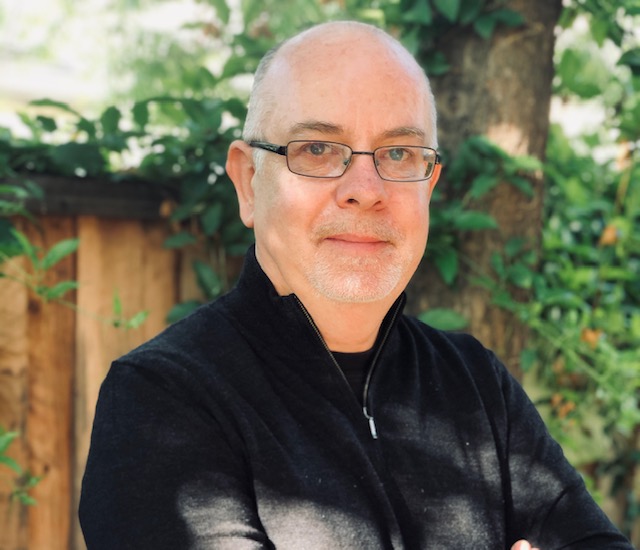}}]{Chris Dick} joined NVIDIA in 2020 where he is a system architect working on the application of Artificial Intelligence and Machine Learning to 5G and 6G wireless.  
In his 30 years working in signal processing and communications he has delivered silicon and software products for 3G, 4G, and 5G baseband DSP and Docsis 3.1 cable access and vector processor architectures. He has performed research and delivered products for digital front-end (DFE) technology for cellular systems with a particular emphasis on digital pre-distortion for power amplifier linearization. Chris has also worked extensively on silicon architecture and compilers for machine learning and parallel computing architectures. 
Prior to moving to Silicon Valley in 1998, he was a tenured academic in Melbourne Australia for 13 years. He has over 250 publications and 100 patents. From 1998 to 2020 he was a Fellow and the DSP Chief Architect at Xilinx.
In 2018 he was awarded the IEEE Communications Society Award for Advances in Communication for research in the area of full-duplex wireless communication.
\end{IEEEbiography}

\vskip -3\baselineskip plus -1fil

\begin{IEEEbiography}[{\includegraphics[width=1in,height=1.25in,clip,keepaspectratio]{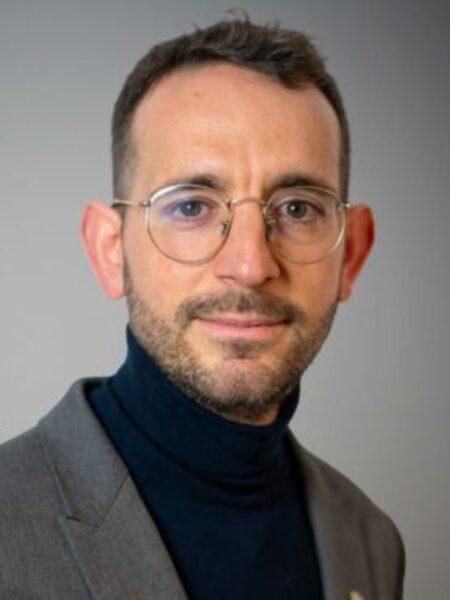}}]{Eduardo Baena} is a postdoctoral research fellow at Northeastern University. Holding a Ph.D. in Telecommunication Engineering from the University of Malaga, his experience spans various roles within the international private sector from 2010 to 2017. Later he joined UMA as a lecturer and researcher contributing to several H2020 projects and as a Co-IP of national and regional funded projects. 
\end{IEEEbiography}

\vskip -3\baselineskip plus -1fil

\begin{IEEEbiography}[{\includegraphics[width=1in,height=1.25in,clip,keepaspectratio]{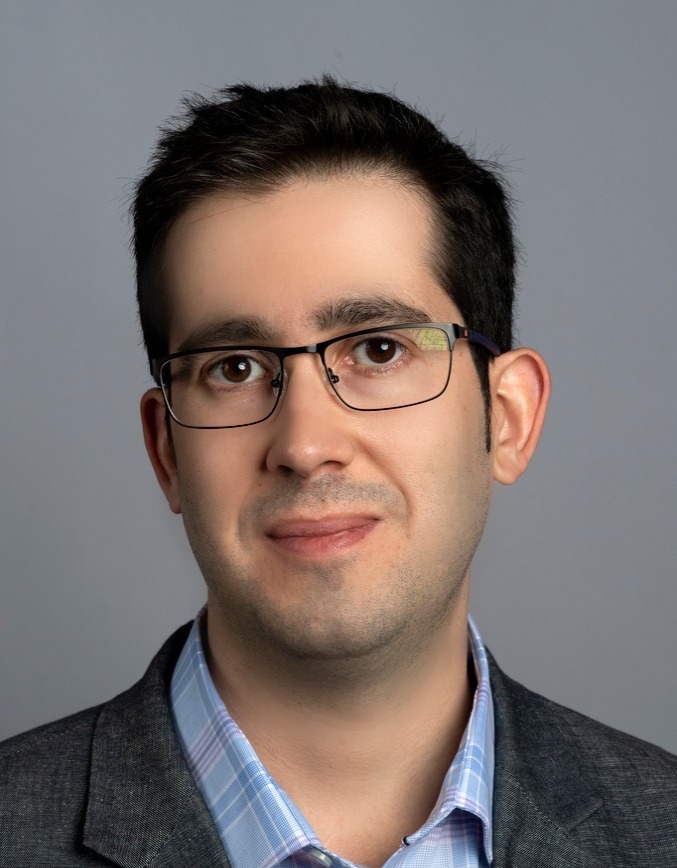}}]{Josep M. Jornet} (M'13--SM'20--F'24) is a Professor in the Department of Electrical and Computer Engineering, the director of the Ultrabroadband Nanonetworking (UN) Laboratory, and the Associate Director of the Institute for the Wireless Internet of Things at Northeastern University (NU). He received his Ph.D. degree in Electrical and Computer Engineering from the Georgia Institute of Technology, Atlanta, GA, in August 2013. He is a leading expert in terahertz communications, in addition to wireless nano-bio-communication networks and the Internet of Nano-Things. In these areas, he has co-authored over 250 peer-reviewed scientific publications, including one book, and has been granted five US patents. His work has received over 17,000 citations (h-index of 61 as of June 2024). He is serving as the lead PI on multiple grants from U.S. federal agencies including the National Science Foundation, the Air Force Office of Scientific Research, and the Air Force Research Laboratory as well as industry. He is the recipient of multiple awards, including the NSF CAREER Award in 2019, the 2022 IEEE ComSoc RCC Early Achievement Award, and the 2022 IEEE Wireless Communications Technical Committee Outstanding Young Researcher Award, among others, as well as four best paper awards. He is a Fellow of the IEEE and an IEEE ComSoc Distinguished Lecturer (2022-2024). He is also the Editor-in-Chief of the Elsevier Nano Communication Networks journal and Editor for IEEE Transactions on Communications and Nature Scientific Reports.
\end{IEEEbiography}

\vskip -3\baselineskip plus -1fil

\begin{IEEEbiography}[{\includegraphics[width=1in,height=1.25in,clip,keepaspectratio]{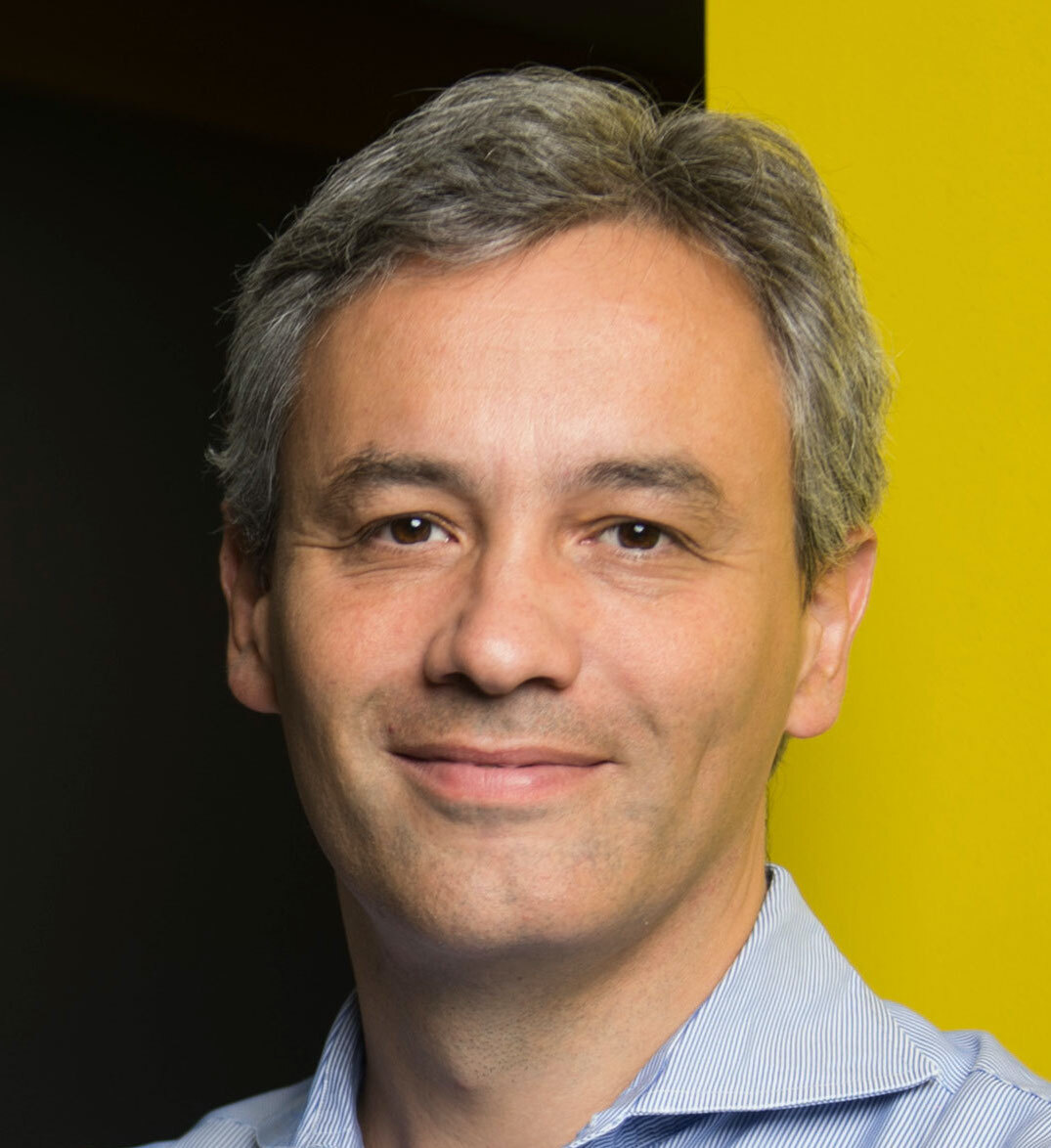}}]{Tommaso Melodia}
is the William Lincoln Smith Chair Professor with the Department of Electrical and Computer Engineering at Northeastern University in Boston. He is also the Founding Director of the Institute for the Wireless Internet of Things and the Director of Research for the PAWR Project Office. He received his Ph.D. in Electrical and Computer Engineering from the Georgia Institute of Technology in 2007. He is a recipient of the National Science Foundation CAREER award. Prof. Melodia has served as Associate Editor of IEEE Transactions on Wireless Communications, IEEE Transactions on Mobile Computing, Elsevier Computer Networks, among others. He has served as Technical Program Committee Chair for IEEE INFOCOM 2018, General Chair for IEEE SECON 2019, ACM Nanocom 2019, and ACM WUWnet 2014. Prof. Melodia is the Director of Research for the Platforms for Advanced Wireless Research (PAWR) Project Office, a \$100M public-private partnership to establish 4 city-scale platforms for wireless research to advance the US wireless ecosystem in years to come. Prof. Melodia's research on modeling, optimization, and experimental evaluation of Internet-of-Things and wireless networked systems has been funded by the National Science Foundation, the Air Force Research Laboratory the Office of Naval Research, DARPA, and the Army Research Laboratory. Prof. Melodia is a Fellow of the IEEE and a Distinguished Member of the ACM.
\end{IEEEbiography}

\vskip -3\baselineskip plus -1fil


\begin{IEEEbiography}[{\includegraphics[width=1in,height=1.1in,clip,keepaspectratio]{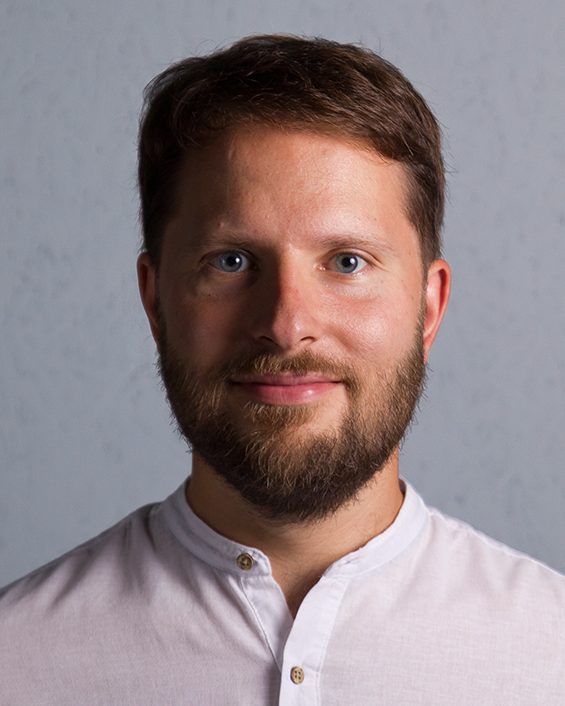}}]{Michele Polese} is a Research Assistant Professor at the Institute for the Wireless Internet of Things, Northeastern University, Boston, since October 2023. He received his Ph.D. at the Department of Information Engineering of the University of Padova in 2020. He then joined Northeastern University as a research scientist and part-time lecturer in 2020. During his Ph.D., he visited New York University (NYU), AT\&T Labs in Bedminster, NJ, and Northeastern University.
His research interests are in the analysis and development of protocols and architectures for future generations of cellular networks (5G and beyond), in particular for millimeter-wave and terahertz networks, spectrum sharing and passive/active user coexistence, open RAN development, and the performance evaluation of end-to-end, complex networks. He has contributed to O-RAN technical specifications and submitted responses to multiple FCC and NTIA notice of inquiry and requests for comments, and is a member of the Committee on Radio Frequency Allocations of the American Meteorological Society (2022-2024). He is PI and co-PI in research projects on 6G funded by the NTIA, the O-RAN ALLIANCE, U.S. NSF, OUSD, and MassTech Collaborative, and was awarded with several best paper awards and the 2022 Mario Gerla Award for Research in Computer Science. Michele is serving as TPC co-chair for WNS3 2021-2022, as an Associate Technical Editor for the IEEE Communications Magazine, as a Guest Editor in an IEEE JSAC Special Issue on Open RAN, and has organized the Open 5G Forum in Fall 2021 and the NextGenRAN workshop at Globecom 2022.
\end{IEEEbiography}

\vskip -3\baselineskip plus -1fil

\begin{IEEEbiography}[{\includegraphics[width=1in,height=1.25in,clip,keepaspectratio]{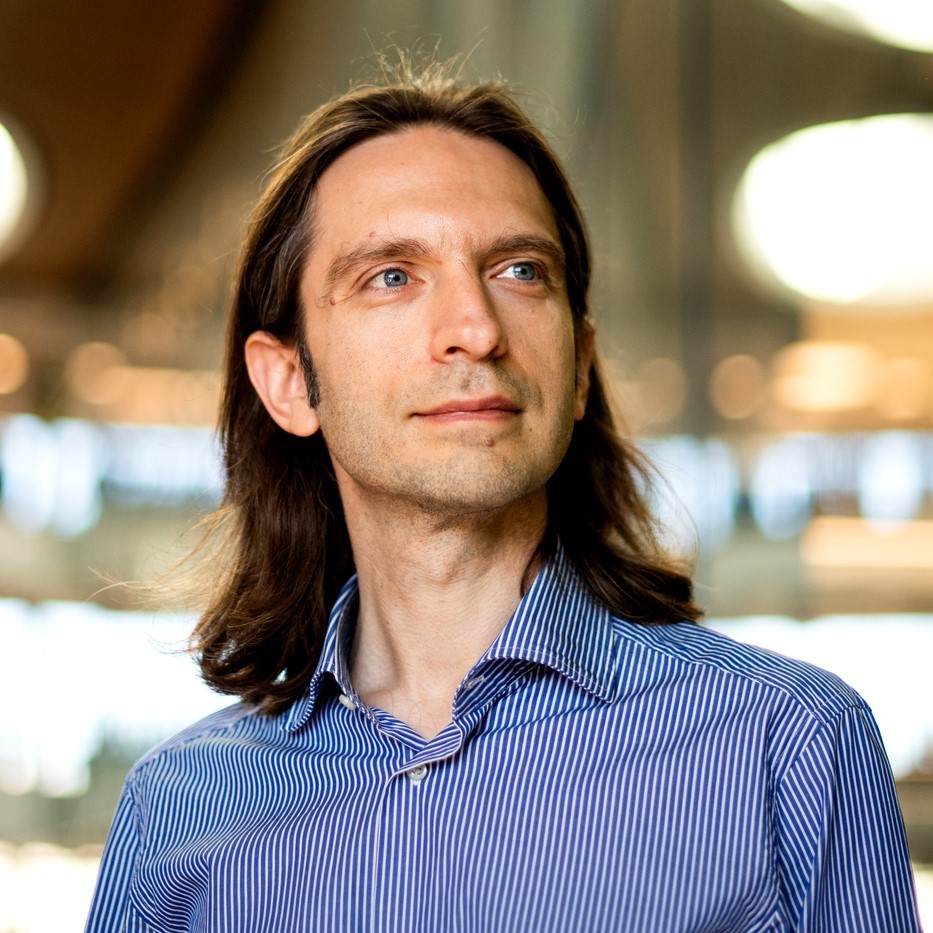}}]{Dimitrios Koutsonikolas} is an Associate Professor in the Department of Electrical and Computer Engineering and a member of the Institute for the Wireless Internet of Things at Northeastern University. Between January 2011 and December 2020, he was in the Computer Science and Engineering Department at the University at Buffalo, first as an Assistant Professor (2011-2016) and then as an Associate Professor (2016-2020) and Director of Graduate Studies (2018-2020). He received his PhD in Electrical and Computer Engineering from Purdue University in 2010. 

His research interests are broadly in experimental wireless networking and mobile computing, with a current focus on 5G networks and latency-critical applications (AR, VR, CAVs) over 5G, millimeter-wave networking, and energy-aware protocol design for smartphones. He has served as the General Co-Chair for IEEE LANMAN 2024, IEEE WoWMoM 2023, and ACM EWSN 2018, and TPC Co-Chair for IEEE LANMAN 2023, IEEE HPSR 2023, IEEE DCOSS 2022, IEEE WoWMoM 2021, and IFIP Networking 2021. He received the IEEE Region 1 Technological Innovation (Academic) Award in 2019 and the NSF CAREER Award in 2016. He is a senior member of the IEEE and the ACM and a member of USENIX.

\end{IEEEbiography}

\end{document}